\documentclass[12pt]{article}

\usepackage{latexsym}
\usepackage{amsmath}
\usepackage{amssymb}
\usepackage{amsfonts}
\usepackage{mathrsfs}
\usepackage{lscape}
\usepackage{dsfont}
\usepackage{lmodern}
\usepackage[colorlinks=true,linkcolor=black,citecolor=black,urlcolor=black,filecolor=black]{hyperref}

 \csname
@addtoreset\endcsname{equation}{section}

\def\a{\alpha}
\def\b{\beta}
\def\g{\gamma}

\def\d{\delta}

\def\e{\epsilon}

\def\h{\eta}
\def\th{\theta}

\def\l{\lambda}
\def\L{\Lambda}
\def\m{\mu}
\def\n{\nu}
\def\x{\xi}

\def\p{\pi}

\def\r{\rho}

\def\s{\sigma}

\def\f{\phi}

\def\vf{\varphi}

\def\o{\omega}

\def\cL{{\cal L}}
\def\cM{{\cal M}}
\def\cN{{\cal N}}
\def\cO{{\cal O}}

\def\cQ{{\cal Q}}

\def\cU{{\cal U}}

\def\cW{{\cal W}}

\def\be{\begin{equation}}
\def\ee{\end{equation}}
\def\bali{\begin{align}}
\def\eali{\end{align}}
\def\bea{\begin{eqnarray}}
\def\eea{\end{eqnarray}}
\def\ba{\begin{array}}
\def\ea{\end{array}}
\def\ben{\begin{enumerate}}
\def\een{\end{enumerate}}
\def\bi{\begin{itemize}}
\def\ei{\end{itemize}}

\def\nn{\nonumber}

\def\tr{\text{tr}}

\def\ww{\wedge}

\def\bra{\langle \,}
\def\ket{\, \rangle}

\def\pe{\prime}
\def\12{\frac{1}{2}}
\def\pr{\partial}

\newcommand{\bin}[2]{\binom{#1}{#2}}

\begin{document}

\begin{flushright}
AEI-2011-041
\end{flushright}

\vspace{20pt}

\begin{center}


{\Large\bf Asymptotic $\cW$-symmetries in three-dimensional \\[10pt]higher-spin gauge theories}


\vspace{25pt}
{\sc A.~Campoleoni, S.~Fredenhagen and S.~Pfenninger}

\vspace{10pt}
{\sl\small
Max-Planck-Institut f{\"u}r Gravitationsphysik\\
Albert-Einstein-Institut\\
Am M{\"u}hlenberg 1\\
14476 Golm,\ GERMANY\\[5pt]

\vspace{10pt}

{\it andrea.campoleoni@aei.mpg.de, stefan.fredenhagen@aei.mpg.de,\\  stefan.pfenninger@aei.mpg.de} 
}

\vspace{70pt} {\sc\large Abstract}\end{center}

We discuss how to systematically compute the asymptotic symmetry
algebras of generic three-dimensional bosonic higher-spin gauge theories
in backgrounds that are asymptotically AdS. We apply these techniques
to a one-parameter family of higher-spin gauge theories that can be
considered as large $N$ limits of $SL(N) \times SL(N)$ Chern-Simons
theories, and we provide a closed formula for the structure constants
of the resulting infinite-dimensional non-linear $\cW$-algebras. Along
the way we provide a closed formula for the structure constants of all
classical $\cW_N$ algebras. In both examples the higher-spin
generators of the $\cW$-algebras are Virasoro primaries. We eventually discuss how to relate our basis
to a non-primary quadratic basis that was previously discussed in literature.

\newpage


{\linespread{0.8}
\tableofcontents}

\section{Introduction}\label{sec:intro}

In a three-dimensional space-time the little group of massless
particles does not admit representations with arbitrary
helicity. Nevertheless, one can still consider the analogues of the
field equations that for $D > 3$ describe the free propagation of
spin-$s$ massless particles, although in $D=2+1$ they do not propagate
any local degree of freedom for $s > 1$. Any non-linear completion of
these field equations defines a three-dimensional higher-spin (HS)
gauge theory, at least as a classical theory. The first member of this
class is just Einstein gravity, which does not admit wave solutions in
$D=2+1$. Despite the lack of propagating degrees of freedom, these toy
models already display many features of their higher-dimensional
counterparts. They thus often provide a manageable testing ground for
various ideas on field theories involving higher spins.

A marked simplification arises when one considers pure HS gauge
theories without matter: in $D=2+1$ the coupling of massless
higher-spin fields to gravity can be described by a Chern-Simons (CS)
action for any value of the cosmological constant \cite{Blencowe}. On
the other hand, in higher space-time dimensions the non-linear field
equations of Vasiliev \cite{vas-int} require a non-vanishing
cosmological constant (see \cite{vas-rev,exact_review} for a review),
while on flat backgrounds a classification of cubic vertices is now
available \cite{vertices}, but a complete interacting theory is still
lacking (see \cite{rev-int} for an account of the state of the art).

Even if it is not needed to handle a full interacting theory, a
negative cosmological constant plays an important role in $D=2+1$. In
the gravitational sector it allows for a richer space of solutions,
still without gravitational waves but with black holes \cite{BTZ}. It
also allows to build three-dimensional counterparts of Vasiliev's
models \cite{vas-prok}, describing the coupling of two or four scalars
to a HS gauge sector. The latter toy models provide a natural arena to
test -- with suitable extrapolations -- various proposals that relate
holographically Vasiliev's theory to conformal field theories. This
interplay was first investigated by Sezgin and Sundell \cite{Per} (see
also \cite{pre_AdS/CFT} for earlier suggestions), while recently
various groups independently elaborated upon a conjecture by Klebanov
and Polyakov, that links Vasiliev's theory on $AdS_4$ to the large $N$
limit of the three-dimensional critical $O(N)$ vector model
\cite{KP,AdS/CFT}.

A first step toward the characterisation of possible two-dimensional
CFT duals was performed in \cite{HR,spin3}, with an analysis of the
asymptotic symmetries of some three-dimensional classical HS gauge
theories. The outcome generalises an earlier result by Brown and
Henneaux \cite{BH}, that defined the class of asymptotically Anti de
Sitter solutions of three-dimensional Einstein's equations such that it
contains all physically relevant solutions and all its elements have
finite boundary charges. In $D=2+1$ these conditions allow for an
enhancement of asymptotic symmetries from the AdS algebra to two
copies of a centrally extended Virasoro algebra, with a central charge
that grows with the AdS radius. Adding massless higher-spin fields
maintains the conformal symmetry and actually extends it: each
Virasoro algebra is replaced by a centrally extended non-linear
$\cW$-algebra, with the same central charge as in pure gravity (see
\cite{W} for an introduction to $\cW$-algebras). The classical
asymptotic $\cW$-symmetries of \cite{HR,spin3} were then shown to
survive even at the quantum level in \cite{quantum_W}, and these
observations led Gaberdiel and Gopakumar to conjecture in
\cite{minimal} a duality between suitable large $N$ limits of minimal
model CFT's with $\cW_N \times \cW_N$ symmetry and the Vasiliev-like
models of \cite{vas-prok} (see also \cite{GH,minimal2,Yin} for some
recent additional checks of this conjecture and
\cite{Kiritsis,minimal_ext} for its extension to other classes of
minimal models).

In \cite{HR,spin3,GH} the identification of asymptotic symmetries
rests heavily upon the CS formulation of the HS dynamics, somehow
following the derivation of \cite{Henneaux_vanDriel} of the original
Brown-Henneaux result. On the one hand, the CS action enables one to
define quite straightforwardly the boundary charges \cite{Banados}. On
the other hand, CS theories defined on a manifold with boundary admit
a non-trivial boundary dynamics, generically described by a
Wess-Zumino-Witten (WZW) action (see, for instance,
\cite{Banados_review} and references therein). As a result, the
Poisson structure on the phase space of boundary excitations is
generically an affine algebra. However, even if the action of
three-dimensional HS gauge theories can be cast in a CS form, not all
solutions of the CS theory are admissible classical higher-spin
configurations. Selecting the class of asymptotically AdS solutions
imposes a constraint on the phase space of the boundary theory and
$\cW$-algebras emerge as Dirac-bracket algebras on the constrained
phase space.

In the mathematical literature this way of constructing classical
$\cW$-algebras out of affine algebras is known as Drinfeld-Sokolov
(DS) reduction (see \cite{Dickey} for a review), and it was first
applied to WZW theories at the end of the eighties
\cite{principal}. It associates a classical centrally extended
$\cW$-algebra to any semisimple Lie algebra, independent of whether it
is the gauge algebra of a sensible toy model for higher-spin
interactions or not. In this paper we propose a procedure to compute
the structure constants of any $\cW$-algebra that can be obtained by a
DS reduction, and we apply it to a class of algebras that are relevant
to the study of higher spins.

In Section \ref{sec:reduction} we begin by recalling the interplay
between the DS reduction and the asymptotic symmetries of
three-dimensional HS gauge theories. Then we present our procedure to
``reduce'' a generic affine algebra. We eventually use this tool to
shed light on some general properties of the resulting $\cW$-algebras
and to analyse some examples that lay outside of the class of algebras
considered in the following sections. Let us already mention that the
DS construction rests upon a gauge choice: different choices lead to
different bases for the $\cW$-algebra. Our analysis is focused on the
so called ``highest-weight gauge'', thus providing an alternative to
similar results that were previously obtained in the so called
``$U$-gauge'' (see chapter 9 of \cite{Dickey}). The highest-weight
choice gives a $\cW$-algebra where all generators are primaries with
respect to the lowest-spin ones. In the absence of spin-$1$ generators
all of them are thus Virasoro primaries. In the general case one can
easily recover a basis with this property by shifting the Virasoro
current with the Sugawara energy-momentum tensor built from spin-1
currents.

In Section \ref{sec:W} we return to three-dimensional HS gauge
theories. Besides them there is a one-parameter family that plays a
distinguished role, and we focus on it. It was first discussed in
\cite{BBS,Pope,Vasiliev,FL,hoppe_proof} in the late eighties, when the
subject was in its infancy. All its members are bosonic theories
describing, for generic values of a parameter $\l$, the coupling to
gravity of a set of massless fields where each integer spin from 3
onwards appears once. The interest in this family is twofold: on the
one hand, for integer $\l = N$ the CS action reduces to that of a
$SL(N) \times SL(N)$ theory, while all other values of $\l$ provide a
sort of large $N$ limit of this rather natural class of toy models for
HS interactions, out of which the case $N = 2$ coincides with Einstein
gravity \cite{Townsend,Witten}. On the other hand, whenever no
truncations arise, the field content is the same as in the gauge
sector of Vasiliev's models, that are actually built upon the same
gauge algebras \cite{vas-prok}. The corresponding $\cW$-algebras are
thus expected to emerge also in the large $N$ limit of $\cW_N \times
\cW_N$ minimal models, as discussed in \cite{GH,minimal2,Yin}.

Even if for different values of $\l$ the field contents coincide, the
gauge algebras are inequivalent \cite{BBS,hoppe_proof}. Therefore, as
already pointed out in \cite{GH}, different $\l$ lead to inequivalent
asymptotic symmetries. These are given by two copies of an
infinite-dimensional non-linear $\cW$-algebra, that we denote by
$\cW_\infty[\l]$ as in \cite{GH}. This family of $\cW$-algebras was
introduced independently in \cite{W_infinity} and \cite{W_inf_math},
in a non-primary basis with at most quadratic non-linearities
appearing in the Poisson brackets. In Section \ref{sec:W} we use the
machinery developed in Section \ref{sec:reduction} to provide a closed
formula for all structure constants of $\cW_\infty[\l]$ in a
Virasoro-primary basis. Our formula reproduces the results for the
first few spins that were computed in \cite{GH}. Setting $\l = N$ also
gives a closed formula for the structure constants of $\cW_N$ in a
Virasoro-primary basis. In Section \ref{sec:quadratic} we eventually
discuss how to relate systematically our presentation of
$\cW_\infty[\l]$ to the quadratic basis of \cite{W_infinity}. The
paper closes with a summary of our results and some
appendices. Appendix~\ref{app:pope} summarises the structure constants
of the one-parameter family of higher spin algebras that we
consider. In Appendix~\ref{app:proofs} we collected the proofs of some
formulae appearing in the main text, in particular the formula for the
structure constants of $\cW_\infty[\l]$. Finally, in
Appendix~\ref{app:nonPrincipalExp} we display the Poisson brackets of the
examples of $\cW$-algebras that we discuss in the main body of the
text.

Let us finally stress that the characterisation of asymptotically AdS
solutions that triggers our analysis goes beyond the identification of
proper fall-off conditions at spatial infinity. Rather, it selects
exact solutions of the field equations. The study of exact solutions
is another interesting arena where one may take advantage of the
simplicity of three-dimensional toy models to extract information on
their higher-dimensional counterparts. Exact solutions of Vasiliev's
models in $D > 3$ were first obtained in \cite{exact} (see
\cite{exact_review} for a review), while recently solutions displaying
various similarities with gravity black holes were presented in
\cite{sol_gen}. On the three-dimensional side various issues on exact
solutions were discussed in
\cite{vas-prok,BTZ_hs,spin3,Maloney,GK,GK_new}. In Section
\ref{sec:W_metric} we discuss how one could extend an earlier proposal
of \cite{spin3} in order to express our exact solutions in an
alternative form involving only Lorentz-invariant metric-like fields
(see \cite{rev-symm} for a review of the metric-like
formulation). Aside from making more transparent the identification
between CS theories and HS gauge theories, we hope that this interplay
between alternative approaches provides useful tools to better
understand the exact solutions already discussed in literature, such
as the intriguing HS generalisations of gravity black holes
\cite{sol_gen,BTZ_hs,GK,GK_new}.

\section{Asymptotic symmetries from Drinfeld-Sokolov reduction}\label{sec:reduction}

A three-dimensional pure higher-spin (HS) gauge theory coupled to
gravity in backgrounds that are asymptotically AdS can be described by
a Chern-Simons (CS) theory supplemented by suitable boundary
conditions \cite{HR,spin3}. These translate into the Drinfeld-Sokolov
(DS) constraint on the centrally extended loop algebra that
appears on the boundary of a CS theory. Therefore, asymptotic
symmetries are described by the $\cW$-algebras that arise from the DS
reduction. In this section we first review the DS reduction in the
context of HS gauge theories. We then provide an algorithm to perform
it in the highest-weight gauge, from which one obtains $\cW$-algebras
in a basis where all fields are primaries with respect to the lowest
spin ones.

\subsection{Higher-spin gauge theories in $D=2+1$}\label{sec:red_bulk}

In $D=2+1$ Einstein gravity with a negative cosmological constant is equivalent to a $SL(2,\mathbb{R}) \times SL(2,\mathbb{R})$ Chern-Simons theory \cite{Townsend,Witten}. In fact, up to boundary terms, one can rewrite the Einstein-Hilbert action  as
\be \label{EH}
S \,=\, S_{CS}[A] \,-\, S_{CS}[\widetilde{A}] \, ,
\ee
with
\be \label{CS}
S_{CS}[A] \,=\, \frac{k}{4\pi} \int \tr \left(\, A \ww dA \,+\, \frac{2}{3}\, A \ww A \ww A \,\right) .
\ee
The fields $A$ and $\widetilde{A}$ are $sl(2,\mathbb{R})$-valued differential forms so that, for instance, $A = A_\m{}^i\, J_i\, dx^\m$, where the $J_i$ generate the $sl(2,\mathbb{R})$ algebra. We normalise the invariant form entering the CS action such that
\be
\tr \left(J_i J_j\right) \,=\, \frac{1}{2}\, \h_{\,ij} \quad \Rightarrow \quad k \,=\, \frac{l}{4G} \ ,
\ee 
where $l$ denotes the AdS radius and $G$ is Newton's constant. The standard first-order formulation of gravity is recovered by considering the combinations
\be \label{viel}
e \,=\, \frac{l}{2} \left(\, A - \widetilde{A} \,\right) \, , \qquad \qquad \o \,=\, \frac{1}{2} \left(\, A + \widetilde{A} \,\right) \, ,
\ee
that identify the dreibein and the spin connection.

In a similar fashion, the first-order formulation of the free dynamics
of massless bosonic symmetric fields $\vf_{\m_1 \ldots\, \m_s}$ with
$s \geq 2$ involves a vielbein-like 1-form and an auxiliary
\mbox{1-form} which generalises the spin connection
\cite{vas-frame,lopatin-vas}. In $D=2+1$ these two differential forms
have the same structure (i.e.~the same fibre indices), and one can
consider linear combinations of them as in \eqref{viel}. This
eventually allows one to build bosonic massless higher-spin gauge
theories out of $G \times G$ Chern-Simons theories
\cite{Blencowe}. The action has the same form as \eqref{EH}, but now
the gauge fields $A$ and $\widetilde{A}$ take values in a (possibly
infinite-dimensional) Lie algebra $\mathfrak{g}$ admitting a
non-degenerate bilinear invariant form. Vielbeine and spin connections
are identified through \eqref{viel}, while the invariance of the
action under
\be \label{transfA}
\d A \,=\, d A \,+\, [\, A \,,\, \l \,] \, , \qquad\qquad \d \widetilde{A} \,=\, d \widetilde{A} \,+\, [\, \widetilde{A} \,,\, \widetilde{\l} \,]
\ee
leads to two different kinds of gauge transformations generated by the parameters
\be \label{par}
\x \,=\, \frac{l}{2} \left(\, \l - \widetilde{\l} \,\right) \, , \qquad \qquad \L \,=\, \frac{1}{2} \left(\, \l + \widetilde{\l} \,\right) \, .
\ee
Those generated by $\x$ correspond to local translations in pure gravity (that in $D=2+1$ are equivalent to diffeomorphisms \cite{Witten}), while those generated by $\L$ extend the usual local Lorentz transformations:
\begin{subequations} \label{gauge_fields}
\begin{align}
& \d e \,=\, d\x \,+\, [\, \o \,,\, \x \,] \,+\, [\, e \,,\, \L \,] \, , \label{gauge_viel} \\
& \d \o \,=\, d\L \,+\, [\, \o \,,\, \L \,] \,+\, \frac{1}{l^2}\, [\, e \,,\, \x \,] \, . \label{gauge_spin}
\end{align}
\end{subequations}

Even if no local fluctuations are present, one can define the ``spectrum'' of the theory by looking at the transformation properties of the fields under Lorentz transformations. It is thus fixed by the choice of a $sl(2,\mathbb{R})$ subalgebra in $\mathfrak{g}$ that -- together with the corresponding one coming from the second copy of $\mathfrak{g}$ -- identifies the gravitational sector. Once this selection is made one can consider the branching of $\mathfrak{g}$ under the adjoint action of the ``gravitational'' $sl(2,\mathbb{R})$, so that
\be \label{branching}
\mathfrak{g} \,=\, sl(2,\mathbb{R}) \, \oplus \left(\, \bigoplus_{\ell\,,\,a} \, \mathfrak{g}^{(\ell,a)} \,\right) \, .
\ee
Each $\mathfrak{g}^{(\ell,a)}$ has dimension $2\ell+1$ with $2\ell \in \mathbb{N}$, while the index $a$ accounts for possible multiplicities. For infinite-dimensional algebras we thus discard by hypothesis $sl(2)$ embeddings that would bring on infinite-dimensional irreducible representations in \eqref{branching}. The branching of $\mathfrak{g}$ induces the decomposition
\be \label{dec_A}
A(x) \,=\, A_\m{}^i(x)J_i\, dx^\m \,+\ \sum_{\ell\,,\,a} \sum_{m\,=\,-\,\ell}^\ell A^{[a]\,\ell,m}_\m(x)\, (W^{\ell}_m)_{[a]} \, dx^\m \, ,
\ee
and a similar one for $\widetilde{A}$. Here the $(W^{\ell}_m)_{[a]}$
generate $\mathfrak{g}^{(\ell,a)}$, while the $J_i$ generate
$sl(2,\mathbb{R})$ as in pure gravity. Let us now focus for a while on
$sl(2)$ embeddings that do not involve any half-integer $\ell$. In
this case the dimension of each $\mathfrak{g}^{(\ell,a)}$ equals the
number of independent off-shell fiber components of the vielbein or of
the spin connection associated to a fully symmetric tensor
$\vf^{\,[a]}{}_{\m_1 \ldots\, \m_{\ell+1}}$
\cite{vas-frame,lopatin-vas}. As a result, for any integer $\ell$ the
1-forms
\be \label{viel_lm}
e_\m^{[a]\,\ell,m} \,=\, \frac{l}{2} \left(\, A_\m^{[a]\,\ell,m} - \widetilde{A}_\m^{[a]\,\ell,m} \,\right) \, , \qquad\quad \o_\m^{[a]\,\ell,m} \,=\, \frac{1}{2} \left(\, A_\m^{[a]\,\ell,m} + \widetilde{A}_\m^{[a]\,\ell,m} \,\right)
\ee 
can be identified with the vielbein and the spin connection of a spin-$(\ell+1)$ field. The spectrum is thus specified by the ``spins'' and the multiplicities appearing in \eqref{branching}. Additional comments on this identification will be presented in Section \ref{sec:W_metric}.

The choice of the principal $sl(2,\mathbb{R})$ embedding in a finite-dimensional non-compact simple Lie algebra $\mathfrak{g}$ always fits into this scheme. In this case each $\ell$ in \eqref{branching} corresponds to one of the exponents of $\mathfrak{g}$, so that all values of $\ell$ are integers and greater or equal to 1. The simplest examples in this class are $SL(N) \times SL(N)$ CS theories with a principally embedded gravitational sector. Besides the graviton, they involve fields with spin $(\ell+1) = 3,\ldots,N$. An infinite-dimensional counterpart of the latter HS gauge theories, with a similar but unbounded spectrum, will be studied in Section \ref{sec:W}. An ampler discussion of $SL(N) \times SL(N)$ theories can be found in \cite{spin3}, while here we would like to briefly discuss the subtleties that the choice of alternative embeddings entails. 

First of all, in general \eqref{branching} also contains half-integer values of $\ell$. The corresponding 1-forms in \eqref{viel_lm} cannot be associated to any tensorial field. On the other hand, for each half-integer $\ell$ their number equals the number of off-shell components of a \emph{spinorial} field $\psi^\a{}_{\m_1 \ldots \, \m_{\ell+1/2}}$. Since no local degrees of freedom are involved, the spinorial nature of the resulting fields does not prevent one from applying the previous construction even starting from a purely bosonic gauge algebra. A slight generalisation is also possible, since spinorial fields do not require a spin-connection in the first-order formulation \cite{vas-frame,aragone-ferm,vas-ferm}. The doubling of $\mathfrak{g}$ would then lead to a doubling of the number of these fields in the spectrum. The same is true for spin-1 fields, corresponding to $sl(2,\mathbb{R})$ singlets in \eqref{branching}. As a result, in general one could even consider $G \times \widetilde{G}$ Chern-Simons theories, provided that $\widetilde{\mathfrak{g}}$ is the subalgebra of $\mathfrak{g}$ needed to reconstruct vielbeine and spin connections for the fields of integer spin $s \geq 2$.\footnote{An analogue construction is standard in the context of pure supergravity theories in $D=2+1$ where, besides the graviton, only fields with spin $s < 2$ appear. This allows one to consider super-CS theories built upon generic $G \times \widetilde{G}$ supergroups or simply upon $SL(2,\mathbb{R}) \times G$ \cite{Townsend}.} The basic building blocks of a toy model for HS interactions would still be available: one has a set of fields that do not propagate any local degree of freedom and that transform under Lorentz transformations as those used to describe the free propagation of ``higher spins'', in the sense specified at the beginning of the Introduction.

However, we also know the action of these theories, and we can check if it also displays the features that one would like to associate to a sensible toy model for HS interactions. The simplest feature to analyse is the structure of kinetic terms, that is dictated by the structure of the invariant form appearing in \eqref{CS}. To study it, we have to fix our notation. We denote the generators of $sl(2,\mathbb{R})$ by $J_{\pm},J_{0}$ and we choose the convention 
\begin{equation}\label{commutation-sl2}
[\,J_{+}\,,\,J_{-}\,]\,=\,2J_{0} \quad  ,\quad [\,J_{\pm}\,,\,J_{0}\,]\,=\,\pm\, J_{\pm} \ .
\end{equation}
We also choose a basis of $\mathfrak{g}^{(\ell,a)}$ such that
\be \label{primary}
[\, J_i\,,\, (W^{\ell}_m)_{[a]} \,] \,=\, (i\ell-m)\, (W^{\ell}_{i+m})_{[a]} \, ,
\ee
where $i=0,\pm 1$, and we identified $J_{\pm 1}\equiv J_{\pm}$.
Eq.~\eqref{primary} forces the Killing form to satisfy
\be \label{kill}
\tr \left(\,(W^{k}_m)_{[a]}\, (W^{\ell}_n)_{[b]}\,\right) \,=\, (-1)^{\ell-m}\, \frac{(\ell+m)!(\ell-m)!}{(2\ell)!}\ \d^{k,\ell}\, \d_{m+n,0} \, (N_\ell)_{ab} \, ,
\ee
with 
\be \label{M_kill}
(N_\ell)_{ab} \,=\, \tr \left(\,(W^{\ell}_\ell)_{[a]} (W^{\ell}_{-\ell})_{[b]}\,\right) \, .
\ee

Eq.~\eqref{kill} implies $(N_\ell)_{ab} = (-1)^{2\ell}
(N_\ell)_{ba}$, so that the matrix $N_\ell$ is symmetric for integer values of $\ell$ and skew symmetric for half-integer values of $\ell$. This leads respectively to symmetric or skew symmetric kinetic terms. While unfamiliar, the latter are precisely as pertains to the bosonic nature of these spinorial fields, and are instrumental in order to attain a non-trivial kinetic term for them, unless further prescriptions are introduced, like a grading of the gauge algebra. At any rate, the models that we are considering involve more than one field, so that an issue should be checked: the relative sign between different kinetic terms. Even if no local fluctuations are available in three dimensions, one could still require that no sign differences are present, as it is crucial in
higher space-time dimensions. This is not the case for a generic
choice of $\mathfrak{g}$ and of a $sl(2)$ embedding in it\footnote{While this paper was in preparation, HS gauge theories based on a non-principally embedded
gravitational sector were discussed in \cite{GK_new} (see also
\cite{spin3} for previous comments). Its authors proposed to consider all possible embeddings in a given gauge algebra as different phases of a common theory, related by a breaking of the Lorentz-like symmetries of~\eqref{gauge_fields}. The opportunities opened by this observation could well overcome our reservations, but still any attempt to extrapolate possible results to higher dimensions should face the subtleties that we remarked here.}. The
relative signs between kinetic terms are also affected by the choice
of a real form for $\mathfrak{g}$. For instance, as we shall discuss
in Section \ref{sec:W_hs}, these considerations select the real form
$sl(N,\mathbb{R})$ in the case of $SL(N) \times SL(N)$ CS theories
that we mentioned before.

\subsection{Asymptotic symmetries}\label{sec:red_sym}

We are now going to discuss the asymptotic symmetries of
asymptotically-AdS configurations. Therefore, our CS theories have to
be defined on manifolds $\cM$ with a cylindrical boundary $\pr \cM$
parameterised by a time-coordinate $t$ and an angular coordinate
$\theta$. In order to fix our notation, in this section we first
briefly recall the main features of CS theories on manifolds of this
type following the reviews \cite{Banados_review,Carlip}. Then,
following \cite{spin3}, we discuss how the conditions selecting
asymptotically-AdS configurations translate into the Drinfeld-Sokolov
constraint.

Let us begin by focusing on a single chiral sector, say the one involving $A$. As reviewed in \cite{Banados_review,spin3}, it is always possible to choose the gauge 
\begin{align} \label{gaugefixing}
A_\r \,=\, b^{-1}(\r)\, \pr_\r \, b(\r) \, ,
\end{align}
where $\r$ is a radial coordinate and $b(\r)$ is an arbitrary function taking values in the gauge group $G$. The gauge \eqref{gaugefixing} is preserved by residual gauge transformations with parameters
\be \label{par_fix}
\L \,=\, b^{-1}(\r)\,\l(t,\theta)\,b(\r) \, ,
\ee
and on shell it implies
\be
A_\theta \,=\, b^{-1}(\r)\,a(t,\theta)\,b(\r) \, .
\ee
Here, $\l(t,\theta)$ and $a(t,\theta)$ are arbitrary
$\mathfrak{g}$-valued functions. One can then impose the boundary
condition
\be \label{bnd_cond}
\left. \left( \frac{A_t}{l} - A_\theta \right) \right|_{\pr \cM} \,=\, 0
\ee
that cancels the boundary term appearing in the variation of the
action. In pure AdS gravity, \eqref{bnd_cond} is satisfied by all BTZ
backgrounds, so that we can safely use it to select the space of
asymptotically-AdS solutions. Requiring \eqref{bnd_cond} on the
boundary forces $A_{t} = l\, A_\theta$ everywhere in the bulk and
removes the gauge invariance, because both $a$ and $\l$ must depend on
$(\frac{t}{l} - \theta)$ so that there is no more an arbitrary time
dependence.

We are left with the $\mathfrak{g}$-valued function $a(\theta)$ on which the gauge transformations generated by \eqref{par_fix} act as
\begin{equation}\label{gaugetransformation}
\delta_{\lambda}a(\theta) \,=\, \lambda ' (\theta) \,+\, [\,a (\theta) \,,\, \lambda (\theta) \,] \, , 
\end{equation}
where a prime denotes a derivative in $\theta$. These are not proper gauge transformations \cite{Banados_review,Carlip}, but rather global symmetries generated by the boundary charges
\begin{equation} \label{charges_gen}
Q (\lambda) = - \, \frac{k}{2\pi} \int d\theta \, \tr \left(\lambda
(\theta)a (\theta) \right) \, ,
\end{equation}
where $k$ times the trace denotes the invariant bilinear form that is used to define the CS action. The latter observation suffices to fix the canonical structure of the boundary theory since
\be \label{var_can}
\d_\l a(\theta) \,=\, \{ Q(\l) , a(\theta)\}
\ee
implies
\begin{equation} \label{poissonCS}
\{Q (\lambda),Q (\eta) \} = - \, \frac{k}{2\pi}\int d\theta \, \tr
\left(\eta (\theta)\delta_{\lambda}a (\theta) \right) \, .
\end{equation}
This Poisson algebra is the centrally
extended loop algebra of $\mathfrak{g}$ (see, for instance, \cite{Banados_review,Carlip}), and it induces an analogue Poisson structure on the space of on-shell configurations $a(\theta)$, that accounts for the boundary degrees of freedom.

The other chiral sector, involving $\widetilde{A}$, can be treated in a similar fashion, but with some small variations needed to ensure the invertibility of the dreibein. This is guaranteed if one reaches the following on-shell parameterisation,
\be \label{summary_cond}
l^{-1} \widetilde{A}_t \,=\, -\, \widetilde{A}_\theta \,=\, b(\r)\,\tilde{a}(t,\theta)\,b^{-1}(\r) \, , \qquad \widetilde{A}_\r \,=\, b(\r)\, \pr_\r \, b^{-1}(\r) \, ,
\ee
and restricts the $b(\r)$ appearing both in \eqref{gaugefixing} and
\eqref{summary_cond} to take values in the ``gravitational'' subgroup
of $G$. Even if the dreibein is always invertible, in \cite{spin3}
(see also \cite{HR,GK}) we argued that \eqref{summary_cond} and the
corresponding condition for $A$ do not provide a satisfactory on-shell
parameterisation of the space of asymptotically-AdS configurations. We
thus proposed to also require a finite difference between them and the
AdS solution at the boundary,
\be \label{ads_cond}
\left( A - A_{AdS}\right) \Big|_{\pr \cM} \,=\, \cO(1) \, ,
\ee 
and similarly for $\widetilde{A}$. Eq.~\eqref{ads_cond} translates into the Drinfeld-Sokolov condition on $a(\theta)$ and thus recovers the conformal symmetry discovered by Brown and Henneaux in the metric formulation of gravity \cite{BH}. A similar characterisation of the space of asymptotically-AdS solutions was discussed in the context of (super)gravity theories in \cite{Henneaux_vanDriel,sugra,extended_SC}.

To display the consequences of \eqref{ads_cond} on $a(\theta)$ it is convenient to denote the adjoint action of the $sl(2,\mathbb{R})$ generators $J_{i}$ on the elements of $\mathfrak{g}$ by $L_{i}$,
\begin{equation} \label{adjoint}
L_{i}\, x :=\, [\,J_{i}\,,\,x\,] \quad \text{for}\ x\in \mathfrak{g} \ .
\end{equation}
The $L_{i}$ satisfy the same commutation relations
\eqref{commutation-sl2} as the $J_{i}$. Since by hypothesis we
consider only $sl(2)$ embeddings that branch $\mathfrak{g}$ into a
(possibly infinite) sum of finite-dimensional $sl(2)$-irreducible
representations, the eigenvalues of $L_{0}$ are half integers. We
can thus decompose the gauge algebra into eigenvectors of $L_{0}$ of
negative, zero or positive eigenvalues,
\begin{equation}\label{firstdecomposition}
\mathfrak{g} \,=\, \mathfrak{g}_{<} \oplus \mathfrak{g}_{0} \oplus
\mathfrak{g}_{>} \ .
\end{equation} 
The Drinfeld-Sokolov condition amounts to the constraint that 
$a(\theta)-J_{+}$ has no components corresponding to the negative spectrum
of $L_{0}$,
\begin{equation} \label{a_cond}
a (\theta)-J_{+} \in\, \mathfrak{g}_{0} \oplus \mathfrak{g}_{>} \ ,
\end{equation}
or equivalently,
\begin{equation}\label{DrinfeldSokolov}
P_{\mathfrak{g}_{<}}\left( a (\theta) -J_{+} \right) \,=\, 0 \ ,
\end{equation}
where $P_{\mathfrak{g}_{<}}$ is the orthogonal projector onto
$\mathfrak{g}_{<}$. In HS gauge theories one has to impose an
analogous condition also on $\tilde{a}(\theta)$, but the analysis
proceeds along the same lines as the one for $a(\theta)$. Therefore,
in the following we continue to focus on $a(\theta)$.

The constraints~\eqref{DrinfeldSokolov} are in general of first class\footnote{When $L_0$ admits half-integers eigenvalues some constraints are second class. However, this does not affect the possibility to reach the gauge \eqref{aconstraint} \cite{non_principal}.} and they generate gauge transformations. To get to the reduced phase space, one has to fix the gauge. A natural gauge choice is the so-called highest-weight gauge
\begin{equation}\label{aconstraint}
a (\theta) \,=\, J_{+} \,+\, a_{-} (\theta) \ ,
\end{equation}
where $a_{-}$ satisfies\footnote{Note that $L_{-}$ increases the $L_{0}$ eigenvalue, so that $a_-(\theta)$ can be expanded in a set of highest weight eigenvectors for $L_0$. This rather unintuitive association follows from the convention~\eqref{commutation-sl2} that we chose for the $sl(2,\mathbb{R})$ algebra.}
\begin{equation}
L_{-}\,a_{-} (\theta) \,=\, 0 \ .
\end{equation}
This choice fixes the gauge completely.  In the restricted class of
solutions satisfying \eqref{ads_cond} the boundary degrees of freedom
are thus described by $a_{-}(\theta)$
\cite{principal,non_principal}. We shall now analyse the symmetries of
the constrained boundary theory, which correspond to the asymptotic
symmetries of the HS models that we introduced in Section
\ref{sec:red_bulk}.

\subsection{Drinfeld-Sokolov reduction in highest-weight gauge}\label{sec:red_red}

To find the symmetries of the constrained theory, we look at the set of symmetry
transformations~\eqref{gaugetransformation}
that leave the form~\eqref{aconstraint} of $a(\theta)$ invariant,
\begin{equation}
L_{-} (\delta_{\lambda}a) \,=\, 0 \ .
\end{equation}
This condition translates into 
\begin{equation}\label{aconstraint2}
L_{-} \left(\, \partial_{\theta}+[\,a_{-} (\theta),\, \cdot\, ] \,\right) \lambda (\theta) \,+\,
L_{-}L_{+}\lambda (\theta) \,=\, 0 \ .
\end{equation}
The operator $L_{-}L_{+}$ can be rewritten as
\begin{equation}
L_{-}L_{+} \,=\, -\, \Delta \,+\, L_{0} (L_{0}-1) \ ,
\end{equation} 
where we introduced the quadratic Casimir
\be
\Delta \,=\, L_{0}^2 \,-\, \frac{1}{2} \left(\, L_{+}L_{-}+L_{-}L_{+} \,\right) \, .
\ee
In the basis of $\mathfrak{g}$ introduced in \eqref{primary} the operator $(\Delta - L_{0} (L_{0}-1))$ acts as
\begin{equation}
\left(\, \Delta - L_{0} (L_{0}-1)\,\right) (W^\ell_m)_{[a]} \,=\, (\ell-m)(\ell+m+1) \, (W^\ell_m)_{[a]}\ ,
\end{equation}
i.e. by multiplication with a number that is non-zero for $m\not=
\ell$.  We denote by $\mathfrak{g}_{-}$ ($\mathfrak{g}_{+}$) the space
of highest (lowest) weight states,
\be
x \in \mathfrak{g}_- \ \Leftrightarrow \ L_- x = 0 \, , \qquad x \in \mathfrak{g}_+ \ \Leftrightarrow \ L_+ x = 0 \, .
\ee
In general, $\mathfrak{g}_{-}$ and $\mathfrak{g}_{+}$ can have a
non-trivial intersection which contains the $sl (2)$ singlets. We also
introduce the projection operators $P_{\pm}$ onto $\mathfrak{g}_{\pm}$,
respectively. The operator $(\Delta -L_{0} (L_{0}-1))$ is invertible
on the orthogonal complement of $\mathfrak{g}_{+}$, and we define
\begin{equation}\label{Roperator}
R:=\,- \, \frac{1}{\Delta -L_{0} (L_{0}-1)}\, \left(\,1-P_{+}\,\right) \, .
\end{equation}
In particular we have
\begin{equation} \label{comb1}
R\, L_{-}L_{+} \,=\, L_{-}L_{+}\, R \,=\, 1-P_{+}\ .
\end{equation}
Furthermore we introduce the covariant derivative
\begin{equation}\label{covDerivative}
D_{\theta} :=\, \partial_{\theta} \,+\, [\,a_{-} (\theta),\, \cdot \, ] \ ,
\end{equation}
which commutes with $L_{-}$, because $L_{-}a_{-}=0$. 

Applying $R$ to \eqref{aconstraint2} and taking into account \eqref{comb1}, we eventually obtain
\begin{equation}\label{recursion}
\lambda (\theta) \,=\, \lambda_{+} (\theta) \,-\, RL_{-}D_{\theta} \lambda
(\theta) \ .
\end{equation}
Here, $\lambda_{+} (\theta)= P_{+}\lambda (\theta)$ is the
lowest-weight part of $\lambda$. Eq.\ \eqref{recursion} is solved by
\begin{equation}\label{solution-lambda}
\lambda (\theta) \,=\, \frac{1}{1+RL_{-}D_{\theta}}\, \lambda_{+} (\theta) \ ,
\end{equation}
which expresses the gauge parameter $\lambda$ in terms of its lowest-weight part.\footnote{A similar formula appears in \cite{deBoer}.}
Inserting the solution~\eqref{solution-lambda} into the
expression~\eqref{gaugetransformation} for $\delta_{\lambda} a$, we
find
\begin{equation}\label{globalsymmetry}
\delta_{\lambda} a (\theta) \,=\, P_{-}\, \frac{1}{1+D_{\theta}RL_{-}}\,
D_{\theta}\lambda_{+} (\theta) \,=\, P_{-} \sum_{n\,=\,0}^{\infty} \big(- D_{\theta}RL_{-}\big)^{n} D_{\theta}\lambda_{+} (\theta) \ .
\end{equation}
This finally expresses $\delta_{\lambda } a$ in terms of $\lambda_{+}$. One might be worried about the infinite series appearing in~\eqref{globalsymmetry}. For a gauge parameter with definite $sl (2)$ quantum numbers as $\l_+(\theta) = \e(\theta) W^\ell_\ell$, however, the series expansion
in~\eqref{globalsymmetry} stops at the term with $n=2\ell$. This is
because each term $D_{\theta}RL_{-}$ involves the application of
$L_{-}$, and $(L_{-})^{2\ell+1}W^\ell_\ell=0$, while $L_{-}$
commutes with $D_{\theta}$ and $R$ does not change the $sl (2)$ quantum numbers. The indices $[a]$ of \eqref{dec_A} do not play any role in this argument and thus we omitted them for simplicity.

In order to identify the Poisson structure on the reduced phase space
one can then substitute \eqref{globalsymmetry} in
\eqref{poissonCS}. It is also possible to display the Poisson brackets
between fields of defined conformal spin. To this end one can expand
$a_-(\theta)$ and the independent part of the gauge parameter,
encoded in $\l_+(\theta)$, in the basis \eqref{primary}:
\begin{subequations}
\begin{align}
& a_-(\theta) \,=\, \frac{2\pi}{k} \left(\, \cL(\theta) J_- \,+\, \sum_{\ell\,,\,a} \, \cW^{[a]}_\ell(\theta)\, (W^{\ell}_{-\ell})_{[a]} \,\right) \, , \label{exp_a} \\
& \l_+(\theta) \,=\, \e(\theta) J_+ \,+\, \sum_{\ell\,,\,a} \, \e^{[a]}_\ell(\theta)\, (W^{\ell}_{\ell})_{[a]} \, .  \label{exp_l}
\end{align}
\end{subequations}
Here $[a]$ is a colour index, while $\ell$ is a $sl(2)$ quantum number. The charges \eqref{charges_gen} which generate the transformations \eqref{globalsymmetry} then read
\be \label{charge_fix}
Q(\l_+) \,=\, \int\! d\theta\, \e(\theta)\, \cL(\theta) \,-\, \sum_{\ell,\,a,\,b} \, (N_\ell)_{ab} \int\! d\theta\, \e^{[a]}_\ell(\theta)\, \cW^{[b]}_\ell(\theta) \, ,
\ee
with the matrices $(N_\ell)_{ab}$ defined in \eqref{M_kill}. By
substituting \eqref{charge_fix} in \eqref{var_can} one can eventually
read off the Poisson brackets $\{\cW^{[a]}_i(\theta) ,
\cW^{[b]}_j(\theta')\}$. If all values of $\ell$ are integers, one can
diagonalise $(N_\ell)_{ab}$ and thus determine all Poisson brackets
involving $\cW^{[a]}_\ell$ by looking at the gauge transformations
generated by $\e^{[a]}_\ell$. If some half-integer values of $\ell$
appear in \eqref{branching} one can at most make $(N_\ell)_{ab}$
block-diagonal, with a sequence of $2 \times 2$ blocks. This means
that the Poisson brackets of a given field can be extracted from the
gauge transformations generated by the gauge parameter with the
``partner'' colour charge.

The determination of the Poisson brackets can also formulated
covariantly in colour indices. To this end it is convenient to denote
the inverse of the matrix $(N_\ell)_{ab}$ by $(N_\ell)^{ab}$, and to
consider
\be
\d_{i\, [a]}\, \cW^{[b]}_j(\theta) =\, \sum_c\, (N_j)^{bc}\,
\tr\left((W^{j}_{j})_{[c]} \, \frac{1}{1+D_{\theta}RL_{-}}\,
D_{\theta}\,\e (\theta) (W^i_i)_{[a]}\right) \,,
\ee
where $\sum_c (N_j)^{bc}\,\tr (W^{j}_{j[c]}\ \cdot\ )$ selects the coefficient
in front of the generator $(W^{j}_{-j})_{[b]}$, and the variation is
taken with respect to $\lambda_{+} (\theta)=\e (\theta)
(W^{i}_{i})_{[a]}$. From~\eqref{var_can} we have
\be
\d_{i\, [a]}\, \cW^{[b]}_j (\theta ')\,=\, -\, \sum_{c}\, (N_i)_{ac} \int d\th''
\e (\theta '')\, \big\{ \cW^{[c]}_i(\th'') \,,\, \cW^{[b]}_j(\th') \big\} \, ,
\ee
so that we can obtain the Poisson brackets from the variation by
\be
\big\{ \cW^{[a]}_i(\th) \,,\, \cW^{[b]}_j(\th') \big\} \,=\, - \,
\sum_{c}\, (N_{i})^{ac} \d_{i\, [c]} \, \cW^{[b]}_j (\theta ') 
\Big|_{\e (\theta ')\,=\,\delta (\theta -\theta ')} \, . \label{covariant_bracket}
\ee

\subsection{General properties}\label{sec:red_gen}

The Poisson algebra that we obtained in the last section by the DS
reduction of a centrally extended loop algebra of course depends very
much on the detailed structure of the algebra $\mathfrak{g}$ that we
started with and on the choice of a $sl(2,\mathbb{R})$ embedding. On
the other hand, there are a few general properties that we want to
discuss here.

\subsubsection{Primary basis}\label{subsec:primary}

The DS reduction in highest-weight gauge always leads to a
presentation of the resulting $\cW$-algebra in a basis where all
generators are primaries with respect to the lowest spin ones. If no
spin-1 fields are present -- as in the case of principal
$sl(2,\mathbb{R})$ embeddings -- all generators are thus automatically
Virasoro primaries.

If spin-1 fields are present, corresponding to $sl(2)$ singlets in
\eqref{branching}, we can compute their Poisson brackets with the other
fields by evaluating the gauge transformations
generated by $\l_+ = \e(\theta) (W^{0}_0)_{[a]}$. From the previous
discussion we know that we only have to evaluate the term at zeroth order
in the series \eqref{globalsymmetry},
\be \label{cov_zero}
D_\theta \l_+ =\, \e^{\,\pe} \, (W^{0}_0)_{[a]} \,-\, \e\,\cQ_{[a]}\, a_- \, .
\ee
Here, in analogy with \eqref{adjoint}, we denoted by $\cQ_{[a]}$ the adjoint action of $(W^{0}_0)_{[a]}$ on the elements of the algebra,
\be
\cQ_{[a]}\, x :=\, [\, (W^{0}_0)_{[a]} ,\, x \,] \quad \text{for}\ x\in \mathfrak{g} \ .
\ee
These operators cannot modify the $sl(2)$ quantum numbers, so that we can describe their action by
\be
\cQ_{[a]} (W^{\ell}_m)_{[b]} \,=\, \sum_c\, (f_\ell)^c{}_{ab}\, (W^{\ell}_m)_{[c]} \, .
\ee
Therefore, all terms in \eqref{cov_zero} belong to $\mathfrak{g}_-$ and the projector $P_-$ does not induce any modification. Expanding $a_-(\theta)$ as in \eqref{exp_a} we eventually get
\be \label{gauge_zero}
\d_{[a]} \cW^{[b]}_\ell \,=\, -\, \e\, \sum_c\, (f_\ell)^b{}_{ac}\, \cW^{[c]}_\ell \,+\, \frac{k}{2\pi}\, \e^{\,\pe}\, \d_a{}^b\, \d_{0,\ell} \, .
\ee

In the singlet sector one can always diagonalise the matrix $(N_\ell)_{ab}$ appearing in \eqref{charge_fix}, but we can also substitute \eqref{gauge_zero} in \eqref{covariant_bracket} to get
\begin{subequations}
\begin{align}
& \big\{ \cW^{[a]}_0(\theta) , \cW^{[b]}_0(\theta') \big\} \,=\, \d(\theta-\theta') \sum_c (f_0)^{ba}{}_{c}\, \cW^{[c]}_0(\theta') \,+\, \frac{k}{2\pi}\, (N_0)^{ab}\, \pr_\theta \d(\theta-\theta') \, , \label{kac-moody}\\[2pt]
& \big\{ \cW^{[a]}_0(\theta) , \cW^{[b]}_\ell(\theta') \big\} \,=\, \d(\theta-\theta') \sum_c (f_\ell)^{ba}{}_{c}\, \cW^{[c]}_\ell(\theta') \quad \textrm{for}\ \ell \geq 1 \, .
\end{align}
\end{subequations}
In both cases one index of $(f_\ell)^{b}{}_{ac}$ is raised using $(N_0)^{ab}$, that is the inverse of the Killing metric of the subalgebra of $\mathfrak{g}$ spanned by the $sl(2)$ singlets.  The spin-1 fields $\cW^{[a]}_0$ thus generate the centrally extended Kac-Moody subalgebra \eqref{kac-moody}. They always appear when non-principal $sl(2)$ embeddings are considered, so that the presence of a Kac-Moody subalgebra is a neat signature of this class of DS reduction (see also \cite{non_principal} for more comments on this setup). 

Let us now evaluate the Poisson bracket of an arbitrary field with the Virasoro current. In this case it is sufficient to consider the gauge transformations generated by $\lambda_{+} (\theta) = \epsilon(\theta) J_{+}$. In the computation the singlets behave differently with respect to all other fields and it is convenient to split $a_-(\theta)$ as
\be
a_-(\theta) \,=\, \hat{a}_-(\theta) \,+\, \frac{2\pi}{k}\, \sum_a\, \cW^{[a]}_0(\theta) (W^{0}_0)_{[a]} \, .
\ee
To proceed we only have to compute the series in~\eqref{globalsymmetry} up to $n=2$:
\begin{subequations}
\begin{align}
& D_{\theta}\lambda_{+} =\, L_{+} \big( \epsilon ' J_{0} - \epsilon\, \hat{a}_{-}\big) \, , \\[4pt]
& \big(- D_{\theta}RL_{-}\big) D_{\theta}\lambda_{+} =\, -\, \frac{\epsilon ''}{2}\, L_{+}J_{-} +\, \epsilon ' (L_{0}+1)\hat{a}_{-} +\, \epsilon\, \hat{a}_{-}' +\, \frac{2\pi}{k}\,\e\, \sum_a\, \cW^{[a]}_0 \cQ_{[a]} \hat{a}_- \ , \\
& \big(- D_{\theta}RL_{-}\big)^{2} D_{\theta}\lambda_{+} =\, \frac{\epsilon '''}{2}\, J_{-} \ .
\end{align}
\end{subequations}
Summing all contributions that remain after the projection by $P_-$ we eventually find
\begin{equation} \label{transf_prim}
\delta_\l a \,=\, \frac{\epsilon '''}{2}J_{-} \,+\, \epsilon '
(L_{0}+1)a_{-} \,+\, \epsilon\, a_{-}' \,+\, \frac{2\p}{k} \, \e\,\sum_a\, \cW^{[a]}_0 \cQ_{[a]}\, \hat{a}_- \qquad \text{for}\
\lambda_{+}=\epsilon J_{+} \ .
\end{equation}
Expanding $a_-(\theta)$ as in \eqref{exp_a}, then leads to
\begin{subequations} \label{prim_gauge}
\begin{align}
& \d \cL \,=\, \e\, \cL' \,+\, 2\, \e' \cL \,+\, \frac{k}{4\pi}\, \epsilon ''' \, , \\[2pt]
& \d\, \cW_\ell^{(a)} \,=\, \epsilon\, \cW_\ell^{(a)\, \prime}  \,+\, (\ell+1)\, \epsilon '\,
\cW_\ell^{[a]} \,+\, \e \,\frac{2\p}{k} \sum_{b\,,\,c}\, (f_\ell)^a{}_{bc}\, \cW^{[b]}_0 \cW^{[c]}_\ell \quad \textrm{for} \ \ell \geq 1 \, ,
\end{align}
\end{subequations}
while the fields with $\ell = 0$ are left invariant. If no spin-1
fields are present, the terms containing $\cW^{[a]}_0$ disappear, and
the field $\cW_\ell^{[a]}$ transforms as a primary field of
conformal weight $(\ell+1)$. Even in the general case one can easily
express the result in a Virasoro primary basis by redefining the
Virasoro current as
\be
\widehat{\cL} :=\, \cL \,-\, \frac{\pi}{k}\, \sum_{a\,,\,b} \, (N_0)_{ab}
\, \cW_0^{[a]} \, \cW_0^{[b]} \, . \label{SugawaraShift}
\ee
With respect to the improved Sugawara Virasoro current $\widehat{\cL}$ the Poisson brackets read
\begin{subequations}
\begin{align}
& \big\{ \widehat{\cL}(\theta) \,,\, \widehat{\cL}(\theta') \big\} =\, \d(\theta-\theta')\,\widehat{\cL}^{\,\pe}(\theta') \,-\,2\,\pr_\theta\d(\theta-\theta')\, \widehat{\cL}(\theta') \,-\, \frac{k}{4\pi}\, \pr^3_\theta\d(\theta-\theta')\, , \label{virasoro} \\[5pt]
& \big\{ \widehat{\cL}(\theta) \,,\, \cW_{\ell}^{[a]}(\theta') \big\} =\, \d(\theta-\theta')\,\cW^{(a)\pe}(\theta') \,-\,(\ell+1)\,\pr_\theta\d(\theta-\theta')\,\cW_{\ell}^{(a)}(\theta') \, .
\end{align}
\end{subequations}
%

\subsubsection{Central terms}\label{subsec:central}

Eqs.~\eqref{kac-moody} and \eqref{virasoro} display a central term,
but central terms do not arise only in the Kac-Moody or Virasoro
subalgebras. On the other hand, their structure is rather rigid and
does not depend on the particular loop algebra to which one applies
the DS procedure. They can be computed by substituting the covariant
derivative $D_\theta$ with an ordinary derivative in
\eqref{globalsymmetry}. When we consider a transformation by a gauge parameter of a
given conformal spin, say $\l_+(\theta) = \e(\theta)
(W^\ell_\ell)_{[a]}$, we find
\be \label{central}
\left. \d_\l a\, \right|_{D_\theta \, =\, \pr_\theta}\,=\, \frac{(-1)^{2\ell}}{(2\ell)!}\, \e^{(2\ell+1)}\, (W^\ell_{-\ell})_{[a]} \, ,
\ee 
where the exponent between parentheses denotes the action of the
corresponding number of derivatives on $\e$. To reach this result one
just has to combine $L_+ W^\ell_{\ell-(n+1)} = (n+1) W^\ell_{\ell-n}$
(following from \eqref{primary}) with eq.~\eqref{comb1}. In
conclusion, central terms only appear in the Poisson brackets between
fields of the same conformal weight,
\begin{equation}
\big\{ \cW^{[a]}_{\ell}(\theta) \,,\, \cW^{[b]}_{\ell}(\theta') \big\} \,=\, \frac{k}{2\pi (2\ell)!} \,(N_\ell)^{ab}\, \pr_\theta^{2\ell+1} \d(\theta-\theta') \,+\, \ldots \ ,
\end{equation}
where $(N_\ell)^{ab}$ is the inverse of the matrix introduced in \eqref{M_kill}.

\subsubsection{Polynomials in the Virasoro generators}\label{subsec:virasoro}

There is another property of the $\cW$-algebra that does not depend on
the particular Lie algebra $\mathfrak{g}$ chosen as a starting point
for the DS reduction. It is the presence of non-linearities in the
resulting Poisson algebra. This follows from the repeated application
of the covariant derivative in \eqref{globalsymmetry}, but here we
would like to display explicitly a class of polynomial terms that does
not depend on the structure of $\mathfrak{g}$ but only on the
branching \eqref{branching}. These are the polynomials that involve
just the Virasoro generator $\cL$, whose structure depends only on the
commutators \eqref{primary}.

One can compute the polynomials containing only $\cL$ by considering
only the Virasoro part in the covariant derivative,
\be \label{der_vir}
D_\theta \,\to \, \pr_\theta \,+\, \frac{2\pi}{k}\, \cL(\th)\, L_- \, .
\ee
Furthermore, as in \eqref{central}, it is convenient to consider
separately the contributions to $\d_\l a$ coming from gauge parameters
of different spin. Let us for simplicity elide colour indices and
consider $\l_+(\theta) = \e(\theta) W^\ell_\ell$. This leads to
\be \label{var_vir}
\begin{split}
\d_\l a\, \Big|_{D_\theta \,=\, \pr\, + \frac{2\pi}{k} \cL L_-} & =\, \frac{(-1)^{2\ell}}{(2\ell)!}\, \sum_{r\,=\,0}^{\lceil \ell \rceil} \left(\frac{2\pi}{k}\right)^{\!\!r}\, \sum_{\,p_1\,=\,0}^{2(\ell-r)+1}\, \sum_{p_2\,=\,0}^{2(\ell-r)+1-p_1} \ldots \sum_{p_r\,=\,0}^{2(\ell-r)+1-\sum_1^{r-1}p_t} \\[5pt]
& \times\, C[\,\ell\,]_{p_1 \ldots\, p_r}\, \cL^{(p_1)} \ldots\, \cL^{(p_r)}\, \e^{(2(\ell-r)+1-\sum_1^{r}p_t)} \ W^\ell_{-\ell} \, ,
\end{split}
\ee
where the extremum of the sum over $r$ is $\ell$ or $\ell + 1/2$,
depending on whether $\ell$ is integer or half integer. As a result,
fields with half-integer spin also admit pure-Virasoro terms that do
not contain derivatives, while if the spin is integer there is at
least one derivative in all terms of \eqref{var_vir}. The ``structure
constants'' appearing in \eqref{var_vir} read
\be \label{const_vir}
\begin{split}
C[\,\ell\,]_{\,p_1 \ldots\, p_r} & = \sum_{i_1\,=\,p_1}^{2(\ell-r)+1}\, \sum_{i_2\,=\,\langle p_1+p_2-i_1 \rangle_+}^{2(\ell-r)+1-i_1} \ldots \sum_{i_r\,=\,\langle \sum_1^r p_a - \sum_1^{r-1}i_t \rangle_+}^{2(\ell-r)+1-\sum_1^{r-1}i_t} \, \prod_{s\,=\,1}^r \binom{\sum_1^s p_t - \sum_1^{s-1}i_t}{p_s} \\
& \times \left( 2\,(r-s) + 1 + \sum_{t\,=\,1}^{r-s+1} i_t \right) \left( 2\,(l-r+s) - \sum_{t\,=\,1}^{r-s+1} i_t \right)  ,
\end{split}
\ee
where $\langle i \rangle_+ = \max(0,i)$. The details of their computation are presented in Appendix \ref{app:proofs}. 

Note that the tensors $C[\,\ell\,]_{p_1 \ldots\, p_r}$ are not
symmetric for interchanges of the indices $p_s$. As a result a
symmetrisation is needed in order to extract the coefficient appearing
in front of a non-ordered combination of derivatives of $\cL$. Let us
also stress that the terms displayed here can only appear in the
brackets $\{ \cW^{[a]}_{\ell}(\theta) , \cW^{[b]}_{\ell}(\theta') \}$,
precisely as in the discussion of the central terms (that actually
come from the term of order zero in the sum over $r$). This is due to
the proportionality of \eqref{var_vir} to $W^\ell_{-\ell}$ and to the
fact that the adjoint action of $sl(2)$ generators cannot modify the
quantum number $\ell$. This is no longer true if one considers the
full covariant derivative in \eqref{globalsymmetry}, so that the
computation of the remaining terms in the Poisson algebra requires the
knowledge of the detailed structure of $\mathfrak{g}$. In the
following we shall exploit it in a couple of examples, before
presenting the main results of this paper in Section \ref{sec:W},
where we apply this procedure to compute the asymptotic symmetries of
$SL(N,\mathbb{R}) \times SL(N,\mathbb{R})$ HS gauge theories and their
$N \to \infty$ limits.

\subsection{Non-principal embeddings: two examples}\label{sec:red_examples}

%
In this section we present the main features of DS reductions based on
non-principal $sl(2,\mathbb{R})$ embeddings by studying in detail two
examples. The interpretation of the $\cW$-algebras we are going to
discuss as asymptotic symmetries of classical HS gauge theories could be not
completely straightforward due to the comments we presented at the end of Section~\ref{sec:reduction}. On the other hand, we feel it could be instructive to also examine the features of non-principal DS reductions, also in view of possible applications to other classes of minimal models aside from those considered in
\cite{minimal,Kiritsis,minimal_ext}. See also \cite{GK_new} for a
discussion of their possible role in HS gauge theories.

In particular, we perform the DS reductions associated to two
non-principal embeddings of $sl(2,\mathbb{R})$ in
$sl(N,\mathbb{R})$. The algebra $sl(N,\mathbb{R})$ admits a number of
inequivalent $sl(2)$ embeddings equal to the number of partitions of
$N$ \cite{Dynkin}. One of them is just the trivial embedding that
cannot be used to build a HS gauge theory, and whose DS reduction
gives the affine extension of $sl(N,\mathbb{R})$ rather than a
$\cW$-algebra \cite{non_principal}. In the $sl(3,\mathbb{R})$ case
there is thus only one non-trivial non-principal sl(2) embedding and
we present its DS reduction, together with the one associated to the
corresponding embedding of $sl(2,\mathbb{R})$ in
$sl(4,\mathbb{R})$. In fact, inequivalent embeddings can be obtained
by embedding different $n \times n$ representations of
$sl(2,\mathbb{R})$ in the fundamental of $sl(N,\mathbb{R})$. In both
cases we consider the ``next-to-principal'' $sl(2,\mathbb{R})$
embeddings in $sl(N,\mathbb{R})$, where one singles out a $(N-1)
\times (N-1)$ representation of $sl(2,\mathbb{R})$ in the fundamental
of $sl(N,\mathbb{R})$.

\subsubsection{The Polyakov-Bershadsky $\cW_3^{(2)}$ algebra} \label{subsec:non-principalW3}

We consider here the $sl(2,\mathbb{R})$ embedding in $sl(3,\mathbb{R})$ that branches the fundamental representation as
\be \label{branch3}
\mathbf{\underline{8}} \,=\, \mathbf{\underline{3}} \,+\, 2\cdot \mathbf{\underline{2}} \,+\, \mathbf{\underline{1}} \, .
\ee
The three-dimensional representation in \eqref{branch3} corresponds to the $sl(2,\mathbb{R})$ subalgebra used to implement the DS reduction. Accordingly to this decomposition, the $sl(3,\mathbb{R})$ algebra can be realised in terms of the three $sl(2,\mathbb{R})$ generators $W^1_i \ (i
= -1,0,1)$, two sets of generators $\psi^{[a]}_m
\ (m = -\frac{1}{2},\frac{1}{2}\,; \ a=1,-1 )$ of spin $3/2$ and one generator $W^0_0$ of spin $1$:
\begin{subequations}
\begin{align}
&
\big[\,
W^1_i\, , \, W^1_j
\,\big] \,=\, (i-j)\, W^1_{i+j} \ , \\
&
\big[\,
W^1_i\, , \, W^0_0
\,\big] \,=\, 0 \ ,
\\
&
\big[\,
W^1_i\, , \, \psi^{[a]}_m
\,\big] \,=\, \left(\frac{i}{2}-m \right)\psi^{[a]}_{i+m} \ ,\label{prim_ferm}\\
&
\big[\,
W^0_0\, , \, \psi^{[a]}_m
\,\big] \,=\, a\, \psi^{[a]}_{m} \ , \\
&
\big[\,
\psi^{[a]}_m \, , \, \psi^{[b]}_n 
\,\big] \,=\, \frac{a-b}{2}\left(\, W^1_{m+n} 
\, + \, \frac{3}{2} \,  (a-b) \, m (m-n)^2 \, W^0_0
\,\right) \, .
\end{align}
\end{subequations}
The presence of two sets of generators with half-integer $\ell$ allows to consider linear combinations of them without spoiling \eqref{prim_ferm}, and we defined the $\psi^{[a]}_m$ so that they are eigenvectors of the adjoint action of $W^0_0$. On the other hand, in agreement with the discussion in Section~\ref{sec:red_bulk}, this freedom does not suffice to separate their contributions to the Killing metric.

The DS reduction based on the embedding \eqref{branch3} was first performed independently by Polyakov and Bershadsky \cite{Polyakov} and the resulting $\cW$-algebra is usually denoted by $\cW_3^{(2)}$. In order to perform it with our techniques, it is convenient to introduce the notation 
\be
  a_- \, = \, \frac{2\p}{k} \left(\, \cW_0(\th) \, W^0_0 \, + \,
  \cW_{\frac{1}{2}}^{[1]}(\th) \, \psi_{-\frac{1}{2}}^{[1]}  \, + \, 
  \cW_{\frac{1}{2}}^{[-1]}(\th) \, \psi_{-\frac{1}{2}}^{[-1]} \, + \, \cL(\th) \, W^1_{-1} 
  \,\right) ,
\ee  
and to denote the gauge variation with respect to $\l_+ (\th) =
\e(\th)(W^\ell_\ell)_{[a]}$ as $\d_{\ell\,[a]}$. The general procedure
\eqref{globalsymmetry} leads eventually to the following
transformations preserving the highest-weight gauge:
\begin{subequations}\label{W3GaugeTransf}
\begin{align}
  &\d_{0} \cW_0\, =\,  \frac{k}{2\pi} \, \e' 
  \, , \qquad \d_{0} \cW_{\frac{1}{2}}^{[a]} \, =\,  -\,a \, \e \, \cW_{\frac{1}{2}}^{[a]}
  \, , \qquad \d_{0} \cL \, =\, 0 \, ,
  \\[10pt]
  &\d_{1} \cW_0 \, =\, 0 \, , 
  \qquad \d_{1} \cL \, =\, 
  2\, \e' \, \cL \, + \, \e \, \cL' \, + \, \frac{k}{4\pi}\,\e''' \, ,
  \\
  &\d_{1} \cW_{\frac{1}{2}}^{[a]}\, =\, \frac{3}{2} \, \e' \, \cW_{\frac{1}{2}}^{[a]} \, + \, 
  \, \e \, \cW_{\frac{1}{2}}^{[a]\,\pe} 
  \, + \, a \, \frac{2\p}{k} \, \e \, \cW_{\frac{1}{2}}^{[a]}\, \cW_0 \, , \label{nonPrimaryField_W3}
  \\[10pt]
  &\d_{\frac{1}{2}\, [a]} \cW_{\frac{1}{2}}^{[b]} \, =\, \d_{a}{}^{b}\left(  
  \, - \, \e\, \cL \, - \,  2 \, a \, \e'\,\cW_0
  \, - \,  a \, \e\, \cW_0' \, - \,  
  \, \frac{2\pi}{k} \,  \e \, \cW_0 \, \cW_0 
  \, - \, \frac{k}{2\pi}\, \e''\right) \, .  \label{colorGaugeTrans_W3}
\end{align}
\end{subequations}

The resulting $\cW$-algebra is obtained by substituting
\eqref{W3GaugeTransf} in \eqref{covariant_bracket}, and it coincides
with the one in \cite{Polyakov}. We present it in Appendix
\ref{app:nonPrincipalExp}. To compute the Poisson brackets one has to
know the structure of the Killing metric of $sl(3,\mathbb{R})$. In
our basis it is block-diagonal with one block for each type of field,
except for a mixing in the two sets of spin-$\frac{3}{2}$ generators
$\psi^{[a]}_m$.  In particular, the various matrices $(N_\ell)_{ab}$
of \eqref{M_kill} are \footnote{ In \eqref{N_fermi3} we ordered the
generators such that $\tr\left(\psi^{[1]}_{\frac{1}{2}}
\psi^{[-1]}_{-\frac{1}{2}}\right) = 1$, while
$\tr\left(\psi^{[-1]}_{\frac{1}{2}} \psi^{[1]}_{-\frac{1}{2}}\right) =
-1$.  }
\be \label{N_fermi3}
(N_{\frac{1}{2}})_{ab} = \left(
\begin{array}{cc}
0&1\\
-1&0
\end{array}
\right) \ , \qquad N_0 \ = \ \frac{2}{3} \ .  
\ee
Since one block in the Killing form involves generators with different
colour indices, the structure of the gauge transformation
\eqref{colorGaugeTrans_W3} and the corresponding Poisson bracket
\eqref{coloredPB_W3} differ by more than just a numerical factor,
namely also by the colours of the fields that occur.

The fields
$\cW_{\frac{1}{2}}^{[a]}$ are not Virasoro primaries (see \eqref{nonPrimaryField_W3}). As described in general in
\eqref{SugawaraShift}, and for this example already mentioned in
\cite{Polyakov}, a shift 
\be \label{W3SugawaraShift}
\cL \, \rightarrow \, \cL \, - \,
\frac{2\pi}{3k} \left(\cW_0 \right)^2 \, \equiv \, \widehat{\cL}
\ee
leads to a basis where all fields are primaries with respect to $\widehat{\cL}$.

\subsubsection{A non-principal $sl(2,\mathbb{R})$ embedding in $sl(4,\mathbb{R})$}

We consider here the $sl(2,\mathbb{R})$ embedding in $sl(4,\mathbb{R})$ that branches the fundamental representation as
\be \label{branch4}
\mathbf{\underline{15}} \,=\, \mathbf{\underline{3}} \,+\, \mathbf{\underline{5}} \,+\, 2\cdot \mathbf{\underline{3}} \,+\, \mathbf{\underline{1}} \, .
\ee
The first three-dimensional representation in \eqref{branch4} corresponds to the $sl(2,\mathbb{R})$ subalgebra used to implement the DS reduction. In a HS perspective it would thus be associated to the graviton, while the other two would lead to two coloured massless spin-2 fields. Accordingly to this decomposition, the $sl(4,\mathbb{R})$ algebra can be realised in terms of the three $sl(2,\mathbb{R})$ generators $W^1_i \ (i
= -1,0,1)$, two extra sets of spin-2 generators
$\f^{[a]}_i \ \ (i = -1,0,1$\,;\ $ a=1,-1 )$, one set of spin-3
generators $W^2_m \ \ (m = -2,\ldots,2)$ and the $sl(2)$ singlet $W^0_0$ as
\begin{subequations} \label{basis_sl4}
\begin{align}
&
\big[\,
W^1_i\, , \, W^1_j
\,\big] \,=\, (i-j)\,W^1_{i+j} \ ,
\\
&
\big[\,
W^1_i\, , \, \f^{[a]}_j
\,\big] \,=\, (i-j)\,\f^{[a]}_{i+j} \ ,
\\
&
\big[\,
W^1_i\, , \, W^2_m
\,\big] \,= \left(2i-m \right) W^2_{i+m} \ ,
\\
&
\big[\,
W^1_i\, , \, W^0_0
\,\big] \,=\, 0  \ ,
\\[10pt]
&
\big[\,
W^0_0 \, , \, \f^{[a]}_i
\,\big] \, = \, a \, \f^{[a]}_i \ ,
\\
&
\big[\,
W^0_0 \, , \, W^2_i
\,\big] \, = \, 0 \ ,
\\[10pt]
&
\big[\,
W^2_m \, , \, W^2_n
\,\big] \, = \, -\, \frac{1}{12} \, (m-n)\left(2m^2 \, + \,  2n^2 \, - \, mn \, - \, 8  \right) W^1_{m+n} \ ,
\\
&
\big[\,
W^2_m \, , \, \f^{[a]}_i
\,\big] \, = \, \frac{a}{6}\left( m^2 \, + \, 6 i^2 \, - \, 3mi \, - \, 4 \right)\f^{[a]}_{m+i} \ ,
\\
&
\big[\,
\f^{[a]}_i \, , \, \f^{[b]}_j
\,\big] \, = \, \frac{a-b}{2} \, \bigg(  -
\,a\, (i-j)W^1_{i+j} \, + \, 2 \, W^2_{i+j} - \, \frac{4}{3}\left(2-3|i-j|  \right)\d_{i+j,0} \, W^0_0 \bigg) \, . 
\end{align}
\end{subequations}
As in the previous example, we used the freedom in the definition of the $\f^{[a]}_m$ to let them be eigenvectors of the adjoint action of $W^0_0$.

With the convention
\be
a_- \, = \, \frac{2\p}{k} \left(\, \cW_0(\th)  W^0_0 
 + \, 
\cL(\th) W^1_{-1} 
 + \, 
\cW_1^{[-1]}(\th) \f_{-1}^{[-1]} 
 + \, 
\cW_1^{[1]}(\th) \f_{-1}^{[1]} 
 + \, 
\cW_2(\th) W^2_{-2} 
\,\right) , \label{HW-ex}
\ee
the transformations preserving the highest-weight parameterisation of $a_-$ are
\begin{subequations}
\begin{align}
  &\d_{0} \cW_0\, =\,  \frac{k}{2\pi} \, \e'\, , \qquad 
  \d_{0} \cW_{1}^{[a]} \, =\,  -\, a \,  \e \, \cW_{1}^{[a]} \, ,\qquad 
  \d_{0} \cL \, =\, \d_{0} \cW_2 \, =\, 0 \, ,
  \\[10pt]
  &
  \d_{1} \cL\, = \, 2 \, \e' \, \cL \, + \, \e \, \cL' \, + \, \frac{k}{4\p} \, \e''' \, ,
  \\
  &
  \d_{1} \cW_1^{[a]}\, = \, 2 \, \e' \,  \cW_1^{[a]} \, + \, \e \, \left(\cW_1^{[a]} \right)'
  \, + \, 
  \frac{2\p}{k} \, a\, \e \, \cW_0 \, \cW_1^{[a]} \, ,
  \\
  &
  \d_{1} \cW_2\, = \, 3 \, \e' \, \cW_2 \, + \, \e \, \left(\cW_2\right)' \, ,
   \displaybreak[0]
  \\[10pt]
  &
  \d_{1\,[a]} \cW_1^{[b]} \, =\, \d_{a}{}^b 
  \bigg( 
  \,   \frac{3a}{2}  \, \e'' \, \cW_0
  \, + \, 
  \frac{3a}{2}  \, \e' \, \cW_0'
  \, + \, 
  \frac{a}{2}  \, \e \, \cW_0''
  \nn \\
  &
  \phantom{\d_{1\,[a]} \cW_1^{[b]} \, =\, \d_{a,b} \bigg( }
  + \,  2 \, \e' \, \cL \, + \, \e \, \cL' 
  \, + \, 
  \frac{k}{4\p}\, \e''' 
  \, + \, 2\, a \,  \e \, \cW_2 
  \nn\\
  &
  \phantom{\d_{1\,[a]} \cW_1^{[b]} \, =\, \d_{a,b} \bigg( }
  + \, \frac{2\p}{k} \, 2 \, a \, \e \, \cW_0 \, \cL
  \, + \
  \frac{2\p}{k} \, \frac{3}{2} \, \e'  \, \cW_0 \, \cW_0
  \, + \
  \frac{2\p}{k} \, \frac{3}{2} \, \e \,  \cW_0 \, \cW_0'
  \nn \\
  &
  \phantom{\d_{1\,[a]} \cW_1^{[b]} \, =\, \d_{a,b} \bigg( }
  + \
   \left(\frac{2\p}{k}\right)^2  \frac{a}{2} \, \e \,  \cW_0 \, \cW_0 \,  \cW_0
  \bigg) \, ,  
  \\[10pt] 
  &\d_{1\,[a]} \cW_2 \, =\, 
  - \frac{5a}{3} \,  \e'' \, \cW_1^{[-a]}
  \, -  
   \frac{5a}{6} \, \e' \, \left( \cW_1^{[-a]} \right)'
  \, - \
  \frac{a}{6} \, \e \, \left( \cW_1^{[-a]} \right)''
 \nn \\
  &
  \phantom{\d_{1\,[a]} \cW_2 \, =\, }
  - \,  
  \frac{2\p}{k} \, a \, \frac{8}{3} \, \e \, \cL \, \cW_1^{[-a]} 
  \, - \, 
  \frac{2\p}{k} \, \frac{5}{2} \, \e' \, \cW_0 \, \cW_1^{[-a]} 
  \, - \, 
  \frac{2\p}{k} \, \frac{3}{2} \, \e \, \cW_0' \, \cW_1^{[-a]} 
  \nn\\
  &
  \phantom{\d_{1\,[a]} \cW_2 \, =\, } 
   - \, 
   \frac{2\p}{k} \, \frac{a}{2} \, \e \, \cW_0 \, \left(\cW_1^{[-a]}  \right)'
  - \, \left(\frac{2\p}{k}  \right)^2  a \, \e \, \cW_0 \, \cW_0 \, \cW_1^{[-a]} \, ,
  \\[10pt]
  &
  \d_{2} \cW_2 \, =\, \frac{5}{6} \,  \e''' \, \cL 
  \, + \, 
  \frac{5}{4} \,  \e'' \, \cL' 
  \, + \, 
  \frac{3}{4} \,  \e' \, \cL'' 
  \, + \,  
  \frac{1}{6} \,  \e \, \cL''' 
  \, + \,  
  \frac{k}{48\pi}  \, \e^{(5)}
  \nn \\
  &
  \phantom{\d_{\e_2} \cW_2 \, =\, }
  +\frac{2\pi}{k}\, \, \frac{8}{3} \, \e' \, \cL \, \cL 
  \, + \, 
  \frac{2\pi}{k} \, \frac{8}{3} \, \e \, \cL \, \cL'
  \, + \, 
  \frac{2\pi}{k}\, 8 \, \e' \, \cW_1^{[-1]} \, \cW_1^{[1]} \, 
  \nn\\
  &
  \phantom{\d_{\e_2} \cW_2 \, =\, }
  + \
  \frac{2\pi}{k}\, 4 \,  \e \, \left(\cW_1^{[-1]}\right)' \, \cW_1^{[1]} \, 
  \, + \
  \frac{2\pi}{k}\, 4 \,  \e \, \cW_1^{[-1]} \, \left(\cW_1^{[1]}\right)' 
  \, . 
\end{align}
\end{subequations}

In the basis \eqref{basis_sl4} the Killing metric splits into blocks for the different
field types except for a mixing in the two extra sets of spin-2
generators. More concretely, the matrices $(N_\ell)_{ab}$ of \eqref{M_kill} become
\be (N_1)_{ab} \, = \, \left(
\begin{array}{ccc}
0&1\\
1&0
\end{array}
\right) \, , \qquad N_0 \, = \, \frac{3}{16} \, , \qquad N_2 \, = \, 1 \, .
\ee 
In this case the matrix $(N_1)_{ab}$ can be clearly diagonalised, but this would spoil our choice of working with $W^0_0$ eigenvectors. Note, however, that its eigenvalues are $\pm 1$, so that the kinetic terms of the two coloured spin-2 fields would have opposite sign in the action \eqref{EH}.

To obtain a $\cW$-algebra in a Virasoro-primary basis, one can again shift $\cL$ as
\begin{equation}\label{W4SugawaraShift}
\cL \,\rightarrow\,  \cL \, - \,  \frac{3\pi}{16k} \, \cW_0\,  \cW_0
\,\equiv\, \widehat{\cL} \, .
\end{equation}
The corresponding Poisson algebra is presented in Appendix \ref{app:nonPrincipalExp}.

\section{The structure of $\cW_{\infty}[\l]$ in a primary basis}\label{sec:W}

In this section we use the techniques developed in Section \ref{sec:reduction} to study an interesting one-parameter family of higher-spin gauge theories. Their gauge algebras are the direct sum of two copies of the infinite-dimensional Lie algebras called $hs[\l]$. After a brief introduction to these algebras we shall stress their link with higher spins by discussing the relation between a suitable basis of their invariant tensors and Fronsdal's metric-like fields. For any $\l$ we shall then compute the structure constants of the $\cW$-algebra of asymptotic symmetries in a Virasoro-primary basis.

\subsection{The higher-spin algebras $hs[\l] \oplus hs[\l]$}\label{sec:W_hs}

In Section \ref{sec:red_bulk} we have seen that in any
three-dimensional HS gauge theory a crucial role is played by the
gauge subalgebra that describes the gravitational sector of the
model. Instead of choosing it among all possible embeddings in a given
algebra, one can actually proceed in a different direction and build
HS algebras out of products of generators of the ``gravitational''
$sl(2,\mathbb{R}) \oplus sl(2,\mathbb{R})$ gauge algebra.

For instance, following \cite{BBS,Pope,Vasiliev,FL,hoppe_proof,GH}, 
we start with the universal enveloping algebra of
$sl(2,\mathbb{R})$ generated by $J_{\pm}$ and $J_{0}$. We then do the
identification 
\be
C_2 \,:=\, J_0^2 \,-\, \frac{1}{2} \left(\, J_+J_- + J_-J_+ \,\right) \,\equiv\, \m\, \mathds{1} \, ,
\ee
which sets the quadratic Casimir $C_{2}$ to a particular value $\mu$
that we often parameterise as
\be \label{mu}
\m \,=\, \frac{1}{4} \left(\, \l^2 - 1 \,\right) \, .
\ee
The algebra obtained in that way is spanned by the identity
$\mathds{1}$ and the elements
\be \label{gen}
W^\ell_m :=\, (-1)^{\ell-m}\, \frac{(\ell+m)!}{(2\ell)!}\, L^{\,\ell-m}_- J^{\,\ell}_+  \, , \qquad \ell \geq 1 \, , \quad -\,\ell \leq m \leq \ell \, ,
\ee
where $L_i$ denotes the adjoint action of $sl(2,\mathbb{R})$ as in
\eqref{adjoint}. From their definition it is manifest that they
satisfy the commutators
\be \label{primary2}
[\, W^1_m \,,\, W^\ell_n \,] \,=\, (\ell\, m - n)\, W^\ell_{m+n} \, ,
\ee
and we can identify the generators with $\ell = 1$ with the
$sl(2,\mathbb{R})$ ones. The whole set of $W^\ell_m$ generates a Lie
algebra $hs[\lambda]$ whose remaining commutators are fixed by the $sl(2)$
commutators of eq.\ \eqref{commutation-sl2}. It branches as
\be \label{branch_hs}
hs[\l] \,=\, \bigoplus_{\ell\,=\,1}^\infty \, \mathfrak{g}^{(\ell)}
\ee
under the adjoint action of the defining $sl(2)$ subalgebra. Different
values of the parameter $\m$ (related to $\l$ by \eqref{mu}) give algebras that are not isomorphic
\cite{BBS,hoppe_proof}.

This construction shows that one can identify $hs[\l]$ with the
subspace orthogonal to the identity in the quotient of the universal
enveloping algebra of $sl(2,\mathbb{R})$ by the ideal generated by
$(C_2-\m \mathds{1})$ \cite{BBS,Pope,Vasiliev,FL,hoppe_proof,GH},
\be
\frac{\cU(sl(2,\mathbb{R}))}{\bra C_2-\m \mathds{1} \ket} \,=\, hs[\l] \,\oplus\, \mathbb{C} \, .
\ee
The whole quotient is an associative algebra. The product of the $W^\ell_m$ was given in \cite{Pope} (see also \cite{Vasiliev}) as
\be \label{WstarW}
W^k_m \,\star\, W^\ell_n \,=\, \frac{1}{2}\, \sum_{i\,=\,|k-\ell|}^{k+\ell} \,
f_\l \! \left(
\begin{array}{cc|c}
k & \ell & i \\
m & n & m+n
\end{array}
\right)
W^{i}_{m+n} \, ,
\ee
where the identity is denoted by $W^0_0$ and the $\l$-dependent structure constants are defined in Appendix \ref{app:pope}. The product \eqref{WstarW} allows one to realise the Lie product on $hs[\l]$ as a $\star$-commutator: 
\be \label{[W,W]}
[\, W^k_m \,,\, W^\ell_n \,] \,= \sum_{\substack{i\,=\,|k-\ell|+1 \\[2pt] i+k+\ell \ \textrm{odd}}}^{k+\ell-1} 
f_\l \! \left(
\begin{array}{cc|c}
k & \ell & i \\
m & n & m+n
\end{array}
\right)
W^{i}_{m+n} \, .
\ee

The associative product \eqref{WstarW} was also used in \cite{Vasiliev} to define an invariant bilinear form on $hs[\l]$ as
\be \label{lambda_trace}
\tr\left(\, W^k_m W^\ell_n \,\right) :=\, \frac{6}{(\l^2-1)} \left. W^k_m \,\star\, W^\ell_n \,\right|_{\, W^{i}_p \,=\, 0 \ \textrm{for} \ i \,>\, 0} \, ,
\ee
i.e.\ by extracting the term proportional to the identity from the
product. The invariant form \eqref{lambda_trace} allows to define a CS
action based on the algebra $hs[\l] \oplus hs[\l]$ as in
\eqref{EH}. Eq.~\eqref{branch_hs} then shows that -- for a generic
value of $\l$ -- the spectrum of the corresponding HS gauge theory
contains all integer spins from $2$ to $\infty$, and each of them
appears only once.  However, when $\l$ is integer, the invariant form
\eqref{lambda_trace} degenerates as one can see from its explicit
expression:
\begin{subequations} \label{KillingForm} 
\begin{align} 
\tr\left(\, W^k_m W^\ell_n \,\right) \,&=\,
(-1)^{\ell-m}\, N_\ell(\l) \, \frac{(\ell+m)!(\ell-m)!}{(2\ell)!}\ \d^{k,\ell}\, \d_{m+n,0} \, , \\[5pt]
N_\ell(\l) \,&=\, -\, \frac{6\,(\ell !)^{2}}{(2\ell +1)!}\, \prod_{i\,=\,2}^\ell\, (i-\l)(i+\l) \, . \label{N_l}
\end{align}
\end{subequations}
The normalisation factors $N_\ell(\l)$ follow from the
definitions in Appendix \ref{app:pope}. For integer $\l = N$ the CS
action \eqref{EH} thus actually corresponds to that of a
$sl(N,\mathbb{R}) \oplus sl(N,\mathbb{R})$ theory.\footnote{The
normalisation factor that we introduced in \eqref{lambda_trace} thus
plays a double role: on the one hand it gives $\tr\left(\, W^1_1
W^1_{-1} \,\right) = -1$. On the other hand it guarantees that $\l =
1$ still provides a gauge theory involving all spins from 2 to
$\infty$, as in \cite{minimal}.} Another interesting value of the
parameter is $\l = 1/2$ that gives the three-dimensional
Fradkin-Vasiliev algebra \cite{frad-vas}.

Before concluding this review, let us notice that the $W^\ell_m$ with
odd $\ell$ form a subalgebra of $hs[\l]$ that we denote by
$ho[\l]$. As a result, one could well build a HS gauge theory on top
of $ho[\l] \oplus ho[\l]$. For $\l \notin \mathbb{N}$ its spectrum
contains all even integer spins greater than zero.\footnote{A similar
truncation is available also in higher space-time dimensions where it
leads to the so called minimal Vasiliev models (see
e.g. \cite{vas-rev}).} For $\l \in \mathbb{N}$ the truncation to even
spins of the $A_N \oplus A_N$ CS theories leads to $B_N \oplus B_N$
gauge algebras for odd $N$ and to $C_N \oplus C_N$ gauge algebras for
even $N$ \cite{FL}, while the $D$ series of simple Lie algebras cannot
be recovered in this fashion. 

Moreover, commutators of generators with even $\ell$ (corresponding to
fields of odd spin!) can be always expanded in a sum of generators
with odd $\ell$. On the other hand, mixed commutators (one odd and one
even $\ell$) always give only terms with even $\ell$. As a result, it
is possible to rescale by $i$ all generators with even $\ell$ to get a
different real form of $hs[\l]$. As already mentioned, for integer $\l
= N $ the previous construction gives the $sl(N,\mathbb{R})$ real
algebra, i.e. the maximal non-compact real form of $sl(N)$. The
rescaling by $i$ of the generators with even $\ell$ leads to the
next-to-maximal non-compact real form of $sl(N)$,
i.e. $su(\frac{N-1}{2},\frac{N+1}{2})$ for odd $N$ or
$su(\frac{N}{2},\frac{N}{2})$ for even $N$. Additional comments on the
possible real forms for $\l \notin \mathbb{N}$ can be found in
\textsection 3.1 of \cite{Pope}. Note, however, that the rescaling by
$i$ reverses the sign of the normalisation factors $N_{2j}$ in
\eqref{KillingForm} and, in turn, of the kinetic terms of the fields
of odd spin. The maximal non-compact real form seems thus to be
preferred to build a HS gauge theory. As noticed in \cite{spin3},
additional subtleties emerge if one consider CS actions built upon two
different real forms of the same gauge algebra. In the following we
shall avoid all of them by focusing on CS theories based on gauge
algebras $hs[\l] \oplus hs[\l]$, with $hs[\l]$ fixed by \eqref{gen}
and \eqref{[W,W]}.

\subsubsection{Metric-like fields and invariant tensors of $hs[\l]$} \label{sec:W_metric}

The previous discussion about $hs[\l]$ suffices to fully characterise
a classical HS gauge theory for any admissible value of $\l$, in a
form which generalises the frame formulation of Einstein
gravity. However, even sticking to this algebraic framework one can
extract some information on the ``metric-like'' formulation of these
theories, involving Lorentz-invariant symmetric fields $\vf_{\m_1
\ldots\, \m_s}$ (see e.g. \cite{rev-symm} for a review). To this end,
it is crucial to realise that the two classes of gauge transformations
discussed in Section \ref{sec:red_bulk} play a very different
role. Those generated by $\x$ in \eqref{gauge_fields} generalise local
translations, while those generated by $\L$ generalise local Lorentz
transformations. As stressed in \cite{spin3}, all metric-like fields
should be invariant under local Lorentz-like
transformations. Moreover, one can write them in terms of the
vielbeine, since the spin connections are just auxiliary fields. In
the $sl(3) \oplus sl(3)$ CS theory discussed in detail in \cite{spin3}
these conditions suffice to fix the structure of all fields in the
spectrum: the metric $g_{\m\n}$ and the spin-3 field
$\vf_{\m\n\r}$. Collecting all vielbeine in the vector $e^A \equiv
e_\m{}^A dx^\m$ and all generators of the algebra in the vector $T_A$,
up to a normalisation constant they read
\begin{align}
& g \,\sim\, \tr\,(\,e\cdot e\,) \,=\, e^A e^B\, \tr\,(\,T_A T_B\,) \, , \label{metric} \\[5pt]
& \vf_3 \,\sim\, \tr\,(\,e\cdot e\cdot e\,) \,=\, \frac{1}{3!}\, e^A e^B e^C\, \tr\,(\,T_{(A}T_{B}T_{C)}\,) \, , \label{spin3}
\end{align}
where the parentheses denote the symmetrisation of the indices they enclose, with unit normalisation.
The cyclicity of the trace guarantees the extended Lorentz invariance. The same result holds for the trace of an arbitrary power of the vielbeine,
\be
\d_\L e \,=\, [\,e\,,\,\L\,] \quad \Rightarrow \quad \d_\L \tr \,( e^n ) \,=\, n\, \tr\, ( e^{n-1} [\,e\,,\,\L\,] ) \,=\,0 \, .
\ee
As a result, for $s > 3$ the invariance under Lorentz-like transformations does not suffice to fix the structure of $\vf_{\m_1 \ldots\, \m_s}$. For instance, for $s=4$ one can consider both $\tr(e^4)$ and $\tr(e^2)\tr(e^2)$ and one has to single out the linear combination that defines $\vf_{\m\n\r\s}$. This freedom corresponds to the existence of two Lorentz-like invariant combinations of rank-4: $\vf_{\m\n\r\s}$ and $g_{(\m\n}\,g_{\r\s)}$.

The realisation of the Lie algebra $hs[\l]$ as a $\star$-commutator algebra proposed in \cite{Pope,Vasiliev} provides a powerful tool to analyse this problem at least for the first values of the spin. In fact, 
\be
k_{A_1 \ldots\, A_s} \,\equiv\, \frac{1}{s!}\, \tr\,(\,T_{(A_1} \ldots T_{A_s)}\,) \,:=\, \frac{6}{(\l^2-1)\,s!}\, \left. T_{(A_1} \star \,\ldots\, \star\, T_{A_s)}\, \right|_{\,T_A\,=\,0}
\ee
is a symmetric invariant tensor of $hs[\l]$ (which coincides with the
Killing metric \eqref{lambda_trace} for $s=2$). Its contraction with
the vielbeine $e^A$ gives a Lorentz-like invariant tensor of rank
$s$.\footnote{See \cite{observables} for a similar construction for $D
> 3$. \label{general_metric}} Metric-like fields should then result
from the contraction of the vielbeine with the elements of a
particular \emph{basis} of the polynomial ring of invariant tensors of
$hs[\l]$. This is clear for all $A_N \oplus A_N$, $B_N \oplus B_N$ and
$C_N \oplus C_N$ CS theories that can be extracted from the $hs[\l]
\oplus hs[\l]$ one. In fact, their spectra are given by the exponents
of the gauge algebras, and are thus in one to one correspondence with
the ranks of their independent Casimir operators. Each Casimir
operator is, in turn, uniquely associated to a symmetric invariant
tensor (see, for instance, \cite{casimir} and references therein). It
is natural to suppose that the same is true even for non-integer $\l$.

Since the relative coefficients between different invariant tensors of
the same rank cannot be fixed by the extended Lorentz invariance, they
should be fixed by the additional requirements that $\vf_{\m_1
\ldots\, \m_s}$ must satisfy:
\begin{enumerate}
\item it has to be doubly traceless as its linearised counterpart (see e.g. \cite{rev-symm});
\item in the linearised regime its rewriting in terms of the vielbeine
has to reproduce the definition in a free theory. 
\end{enumerate}
To impose the first condition one should invert the general definition of the metric \eqref{metric}. For this reason we refrain from discussing it here, deferring to future work a full discussion of the problem. On the other hand, the second condition is more tractable and already suffices to fix the structure of spin-4 and spin-5 fields for any $\l$. 

The linearised definition of $\vf_{\m_1 \ldots\, \m_s}$ can be most conveniently recalled by describing the vielbein $e^{\,\ell,m}$ of \eqref{viel_lm} as a symmetric traceless tensor $e^{a_1 \ldots\, a_\ell}$ (that has $2\ell+1$ independent components as $e^{\,\ell,m}$). 
Denoting the background vielbein by $\bar{e}_\m{}^{a}$ and the linearised fluctuations by $h_\m{}^{a_1 \ldots\, a_{s-1}}$, for $s > 2$ in a linear regime one has
\be \label{lin_metric}
\vf_{\m_1 \ldots\, \m_s} \sim\, \bar{e}_{(\m_1}{}^{a_1} \ldots\, \bar{e}_{\m_{s-1}}{}^{a_{s-1}} h_{\m_s)\,a_1 \ldots\, a_{s-1}} \, .
\ee 
This means that, differently from $g_{\m\n}$, it is usually assumed that higher-spin fields do not receive any background contribution, while the tracelessness of $h^{a_1 \ldots\, a_{s-1}}$ guarantees the doubly tracelessness of $\vf_{\m_1 \ldots\, \m_s}$.

Imposing the matching of the most general Lorentz-like invariant combination with \eqref{lin_metric} one obtains
\begin{align}
& \vf_4 \,\sim\, \tr\,e^4 \,-\, \frac{1}{10}\, (3 \l^2 - 7) \left(\, \tr\,e^2 \,\right)^2 \, , \label{phi4} \\[5pt]
& \vf_5 \,\sim\, \tr\,e^5 \,-\, \frac{5}{21}\, (3 \l^2 - 13)\ \tr\, e^2\, \tr\,e^3 \, . \label{phi5}
\end{align}
However, starting from $s = 6$ this comparison does not suffice to fix all free coefficients. In fact, the most general Lorentz-like invariant combination reads
\be
\vf_6 \,\sim\, \tr\,e^6 \,+\, \a(\l)\, \tr\,e^2\, \tr\,e^4 \,+\, \b(\l) \left(\,\tr\,e^2\,\right)^3 \,+\, \g(\l) \left(\,\tr\,e^3\,\right)^2 .
\ee
The condition \eqref{lin_metric} gives
\be \label{metric6}
\a(\l) \,=\, -\, \frac{5}{6}\, (\l^2 - 7) \, , \qquad \b(\l) \,=\, \frac{1}{42}\,(6 \l^4 - 71 \l^2 + 125) \, ,
\ee
but the term $\tr(e^3)$ does not admit any background contribution. As
a result, $\left(\,\tr\,e^3\,\right)^2$ does not contribute at first
order as well, and the matching with \eqref{lin_metric} cannot put any
constraint on $\g(\l)$. Some extra information follows from the
observation that all $\vf_s$ we were able to identify vanish in the
$sl(N) \oplus sl(N)$ theories with $N < s$. This is expected because
in these cases there are no fields of spin $s$ in the
spectrum.\footnote{A general expression for a basis of the centraliser
of $sl(N)$ whose elements vanish when their rank exceeds $N$ was
considered in \cite{casimir}. All tensors in this basis are orthogonal
to each other and then, in particular, they are all traceless. On the
other hand, the invariant tensors that give the metric-like fields do
not satisfy this property. Therefore, they coincide with those in
\cite{casimir} only for $\l = N < s$.}  One can thus impose the same
condition on $\vf_6$ and this forces $\g(\l) = (1/3)(\l^2-1)$ for $N =
3, 4, 5$. It is however not clear to us if this condition suffices to
also force the double trace constraint for arbitrary values of $\l$,
while the uniqueness of \eqref{phi4} and \eqref{phi5} should ensure
the double tracelessness of $\vf_4$ and $\vf_5$. An explicit check
would anyway provide a non-trivial consistency check of the whole
construction, and we hope to report on it soon.
 
\subsection{Gauge transformations preserving the highest-weight gauge}\label{sec:W_gauge}

We now take the algebras of the last subsection as starting point for
a DS reduction in the highest-weight gauge, and we determine the
structure of the corresponding family of infinite-dimensional
$\cW$-algebras. Since no $sl(2)$-singlets appear in \eqref{branch_hs},
all generators will be Virasoro primaries. The asymptotic symmetries
of the $hs[\l] \oplus hs[\l]$ CS theories that we just discussed are
given by two copies of the resulting $\cW$-algebras, that we denote by
$\cW_\infty[\l]$ as in \cite{GH}. The cases with $\l \in \mathbb{N}$
and $\l = 1/2$ were already discussed in \cite{spin3}
and \cite{HR}, respectively. A discussion of the general case was also anticipated
in \cite{GH}. In these works, however, the computation of structure
constants was completed only for fields of spin $s \leq 3$ (see also
\cite{W_infinity,W_inf_math} for an earlier treatment of
$\cW_\infty[\l]$ algebras and \cite{Khesin} for an abstract proof that
they are actually related to the DS reduction of $hs[\l]$).

Before displaying the structure constants, let us notice that one can
easily evaluate the maximum order of non-linearity appearing in the Poisson brackets.  
Consider a gauge parameter of definite spin, $\lambda_{+}= \e(\theta) W^\ell_\ell$. When we act on it with the covariant derivative entering \eqref{globalsymmetry}, the term in $W^{\ell_1}_{-\ell_1}$ gives a result with $\ell_{tot} \leq \ell+\ell_1-1$ and $L_0$-eigenvalue $\ell-\ell_1$. If we apply the covariant derivative $r$-times, we arrive at a $L_{0}$-eigenvalue $\sum \ell_{i}-\ell$ in a representation with $\ell_{tot} \leq \sum \ell_{i} + \ell-r$. In addition, the action of $D_\theta$
is accompanied by at least $r-1$ applications of $L_{-}$, that means
the $L_{0}$-eigenvalue is $\sum \ell_{i}-\ell +r-1$. Clearly, if
the $L_{0}$-eigenvalue exceeds $\ell_{tot}$, the expression vanishes,
and this happens if $-\ell+r-1>\ell-r$, i.e.\ $r\geq \ell+1$. In conclusion, in
the Poisson brackets of a field with $sl (2)$ label $\ell$ there can be
at most a non-linearity of order $\ell$, as in the Virasoro polynomials discussed in Section \ref{subsec:virasoro}. Actually, as we shall see, the pure-Virasoro terms are the only ones that saturate this bound. This limitation also accounts, for instance, for the linearity of the Virasoro algebra, that only contains spin-2 fields, and for the quadratic order of non-linearity of the $\cW_3$ algebra.

As in subsections \ref{subsec:central} and \ref{subsec:virasoro} the
structure constants can be computed by considering the gauge variation
induced by a parameter of given spin, say $\l_+ =
\e(\theta)W^i_i$. The details of the evaluation of the series
\eqref{globalsymmetry} are presented in Appendix \ref{app:proofs},
while here we directly present our result. In this case the general
decomposition \eqref{exp_a} can be cast in the form 
\be \label{a_W}
a_-(\theta) \,=\, \frac{2\pi}{k}\, \sum_{j\,=\,1}^\infty \, \cW_j(\theta)\, W^{j}_{-j} \, ,
\ee
where we identified $\cL$ with $\cW_1$ since no ambiguities can arise due to the absence of colour indices. The gauge variation of each $\cW_j(\theta)$ with respect to $\l_+ = \e(\theta)W^i_i$ reads
\be \label{closedFormula1}
\begin{split}
& \d_i \cW_j \,=\, \frac{k}{2\pi(2i)!}\,\e^{(2i+1)}\, \d_{i,j} \\
& +\ \sum_{r\,=\,1}^i \, \sum_{\substack{L\,=\,|i-j|+r \\[2pt] i+j+L+r\ \textrm{even}}}^{i+j-r} \sum_{\{a_t\}} \ \sum_{\{p_t\}}\, C[i,j]_{a_1 \ldots\, a_r;\,p_1 \ldots\, p_r}\, \cW_{a_1}^{(p_1)} \ldots\, \cW_{a_r}^{(p_r)}\, \e^{(\hat{n}-\sum_1^r p_t)} \, .
\end{split}
\ee
Here, as in \eqref{var_vir}, an exponent between parentheses denotes the action of the corresponding number of derivatives on the field, while $\hat{n}$ denotes the total number of derivatives which is
\be
\hat{n} :=\, i+j-L-r+1 \, .
\ee
For each $r$ the sums over $a$'s and $p$'s distribute over these indices the ``total spin'' $L$ and the total number of derivatives $\hat{n}$. They thus read
\begin{subequations} \label{sums}
\begin{align}
& \sum_{\{a_t\}} \,:=\, \sum_{a_1\,=\,1}^L \, \sum_{a_2\,=\,1}^{L-a_1} \ \ldots \sum_{a_{r-1}\,=\,1}^{L-\sum_1^{r-2}a_t} \, \d_{a_r,\,L-\sum_1^{r-1}a_t} \ , \\
& \sum_{\{p_t\}} \,:=\, \sum_{p_1\,=\,0}^{\hat{n}} \, \sum_{p_2\,=\,0}^{\hat{n}-p_1} \ \ldots \sum_{p_{r}\,=\,1}^{\hat{n}-\sum_1^{r-1}p_t} \, .
\end{align}
\end{subequations}
Note that for $i=j$, the terms that saturate the bound on the order of
non-linearity are the pure Virasoro terms of
Section~\ref{subsec:virasoro}. On the other hand, for $j<i$ the upper
extremum of the sum over $r$ cannot be reached due to the collapsing
of the sum over $L$. As a result, all Poisson brackets involving
fields of labels less or equal to $\ell$ contain polynomials of order
strictly lower than $\ell$. The only exception is $\{
\cW_{\ell}(\theta) , \cW_{\ell}(\theta') \}$ that contains a
polynomial of order $\ell$ involving only $\cL$ and its first
derivative. Eq.~\eqref{closedFormula1} also shows that the Poisson
brackets (obtained from~\eqref{special_bracket} below) of
$\cW_{\infty}[\lambda]$ are invariant under the map
\begin{equation}\label{automorphism}
\cW_{i} \to (-1)^{i+1} \cW_{i} \ ,
\end{equation}
which is thus an automorphism of $\cW_{\infty}[\l]$. 

In addition to these structural results, we can even provide a closed
formula for the structure constants:
\begin{align}
& C[i,j]_{a_1 \ldots\, a_r;\,p_1 \ldots\, p_r} =\, \frac{(-1)^{\hat{n}+r-1}}{(2j)!} \left(\frac{2\pi}{k}\right)^{\!\!r-1} \, \sum_{q_1\,=\,p_1}^{\hat{n}}\, \sum_{q_2\,=\,\langle(p_1+p_2)-q_1\rangle_+}^{\hat{n}-q_1} \ldots \sum_{q_r\,=\,\langle \sum_1^r p_t - \sum_1^{r-1} q_t \rangle_+}^{\hat{n}-\sum_1^{r-1} q_t} \nn \\
& \times \sum_{\substack{b_1\,=\,\max\left(|a_{r}-b_0|+1\,,\,M(r-1,\,j)\right) \\[3pt] a_{r}+b_0+b_1 \ \textrm{even}}}^{\min\left(a_{r}+b_0-1 \,,\, \sum_1^{r-1} a_t + j - r + 1\right)} \ \ldots \
\sum_{\substack{b_{r-1}\,=\,\max\left(|a_{2}-b_{r-2}|+1\,,\,M(1,\,j)\right) \\[3pt] a_{2}+b_{r-1}+b_r \ \textrm{even}}}^{\min\left(a_{2}+b_{r-2}-1 \,,\, a_1 + j - 1\right)} \nn \\[3pt]
& \times\, \prod_{s\,=\,1}^r \binom{\sum_1^s q_t - \sum_1^{s-1} p_t}{p_s}\, [\, j + b_s - \sum_1^{r-s}a_t - \sum_1^{r-s+1}q_t - r + s \,]_{a_{r-s+1}-b_{s-1}+b_s} \nn \\[2pt]
& \times\, f_\l \! \left(
\begin{array}{cc|c}
a_{r-s+1} & b_{s-1} & b_s \\[2pt]
-\, a_{r-s+1} & -\, j + \sum_{1}^{r-s+1} a_t + \sum_1^{r-s+1} q_t + r - s & \ldots
\end{array}
\right) \, . \label{ClosedFormula}
\end{align}
Here $[a]_n$ denotes the descending Pochhammer symbol defined in \eqref{poch-}, while $\langle a \rangle_+ = \max(0,a)$ and
\be
M(s,\ell) :=\, 2\max\left(\{a_t\}_{t=1}^s\,,\,\ell\,\right) - \sum_{t\,=\,1}^s a_t - \ell + s \, .
\ee
By substituting $\l = N$ with $N$ integer in \eqref{ClosedFormula} one obtains a closed formula for the structure constants of the classical $\cW_N$ algebra. Note that eq.~\eqref{ClosedFormula} expresses the structure constants of $\cW_\infty[\l]$ in terms of those of $hs[\l]$. The techniques of Appendix \ref{app:proofs} actually allow to express the structure constants of any classical $\cW$-algebra (that can be obtained by a DS reduction) in terms of those of the corresponding Lie algebra.

\subsubsection{Gauge transformations up to spin 4}\label{firstTerms}

In order to better elucidate the structure of our result, we now use the closed formula \eqref{ClosedFormula} to present all gauge transformations $\d_i\cW_j$ with $i,j<4$. Letting $\l = 4$ the following results determine the structure constants of $\cW_4$. We shall focus on variations with $i \leq j$, since those with $j > i$ do not lead to new independent Poisson brackets. Moreover, in each $\d_i\cW_j$ the gauge parameter will be always denoted by $\e$, since no ambiguities can arise. As in \eqref{closedFormula1}, $\d_i\cW_j$ denotes indeed the component along $W^j_{-j}$ of the gauge variation of $a_-(\theta)$ induced by the gauge parameter $\l_+ = \e(\theta) W^i_i$.

The first gauge transformations just display that all fields are
primaries with respect to the Virasoro field $\cL \equiv \cW_1$,
\begin{subequations} \label{W4}
\begin{align}
&\d_1 \cL \, =\,  \e\,\cL' \,+\, 2\,\e'\cL \,+\, \frac{k}{4\pi}\,\e''' \ , \label{vir_W4} \\[5pt]
&\d_1 \cW_\ell \, =\,  \e\,\cW_\ell' \,+\, (\ell+1)\,\e'\,  \cW_\ell \ , \qquad \ell = 2,3,... \ .
\end{align}
Augmenting the conformal weight of the gauge parameter by one we find
\begin{align}
  \d_2 \cW_2 \, &=\, -\, \frac{3}{7}\, (\l^2-9) \left[\,
     \cW_3'\,\e + 2\, \cW_3\,\e' \,\right] \,+\, \frac{1}{12} \left[\, 2\,  \cL'''\e + 9\,   \cL''\e' 
       +15\,  \cL' \e''  +10\, \cL\,\e''' \,\right] \nn \\[2pt] 
  & +\, \frac{16\pi}{3k} \left[\, \cL \cL'\e+ \cL^2 \e' \,\right]
     \,+\, \frac{k}{48\pi}\, \e^{(5)} \, . \label{gaugeTransExp3}
\end{align}    
If one substitutes $\l = 3$ in \eqref{gaugeTransExp3}, the
transformations that we already displayed suffice to build the $\cW_3$
algebra (see e.g. \cite{spin3}) and, as already anticipated, they
contain at most quadratic polynomials in the generators. For generic $\lambda$
one can also look at the component along $W^3_{-3}$ of
the gauge variation induced by $\e(\theta)W^2_2$,
\begin{align}
  \d_2 \cW_3 \, & =\,  -\, \frac{2}{9}\, (\l^2-16) \left[\, 2\, \cW_4' \,  \e + 5\, \cW_4 \,  \e'   \,\right]  \nn \\[2pt]
  & +\, \frac{1}{15} \left[\, \cW_2'''\,\e + 6\, \cW_2''\,\e'+ 14\, \cW_2'\,\e''
    +14\, \cW_2\, \e''' \,\right] \nn \\[2pt]
  & +\, \frac{4\pi}{15k}\left[\, 25\, \cL'\,\cW_2  \,\e
    + 18\, \cL\, \cW_2'\,\e + 52\, \cL\, \cW_2 \,\e'  \,\right] \, . \label{d2W3}
\end{align}
The next higher transformation is
\begin{align}
  \d_3 \cW_3 \, & =\, 
  \frac{5}{33}\, (\l^2-16)(\l^2-25) \left[\, \cW_5'\, \e + 2\  \cW_5\, \e' \,\right] \nn \\
  & -\, \frac{1}{30} \ (\l^2-19) \left[\, \cW_3'''\, \e  
    + 5\, \cW_3''\, \e'
    + 9\, \cW_3\, \e''  + 6\, \cW_3\, \e'''  \,\right]\nn\\
  & +\, \frac{1}{360}\left[\, 3 \, \cL^{(5)} \e  
    + 20\, \cL^{(4)} \e' +56\,  \cL^{(3)}\e''
    +  84 \, \cL'' \e{}^{(3)} + 70 \, \cL' \e{}^{(4)} 
    +28\, \cL\, \e{}^{(5)} \,\right]\nn \\[5pt]
  & -\, \frac{2\pi}{15k}\ (29\l^2-284) 
  \left[\, \cW_2 \cW_2'\, \e + \cW_2^2\,  \e' \,\right] \nn \\
  & -\, \frac{28\pi}{15k}(\l^2-19) \left[\, \cL'\, \cW_3\, \e + \cL\, \cW_3'\,\e + 2\, \cL\, \cW_3\,\e' \,\right]  \nn\\
  & +\, \frac{\pi}{90k}
  \left[\, 177 \,   \cL' \cL''\e  + 78\, \cL \cL''' \e 
    + 295\, \cL'^2 \e' 
    +\, 352\, \cL  \cL''\e'
    + 588\, \cL \cL' \e''     + 196 \cL^2 \e'''
    \,\right] \nn\\[5pt]
  & + \frac{32\pi^2}{5k^2} 
  \left[\, 3\, \cL^2\cL' \e   + 2\,  \cL^3 \e' \,\right]
  \,+\, \frac{k}{1440\pi}\, \e{}^{(7)} \, . \label{fin_W4} 
\end{align}
\end{subequations}
Here a cubic polynomial involving only the Virasoro generators appears, while all other polynomials are at most quadratic, in agreement with the discussion following \eqref{sums}. Setting $\lambda=4$ in this and the previous transformations, one
obtains the whole $\cW_4$ algebra.

The gauge variations (\ref{vir_W4}--\ref{fin_W4}) agree with those
presented in eqs.~(3.25--3.29) of \cite{GH}, where their fields
$L_{s}$ are identified with ours by 
\begin{equation}\label{identificationGH}
L_s(\th) = N_{s-1}(\l)\cW_{s-1}(\th)
\end{equation}
involving the factors $N_{i}$ defined in~\eqref{N_l}. The
normalisation chosen in \cite{GH} leads to Poisson brackets that do
not contain poles in $\l$, so that it is possible to discuss the
truncation to $\cW_N$ directly at that level. The price to pay is the
presence of poles in $\l$ in the definition of
$a_{-}(\th)$. Therefore, strictly speaking, some intermediate steps of
the DS reduction are not well defined for integer $\l$.

\subsection{Poisson bracket algebra}\label{sec:W_poisson}

The Poisson brackets of $\cW_\infty[\l]$ can be obtained from the
gauge transformations \eqref{closedFormula1} following the general
procedure outlined at the end of Section \ref{sec:red_red}. In this
case there are no colour indices so that \eqref{charge_fix} reads
\be \label{charge_W}
Q(\l_+) \,=\, -\, \sum_{\ell\,=\,1}^\infty \, N_\ell(\l) \int d\theta\, \e_\ell(\theta)\, \cW_\ell(\theta) \, ,
\ee
where $N_\ell(\l)$ denotes the normalisation factors that we
introduced in \eqref{N_l}. Eq.~\eqref{covariant_bracket} thus
simplifies to 
\be
\big\{ \cW_i(\th) \,,\, \cW_j(\th') \big\} \,=\, - \, \frac{1}{N_{i} (\lambda)}\,
\d_{i} \, \cW_j (\theta ') 
\Big|_{\e (\theta ')\,=\,\delta (\theta -\theta ')} \, . \label{special_bracket}
\ee
Note that for integer $\l$ a diverging factor can appear in
this expression since $N_i(\l)$ vanishes for $i \geq \l$. However, in
these cases \eqref{special_bracket} is clearly meaningless since the
charges involving $\cW_i$ vanish as well. For $\l \in \mathbb{N}$ the
Poisson algebra is thus generated by the $\cW_i$ with $i < \l$. The
Poisson brackets that one obtains with this procedure have the same
structure as the gauge transformations \eqref{closedFormula1}, barring
the substitution
\be
\e^{(n)}(\theta') \,\to \, (-1)^n\, \pr^{(n)}_\theta\d(\th-\th') \, ,
\ee 
and the normalisation factors $N_{i} (\lambda)$. The Poisson brackets
associated to the explicit gauge transformations \eqref{W4} are
collected in Appendix~\ref{app:nonPrincipalExp}.

\section{A quadratic basis for $\cW_\infty[\l]$}\label{sec:quadratic}

In the last section we have discussed the algebras
$\cW_{\infty}[\lambda]$ that arise from the DS-reduction of
$hs[\lambda]$ in the highest-weight gauge. The corresponding Poisson
brackets $\{\cW_{i},\cW_{j} \}$ are non-linear expressions in
the fields where the degree of non-linearity is bounded by the minimum of $i$ and $j$,
so that if one considers fields of higher and higher spins,
arbitrarily high degrees of non-linearity can appear. On the other
hand, the algebra $\cW_{\infty}[\lambda]$ has a basis such that the
Poisson brackets involve at most quadratic polynomials in the
fields. In this section, we want to derive the explicit basis
transformation that relates theses two bases.

We start by a general discussion of gauge freedom in the
DS-reduction. We then present in Section~\ref{sec:Recursive} a recursive
algorithm that can be used to determine the transformation to the
highest-weight basis for the DS-reduction of any Lie
algebra. In Section~\ref{sec:quadraticWN} a quadratic basis for $\cW_N$ is
reviewed, and in Section~\ref{sec:BasistransWN} the basis transformation from
the highest-weight basis to the quadratic basis is determined. These
results are used in Section~\ref{sec:BasistransWinf} to derive the
corresponding basis transformation for $\cW_{\infty}[\l]$.
\medskip

We have seen in Section~\ref{sec:reduction} that the asymptotic
$\cW$-symmetries arise from the symmetry
transformations~\eqref{gaugetransformation} with the
constraint~\eqref{DrinfeldSokolov} that implements the asymptotic AdS
condition. The gauge transformations generated by these constraints
can be used to choose a certain gauge, e.g.\ the highest-weight
gauge~\eqref{aconstraint} for $a (\theta)$ that we used in the last
sections. This gauge choice is very natural, but in some situations
other choices can be more convenient.

To discuss this gauge freedom we introduce the differential operators 
\begin{equation}
l_{a} = \partial_{\theta} + a (\theta) \ ,
\end{equation}
where we have chosen a representation of the Lie algebra
$\mathfrak{g}$. On such a differential operator we can act with
($\theta$-dependent) group elements $g (\theta)$ by conjugation,
\begin{equation}
l_{a} \mapsto g^{-1}(\th) l_{a} g (\theta) \ ,
\end{equation}
which implies the transformation
\begin{equation}
a (\theta) \mapsto  g^{-1} (\theta)\partial_{\theta}g (\theta) 
+ g^{-1}(\theta)a (\theta)g(\theta)
\end{equation}
on $a (\theta )$. For infinitesimal transformations $g
(\theta)=1+\lambda (\theta)$ this reduces
to~\eqref{gaugetransformation}. The AdS or Drinfeld-Sokolov
condition~\eqref{DrinfeldSokolov} is respected by transformations 
$g(\theta)$ that take values in the subgroup $\cN\subset G$ that is
generated by $\mathfrak{g}_{>}$ (defined in \eqref{firstdecomposition}). On each of the gauge orbits
$g^{-1} l_{a} g$ with $g\in \cN$ we can pick a representative
corresponding to a certain gauge choice.

\subsection{Recursive basis transformation}
\label{sec:Recursive}
Given any such representative $\partial_{\theta}+J_{+} +b(\theta)$, we
now describe a procedure how to recursively find the representative
$\partial_{\theta}+J_{+}+a_{-} (\theta)$ in the highest-weight gauge  
on the same orbit, and the gauge transformation $g (\theta)\in \cN$ that
connects the two representatives,
\begin{equation}\label{basistransform}
g (\theta)^{-1} \left(\partial_{\theta} +J_{+}+b (\theta) \right) g(\theta)
= \partial_{\theta} +J_{+} + a_{-} (\theta) \ .
\end{equation}
We write the representation matrices $g(\theta)$ as $g(\theta
)=1+h(\theta)$, where $h(\theta)$ has an expansion 
\begin{equation}
h(\theta)= h^{1}(\theta) + h^{2}(\theta) + \dotsb \ , 
\end{equation}
with $L_{0}h^{n}=n h^{n}$, $n>0$.
From~\eqref{basistransform} we obtain
\begin{equation}\label{rearrangedbasistransform}
-L_{+} h(\theta) + a_{-}(\theta) = 
\partial h(\theta) + b(\theta) + b(\theta)h(\theta) 
- h(\theta)a_{-}(\theta)\ .
\end{equation}
When we act on this equation by $R L_{-}$, where $R$ is the operator that was defined in \eqref{Roperator}, we arrive at
\begin{equation}\label{recursionh}
h(\theta ) = - R L_{-}\left(\partial h(\theta) + b(\theta)
+ b(\theta)h(\theta) - h(\theta) a_{-}(\theta) \right) \ . 
\end{equation}
On the other hand, when we act on~\eqref{rearrangedbasistransform} by
the projector $P_{-}$, we find
\begin{equation}\label{recursiona}
a_{-}(\theta) = P_{-} \left(\partial h(\theta) + b(\theta) +
b(\theta)h(\theta) - h(\theta)a_{-}(\theta) \right) \ .
\end{equation}
These two equations, \eqref{recursionh} and~\eqref{recursiona}, can be
used to construct the basis transformation recursively. Let
\begin{equation}
b(\theta) = b^{0}(\theta) + b^{1}(\theta) + \dotsb \quad ,\quad 
a_{-}(\theta) = a_{-}^{0}(\theta) + a_{-}^{1}(\theta)+ \dotsb \ ,
\end{equation}
where $L_{0}b_{n} = nb_{n}$, and similarly for $a_{-}^{n}$.
Projecting~\eqref{recursionh} to $L_{0}$ eigenvalue $n$ we
find
\begin{equation}\label{explicitrecursionh}
h^{n}  = - R L_{-}\left(\partial h^{n-1} 
+ b^{n-1}  + \sum_{1 \leq m \leq n-1} b^{n-m-1} h^{m} 
-\sum_{1\leq m \leq n-1} h^{m} a_{-}^{n-m-1} \right)\ .
\end{equation}
Analogously, we can project~\eqref{recursiona} to
$L_{0}$ eigenvalue $n$ to obtain
\begin{equation}\label{explicitrecursiona}
a_{-}^{n}  = P_{-}\left(\partial h^{n}
+ b^{n} + \sum_{1\leq m \leq n} b^{n-m} h^{m}
- \sum_{1\leq m\leq n} h^{m} a_{-}^{n-m}  \right)\ .
\end{equation}
Suppose that $b(\theta)$ is given, and that we have determined
$h^{m}(\theta)$ and $a_{-}^{m}(\theta)$ for $m<n$. We can
use~\eqref{explicitrecursionh} to obtain $h^{n}(\theta)$, and then
determine $a_{-}^{n}(\theta)$ by~\eqref{explicitrecursiona}. 

\subsection{A quadratic basis for $\cW_{N}$}
\label{sec:quadraticWN}
We have seen how to construct in general the basis transformation to
the highest-weight basis recursively in the eigenvalues of $L_{0}$. We
now want to become more specific for the Lie algebras $sl (N)$ and the
corresponding $\cW$-algebras $\cW_{N}$, for which another natural basis
exists. We use the defining representation by square matrices of size
$N$, in which $J_{+}$ is represented as
\begin{equation}\label{defofJplus}
J_{+} = \begin{pmatrix}
0 & & & &\\
-1 & 0 & & &\\
0 & -1 & 0 & &\\
  & \ddots & \ddots & \ddots & \\
 & & 0 & -1 & 0 
\end{pmatrix} \ .
\end{equation}
The Drinfeld-Sokolov condition states that $a (\theta)$ is of the form 
\begin{equation}\label{DSforsln}
a (\theta) = J_{+} + u (\theta) \ ,
\end{equation}
where $u (\theta)$ is upper triangular. The subgroup $\cN$ consists
of all matrices of the form $1+h$ where $h$ is strictly upper
triangular. A natural gauge choice is to demand that $u (\theta)$ only has one
non-zero row (the first one),
\begin{equation}\label{formofu}
u = \begin{pmatrix}0 & u_{1} & \dotsb & u_{N-1}\\
 0 & 0 & \dotsb & 0\\
\vdots & \vdots & & \vdots\\
0 & 0 & \dotsb  & 0
\end{pmatrix}\ .
\end{equation}
With this choice one obtains a basis for the $\cW_{N}$ algebras that has
at most quadratic non-linearities in the Poisson brackets. The nice
feature of this basis is that there is a simple way to express the
gauge transformations and thus the Poisson brackets with the help of
pseudo-differential operators, which we shall briefly review in the
following (for details see e.g.\ \cite{Dickey}).

The infinitesimal transformations $g (\theta)=1+\lambda (\theta)$ that
respect the gauge choice are uniquely determined in terms of the first
column of $\lambda$. We introduce the differential operators
\begin{equation}
\lambda_{i} = \sum_{j\,=\,0}^{N-1} \lambda_{ij}\,\partial^{N-j-1}
\end{equation}
of order at most $N-1$ corresponding to the rows of $\lambda$, the
pseudo-differential operator 
\begin{equation}
\lambda^{0} = \sum_{i\,=\,0}^{n-1} \partial^{-N+i} \lambda_{0j} 
\end{equation}
corresponding to the first column, and the differential operator
\begin{equation}
L= \partial^{N} + u_{1}\partial^{N-2} + \dotsb +u_{N-1}
\end{equation}
associated to the gauge fields.
The matrix $\lambda_{ij}$ is then given by the equations
\begin{equation}
\lambda_{i} = \partial^{N-i-1} (\lambda^{0}L)_{+} - (\partial^{N-i-1} \lambda^{0})_{+}L
\end{equation}
in terms of its first column, where $( \ )_{+}$ denotes the projection of a
pseudo-differential operator to its regular part (the non-negative
powers of the differential). This is the analogue
of~\eqref{solution-lambda} in the highest-weight gauge. The gauge
transformation can also be expressed in such a way,
\begin{equation}\label{Adlermap}
\delta_{\lambda} L = L (\lambda^{0} L)_{+} - (L\lambda^{0})_{+}L \ .
\end{equation}
From here one can easily obtain the Poisson brackets. As $L$ only
appears quadratically on the right-hand side of~\eqref{Adlermap}, the
Poisson brackets are at most quadratic in the fields $u_{i}$.
This is quite different from the highest-weight basis, where the
degree of non-linearity is $N-1$.

\subsection{Basis transformation for $\cW_{N}$}
\label{sec:BasistransWN}
The obvious question arises how the gauge choice of the last
subsection and the highest-weight gauge are related. As the
corresponding Poisson brackets have different orders of
non-linearities, the basis transformation has to be non-linear. We
could use our recursive construction from Section~\ref{sec:Recursive}
to relate these two bases, but there is another more elegant way of
relating them. The operator $l_{a}=\partial_{\theta} + a (\theta)$ can
be used to define a system of $N$ first order differential equations
on some functions $f_{0},\dotsc ,f_{N-1}$,
\begin{equation}
l_{a} (f_{0},\dotsc ,f_{N-1})^{T} = 0 \ .
\end{equation}
Due to the specific form~\eqref{DSforsln} of $a (\theta)$ this system
of differential equations is equivalent to a single $N^{\text{th}}$
order differential equation for $f_{N-1}$. The other functions
$f_{0},\dotsc ,f_{N-2}$ are then given in terms of $f_{N-1}$ by
\begin{equation}
f_{j} = \det \left((\partial + J_{+} + u )_{r,s} \right)_{j+1\leq r,s
\leq N-1} f_{N-1} \ ,
\end{equation}
where the determinant is evaluated with the convention that the
entries of the first row appear to the left of entries of the second
row, and so on. The scalar differential operator $L$ that determines
the differential equation for $f_{N-1}$,
\begin{equation}
L f_{N-1} = 0 \ ,
\end{equation}
can also be expressed as a determinant~\cite{Dickey},
\begin{equation}\label{detformula}
L= \det (\partial + J_{+} + u) \ .
\end{equation}
Two differential operators $l_{a}$ and $l_{a'}$ that are related by a
gauge transformation in $\cN$ give rise to the same scalar
differential operator $L$. If $u$ is of the form~\eqref{formofu} (all
rows vanish except for the first), we have
\begin{equation}\label{detinubasis}
L = \partial^{N} + u_{1}\partial^{N-2} + .. + u_{N-2}\partial + u_{N-1}
\ .
\end{equation}
By evaluating the operator $L$ we can thus relate any choice of
representatives to the $u$-basis.

We now want to work out the basis transformation explicitly for the
highest-weight basis of $sl(N)$. The matrix representing $J_{-}$ reads
\begin{equation}\label{defofJminus}
J_{-} = \begin{pmatrix}
 0& 1\cdot (N-1) & 0      &        & \\
  & 0     & 2\cdot (N-2) &  0     & \\
  &       & \ddots & \ddots &   \\
  &       &        & 0      & (N-1)\cdot 1\\
  &       &        &        & 0
\end{pmatrix} \ .
\end{equation}
The highest-weight vectors in $sl(N)$ are $W^{m}_{-m}=J_{-}^{m}$ with 
$m=1,\dotsc ,N-1$, so that $a_{-}$ can be expanded as
\begin{equation}
a_{-}(\theta) = \frac{2\pi}{k}\left( \cW_{1}(\theta) J_{-} + \dotsb 
+ \cW_{N-1}(\theta) J_{-}^{N-1}\right) \ ,
\end{equation}
or, as matrix,
\begin{equation}
a_{-} = \frac{2\pi}{k}
\begin{pmatrix}
0  &  \cW_{1} P_{1}(1) & \cW_{2} P_{2}(1) & \cW_{3}P_{3}(1) & \dotsb & \cW_{N-1}P_{N-1}(1)\\
0  &  0              & \cW_{1} P_{1}(2) & \cW_{2}P_{2}(2) & \dotsb & \cW_{N-2}P_{N-2}(2)\\
0  &  0              & 0              & \cW_{1}P_{1}(3) & \dotsb & \cW_{N-3}P_{N-3}(3)\\
   &                 &                & \ddots        &        & \\
   &                 &                &               &  0      & \cW_{1}P_{1}(N-1)\\
   &                 &                &               &         & 0
\end{pmatrix} \ .
\end{equation}
Here, the $P_{s}$ are defined in terms of ascending and descending
Pochhammer symbols (see~\eqref{poch+} and~\eqref{poch-}),
\begin{equation}
P_{s}(i) = (i)_{s}[N-i]_{s} \ .
\end{equation}
The determinant~\eqref{detformula} is evaluated by summing
over all permutations of the $N$ columns while respecting the row
ordering ($l=\partial + J_{+} + a_{-}$), 
\begin{equation}
L = \det l = \sum_{\sigma \in S_{N}} l_{1\sigma(1)}l_{2\sigma(2)}
\dotsb l_{N\sigma(N)} \ .
\end{equation}
We can order the terms by their degree of non-linearity in
the $\cW_{m}$. The term without any $\cW_{m}$ is simply $\partial^{N}$. A
linear term comes from the part of the determinant where exactly one
entry of the upper triangle contributes:
\begin{equation}
\begin{pmatrix}
\ddots &          &        &         &              & \\
       & \partial & \dotsb & \dotsb & \cW_{s}P_{s}(i+1) & \\
       & -1       & \ddots &        &  \vdots       & \\
       &          & \ddots & \ddots &  \vdots       & \\
       &          &        & -1     &  \partial     & \\
       &          &        &        &               & \ddots 
\end{pmatrix} \ .
\end{equation}
There is only one permutation such that besides $\cW_{s}P_{s}(i+1)$ only
$\partial$'s and $(-1)$'s appear leading to the contribution
\[
\frac{2\pi}{k}\partial_{\theta}^{i} \cW_{s}P_{s}(i+1) 
\partial_{\theta}^{N-s-i-1} \ .
\]
The total contribution linear in $\cW_{s}$ is then
\begin{align}
L_{W_{s}} & = \frac{2\pi}{k}\sum_{i\,=\,0}^{N-s-1} \partial_{\theta}^{i}\, \cW_{s}P_{s}(i+1)
\partial_{\theta}^{N-s-i-1} \\
& = \frac{2\pi}{k}\sum_{i\,=\,0}^{N-s-1} \sum_{p\,=\,0}^{i} \bin{i}{p} P_{s}(i+1) \cW_{s}^{(p)}
\partial_{\theta}^{N-s-p-1} \\
& = \frac{2\pi}{k}\sum_{p\,=\,0}^{N-s-1}\sum_{i\,=\,p}^{N-s-1} \bin{i}{p} P_{s}(i+1)\cW_{s}^{(p)}\partial_{\theta}^{N-s-p-1}\\
& = \sum_{p\,=\,0}^{N-s-1} C(s,p,N) \cW_{s}^{(p)}\partial_{\theta}^{N-s-p-1} \ .
\end{align}
The coefficients $C(s,p,N)$ can be written as
\begin{align}
C(s,p,N) &= \frac{2\pi}{k}\sum_{i\,=\,p}^{N-s-1}\bin{i}{p} P_{s-1}(i+1)\\
 & = \frac{2\pi}{k}\sum_{j\,=\,0}^{N-s-p-1}\frac{(j+1)_{p}}{p!} (j+p+1)_{s}
 [N-j-p-1]_{s} \\
 & = \frac{2\pi}{k}\sum_{j\,=\,0}^{N-s-p-1} \frac{(j+1)_{p+s}}{p!} (N-s-p-j)_{s}\\
 & = \frac{2\pi}{k} s! \bin{p+s}{p}(N-s-p)_{s} 
\sum_{j\,=\,0}^{N-s-p-1}\frac{(s+p+1-N)_{j}(p+s+1)_{j}}{(1+p-N)_{j}j!}\\
 & = \frac{2\pi}{k} (s!)^{2} \bin{p+s}{p} \bin{N-p-1}{s} 
{}_{2}F_{1}(s+p+1-N,s+p+1;1+p-N;1) \ .
\end{align}
The hypergeometric function $_{2}F_{1}$ at argument $1$ is given by
\begin{equation}
{}_{2}F_{1}(s+p+1-N,s+p+1;1+p-N;1) =
\frac{(N-p)_{s+p+1}}{(s+1)_{s+p+1}} \ ,
\end{equation}
so that after some manipulations we find
\begin{equation}\label{linearC}
C(s,p,N) = \frac{2\pi}{k} (s!)^{2} \bin{p+s}{p} \bin{N+s}{2s+p+1} \ .
\end{equation}
The linear contribution to the determinant is thus
\begin{equation}
L_{\text{linear}}= \frac{2\pi}{k}\sum_{s}(s!)^{2}\sum_{p} \bin{p+s}{p}
\bin{N+s}{2s+p+1} \cW_{s}^{(p)}\partial_{\theta}^{N-s-p-1} \ . 
\end{equation}

The higher order terms appear as products of such linear blocks, and
the total determinant is given by
\begin{align}
L & = \sum_{r\,=\,0}^{\lfloor \frac{N}{2}\rfloor} 
\left(\frac{2\pi}{k}\right)^{r} \sum_{s_{1},\dotsc
,s_{r}} \sum_{i_{1},\dotsc
,i_{r}}\partial_{\theta}^{i_{1}} \cW_{s_{1}}p_{s_{1}}(i_{1}+1) \nonumber\\
& \quad \times \partial_{\theta}^{i_{2}}
\cW_{s_{2}}P_{s_{2}}(i_{1}+i_{2}+s_{1}+2) \dotsb
\cW_{a_{r}}P_{a_{r}}(i_{1}+\dotsb +i_{r}+s_{1}+\dotsb
+s_{r-1}+r)\partial_{\theta}^{i_{r+1}} \ .
\end{align}
Here, $i_{r+1}=N-r-\sum_{1}^{r}i_{j} -\sum_{1}^{r}s_{j}$, and all
$i_{j}\geq 0$, $s_{j}\geq 1$. As in the linear example above, we
commute all differentials to the right, which produces derivatives of
the fields $\cW_{s_{j}}$, e.g.\ for the first field $\cW_{s_{1}}$ we have
\begin{equation}
\partial_{\theta}^{i_{1}} \cW_{s_{1}}(\theta) =
\sum_{m_{1}\,=\,0}^{i_{1}}\bin{i_{1}}{m_{1}}\cW_{s_{1}}^{(m_{1})}(\theta)
\partial_{\theta}^{i_{1}-m_{1}} \ .
\end{equation}
For the second field $\cW_{s_{2}}$ we find
\begin{equation}
\partial_{\theta}^{i_{1}+i_{2}-m_{1}} \cW_{s_{2}}(\theta) = 
\sum_{m_{2}\,=\,0}^{i_{1}+i_{2}-m_{1}} \bin{i_{1}+i_{2}-m_{1}}{m_{2}}
\cW_{s_{2}}^{(m_{2})}(\theta)\partial_{\theta}^{i_{1}+i_{2}-m_{1}-m_{2}} \ ,
\end{equation}
and we introduce similar summation variables $m_{j}$ counting the
derivatives of the other fields $\cW_{s_{j}}$. We also introduce the
variables 
\begin{equation}
k_{l} = \sum_{j\,=\,1}^{l}i_{j} \quad ,\quad  p_{j}= \sum_{j\,=\,1}^{l} m_{j} \ .
\end{equation}
The determinant $L$ then becomes
\begin{align}
L &= \sum_{r} \left(\frac{2\pi}{k} \right)^{r}
\sum_{s_{1},\dotsc ,s_{r}} \ \sum_{0\leq k_{1}\leq \dotsb
\leq k_{r}\leq N-r-S} \
\sum_{0\leq p_{1}\leq \dotsb \leq p_{r}, p_{j}\leq k_{j}} \nonumber \\
& \quad  \cW_{s_{1}}^{(p_{1})} \cW_{s_{2}}^{(p_{2}-p_{1})} \dotsb
\cW_{s_{r}}^{(p_{r}-p_{r-1})} \partial_{\theta}^{N-r-p_{r}-S}\nonumber \\
& \quad  \times \bin{k_{1}}{p_{1}}\bin{k_{2}-p_{2}}{p_{2}-p_{1}} \dotsb
\bin{k_{r}-p_{r-1}}{p_{r}-p_{r-1}}\nonumber \\
& \quad  \times P_{s_{1}}(k_{1}+1) P_{s_{2}} (k_{2}+s_{1}+2) \dotsb 
P_{s_{r}} (k_{r}+ \sum_{j\,=\,1}^{r-1}a_{j} + r)\ ,
\end{align}
where $S=\sum_{j=1}^{r}s_{j}$.
This means that a given term $\cW_{s_{1}}^{(p_{1})}\dotsb
\cW_{s_{r}}^{(p_{r}-p_{r-1})}\partial_{\theta}^{N-r-p_{r}-S}$
appears with a coefficient  
\begin{align}
C (\{s_{j} \},\{ p_{j}\},N) &=\left(\frac{2\pi}{k} \right)^{r}
\sum_{k_{r}\,=\,p_{r}}^{N-S-r}
\sum_{k_{r-1}=\,p_{r-1}}^{k_{r}} \dotsb \sum_{k_{1}\,=\,p_{1}}^{k_{2}}
\bin{k_{1}}{p_{1}}\dotsb \bin{k_{r}-p_{r-1}}{p_{r}-p_{r-1}}\nonumber\\
&\quad  \times (k_{1}+1)_{s_{1}}  \dotsb 
(k_{r}+\sum_{j=1}^{r-1}s_{j}+r)_{s_{r}} \nonumber\\
&\quad \times (N-k_{1}-s_{1})_{s_{1}} \dotsb (N-k_{r}
-\sum_{j=1}^{r}s_{j}-r+1)_{s_{r}} \ .
\label{btcoefficient}
\end{align}
By comparison with~\eqref{detinubasis}, these coefficients determine
the basis transformation,
\begin{equation}
\label{ubasistransform}
u_{q}\, =\,  \sum_{r=1}^{\lfloor \frac{q+1}{2}\rfloor} 
\sum_{\substack{\{s \}\\ S+r \,\leq\, q+1}}  \
\sum_{\substack{\{p \}\\ p_{r}\,=\,q+1-r-S}}
C (\{s_{j} \},\{p_{j} \},N)  \cW_{s_{1}}^{(p_{1})}\dotsb
\cW_{s_{r}}^{(p_{r}-p_{r-1})} \ .
\end{equation}

The coefficients~\eqref{btcoefficient} depend in a simple way on
the number of derivatives, $p_{j}$, and we can combine
the coefficients belonging to a certain set of fields of spins
$s_{1},\dotsc ,s_{r}$ into a generating function with auxiliary
variables $\alpha_{j}$,
\begin{align}
\tilde{C} (\{s_{j} \},\{\alpha_{j}\},N) &= \left(\frac{2\pi}{k} \right)^{r}
\sum_{p_{r}=\,0}^{N-r-S} \sum_{p_{r-1}=\,0}^{p_{r}} \dotsb 
\sum_{p_{1}=\,0}^{p_{2}} C (\{s_{j}\},\{p_{j}\},N)
\alpha_{1}^{p_{1}}\alpha_{2}^{p_{2}-p_{1}} \dotsb 
\alpha_{r}^{p_{r}-p_{r-1}}\\
& = \sum_{k_{r}=\,0}^{N-r-S} \dotsb
\sum_{k_{1}=\,0}^{k_{2}} (k_{1}+1)_{s_{1}} \dotsb 
(k_{r}+\sum_{j\,=\,1}^{r-1}s_{j}+r)_{s_{r}} \nonumber\\
& \quad \times (N-k_{1}-s_{1})_{s_{1}} \dotsb 
(N-k_{r}-\sum_{j\,=\,1}^{r}s_{j}-r+1)_{s_{r}}
\nonumber \\
& \quad \times (1+\alpha_{r})^{k_{r}-k_{r-1}}
(1+\alpha_{r}+\alpha_{r-1})^{k_{r-1}-k_{r-2}} \dotsb 
(1+\alpha_{r}+\dotsb +\alpha_{1})^{k_{1}}\ .
\label{tildeCexpr1}
\end{align}
By going back to the variables $i_{j}=k_{j}-k_{j-1}$, we can write it
as a generalised hypergeometric function,
\begin{align}
& \tilde{C} (\{s_{j} \},\{\alpha_{j} \},N) = 
\left(\frac{2\pi}{k} \right)^{r}(1)_{s_{1}}
(N-s_{1})_{s_{1}} \dotsb (r+\sum_{j\,=\,1}^{r-1}s_{j})_{s_{r}} 
(N-r+1-S)_{s_{r}}\nonumber \\
&\qquad \times \sum_{i_{1},\dotsc ,i_{r}\geq 0} 
\frac{(1+s_{1})_{i_{1}} (1-N+s_{1})_{i_{1}}}{(1)_{i_{1}} (1-N)_{i_{1}}}
\dotsb \frac{(r+\sum_{j=1}^{r}s_{j})_{i_{1}+\dotsb +i_{r}} 
(r-N+\sum_{j=1}^{r}s_{j})_{i_{1}+\dotsb
+i_{r}}}{(r+\sum_{j=1}^{r-1}s_{j})_{i_{1}+\dotsb +i_{r}}
(r-N+\sum_{j=1}^{r-1}s_{j})_{i_{1}+ \dotsb +i_{r}}} \nonumber \\
&\qquad \times (1+\alpha_{r})^{i_{r}} \dotsb (1+\alpha_{r}+\dotsb
+\alpha_{1})^{i_{1}} \ .
\label{tildeC}
\end{align}
The sum is bounded by $i_{1}+\dotsb +i_{r}\leq
N-r-S$ because the last Pochhammer symbol in the
numerator would vanish otherwise.

\subsection{Basis transformation for $\cW_{\infty}[\lambda]$}
\label{sec:BasistransWinf}
In the last subsection we have determined the basis transformation
that relates the highest-weight basis and a quadratic basis for
$\cW_{N}$. A quadratic basis also exists for $\cW_{\infty}[\lambda]$. On
the one hand, \cite{W_infinity} generalised the construction of
$\cW_{N}$ via pseudo-differential operators to a one-parameter family of
infinite dimensional $\cW$-algebras with quadratic non-linearities. On the
other hand, \cite{Khesin} showed that these algebras can be
understood as a Drinfeld-Sokolov reduction of $hs[\lambda]$. In the
work of \cite{Khesin}, they used a realisation of $hs[\lambda]$ in
terms of infinite matrices, in which $J_{+}$ and $J_{-}$ are given by
the infinite analogues of~\eqref{defofJplus} and~\eqref{defofJminus}
with $N$ replaced by $\lambda$. 

From this construction it is obvious that also the basis
transformation of the last subsection can be generalised to 
$\cW_{\infty}[\lambda]$: we have to work out the transformation for
large matrices and then replace the explicit $N$-dependence in the
coefficients by $\lambda$. Of course, the size of the matrices is also
determined by $N$, which means that we first have to take $N$
large, until a given coefficient $C (\{s_{j} \},\{p_{j} \},N)$ of the
basis transformation stabilises to a certain polynomial in $N$, and
then replace $N$ by $\lambda$. 

The main difficulty is now that in the
expression~\eqref{btcoefficient} the number $N$ also appears in the
range of the sums, and thus we cannot directly replace $N$ by
$\lambda$. Instead we would have to perform the sum to get an
expression which manifestly is a polynomial in $N$ such that we can do
the replacement. In the example of the linear contribution this is
easily possible: the sum can be performed and the final
expression~\eqref{linearC} is written manifestly as a polynomial in
$N$ of degree $2s+p+1$.

For the non-linear terms we shall instead encode the coefficients in
terms of generating functions, in which the dependence of the
summation range on $N$ can be avoided. We already introduced the
functions $\tilde{C} (\{s_{j} \},\{\alpha_{j} \},N)$ with the
auxiliary variables $\alpha_{j}$, which are generalised hypergeometric
functions. In the expression~\eqref{tildeC} for $\tilde{C}$ we could
leave out the $N$-dependent restriction in the summation range,
because this restriction is automatically implemented by the vanishing
of one of the Pochhammer symbols, so that we could replace $N$ by
$\lambda$. This, however, does not lead to the correct answer, because
the function obtained in such a way does not depend polynomially on
$\lambda$. Let us illustrate this in the example of the linear
terms. For them, the generating function is
\begin{align}
\tilde{C} (s,\alpha ,N) & = \frac{2\pi}{k} (1)_{s} (N-s)_{s}
\sum_{i\,\geq\, 0} \frac{(1+s)_{i} (1-N+s)_{i}}{(1)_{i} (1-N)_{i}}
(1+\alpha)^{i} \\
& = \frac{2\pi}{k} (1)_{s} (N-s)_{s} 
{}_{2}F_{1} (1+s,1-N+s;1-N;1+\alpha) \ .
\end{align}
The function that one obtains by replacing $N$ by $\lambda$ is not
continuous at $\lambda =N$, but instead one has
\begin{align}
& \lim_{\lambda \to N} (\lambda -s)_{s} \, 
{}_{2}F_{1} (1+s,1-\lambda +s;1-\lambda ;1+\alpha)\nonumber\\ 
& \qquad = (N -s)_{s}  \, 
{}_{2}F_{1} (1+s,1-N +s;1-N ;1+\alpha)\nonumber\\
& \qquad \quad   + (-1)^{s} (1+N)_{s} (1+\alpha)^{N}
{}_{2}F_{1} (1+s,1+N+s;1+N ;1+\alpha) \ .
\end{align}
This suggests to extrapolate the function $\tilde{C} (s,\alpha ,N)$ to
\begin{align}
\tilde{C} (s,\alpha ,\lambda) &= \frac{2\pi}{k} (1)_{s} (\lambda -s)_{s} \, 
{}_{2}F_{1} (1+s,1-\lambda +s;1-\lambda ;1+\alpha) \nonumber\\
& \quad - \frac{2\pi}{k}(1)_{s} (-1)^{s} (1+\lambda)_{s} (1+\alpha)^{\lambda}
{}_{2}F_{1} (1+s,1+\lambda+s;1+\lambda ;1+\alpha) \\
& = \frac{2\pi}{k} (s!) (\lambda -s)_{2s+1} 
\bigg(\frac{\Gamma (\lambda)}{\Gamma (1+s+\lambda)} 
{}_{2}F_{1} (1+s,1+s-\lambda;1-\lambda ;1+\alpha)\nonumber\\
& \quad + \frac{\Gamma (-\lambda)}{\Gamma (1+s-\lambda)} (1+\alpha)^{\lambda}
 {}_{2}F_{1} (1+s,1+s+\lambda;1+\lambda ;1+\alpha)
 \bigg)\ .
\end{align}
An elementary transformation of the hypergeometric function leads to
\begin{equation}\label{linearChyper}
\tilde{C} (s,\alpha ,\lambda) = 
\frac{2\pi}{k} (s!)^{2}\bin{\lambda +s}{2s+1}
{}_{2}F_{1} (1+s,1+s-\lambda ;2+2s;-\alpha) \ .
\end{equation}
The coefficient of each power $\alpha^{p}$ is a polynomial in
$\lambda$, which coincides with $C (s,p,N)$ in~\eqref{linearC} when we
replace $N$ by $\lambda$.

In principle one can apply this procedure also to the general
non-linear terms. One replaces $N$ by $\lambda$ in the generalised
hypergeometric function, and determines the discontinuity at $\lambda
=N$ to find the correct extrapolation to $\lambda \not= N$. The
function one obtains in such a way, however, is not of practical use,
unless one finds a transformation to a (generalisation of a)
hypergeometric function that has an expansion in $\alpha_{i}$. Such a
step might be difficult to perform in general, therefore we shall
pursue another strategy here. Again, this is illustrated in the case
of the linear terms, where we already know the correct answer.

Let us go back to the expression~\eqref{tildeCexpr1} for
$\tilde{C}$. In the linear case it reads
\begin{equation}
\tilde{C} (s,\alpha ,N) = \frac{2\pi}{k}\sum_{i\,=\,0}^{N-s-1}
(i+1)_{s} (N-i-s)_{s} (1+\alpha)^{i} \ .
\end{equation}
Now, $(N-i-s)_{s}$ is the coefficient of $\gamma^{s}$ in
$(s!)(1+\gamma)^{N-i-1}$, and $(i+1)_{s}$ is the coefficient
of $\beta^{s}$ in $(-1)^{s} (s!) (1+\beta)^{-i-1}$.
With the help of the auxiliary variables $\beta$ and $\gamma$ we can
then write $\tilde{C}$ as
\begin{equation}
\tilde{C}(s,\alpha ,N) = \frac{2\pi}{k} (s!)^{2} (-1)^{s}
\sum_{i\,=\,0}^{N-s-1} (1+\gamma)^{N-i-1} (1+\beta)^{-i-1} (1+\alpha)^{i}
\Big|_{\gamma^{s}\beta^{s}} \ ,
\end{equation}
where it is indicated that in an expansion in $\gamma$ and $\beta$ we
only keep the term involving $\gamma^{s}\beta^{s}$. We perform
the geometric sum to obtain
\begin{align}
\tilde{C}(s,\alpha ,N) = \frac{2\pi}{k} (s!)^{2} (-1)^{s}
\frac{(1+\gamma)^{N-1}}{(1+\beta)} 
\frac{\left(\frac{1+\alpha}{(1+\beta) (1+\gamma)}
\right)^{N-s}-1}{\frac{1+\alpha}{(1+\beta) (1+\gamma)}-1} 
\Bigg|_{\gamma^{s}\beta^{s}}\ .
\end{align}
We rewrite the denominator as
\begin{equation}
\left( \frac{1+\alpha}{(1+\beta) (1+\gamma)}-1\right)^{-1} 
= \frac{(1+\beta)(1+\gamma)}{\alpha}
\left[1-\frac{1}{\alpha}\left((1+\beta) (1+\gamma)-1\right) \right]^{-1} \ .
\end{equation}
The term in the square bracket is expanded for small $\beta$ and
$\gamma$, while $\alpha$ is kept fixed. We arrive at
\begin{align}
\tilde{C}(s,\alpha ,N) &= \frac{2\pi}{k} (s!)^{2} (-1)^{s}
\left( (1+\gamma)^{s} (1+\beta)^{s-N} (1+\alpha)^{N-s}- (1+\gamma)^{N} 
\right)\nonumber\\
& \quad \times 
\sum_{m\,\geq\, 0} \alpha^{-m-1} (\beta +\gamma +\beta \gamma)^{m}
\Big|_{\gamma^{s}\beta^{s}}\ .
\end{align}
When we further expand this expression, the coefficients of
$\alpha^{p}\beta^{s}\gamma^{s}$ are polynomials in $N$, and we can
replace $N$ by $\lambda$. We obtain
\begin{equation}
\tilde{C}(s,\alpha ,\lambda ) = \frac{2\pi}{k} (s!)^{2} (-1)^{s}
\sum_{p\,\geq\, 0}\sum_{u,v\,=\,0}^{s}\bin{u+v}{v}\bin{s+u}{u+v}\bin{s-\lambda}{s-u}
\bin{\lambda -s}{1+p+u+v} \alpha^{p} \ . 
\end{equation}
It can be verified that this result agrees with the previously derived
expression~\eqref{linearChyper}.

The strategy just described can easily be applied to the general
non-linear terms. We start from~\eqref{tildeCexpr1}, and introduce
auxiliary variables $\beta_{1},\dotsc ,\beta_{r}$ and
$\gamma_{1},\dotsc ,\gamma_{r}$. We can then write
\begin{align}
& \tilde{C} (\{s_{j} \},{\{ \alpha_{j}\}},N) = \left(\frac{2\pi}{k}\right)^{r} 
\prod_{j\,=\,1}^{r} (s_{j}!)^{2} (-1)^{s_{j}}
\sum_{k_{r}=\,0}^{N-r-S} \dotsb \sum_{k_{1}=\,0}^{k_{2}}\nonumber\\
& \qquad \times (1+\gamma_{r})^{k_{r-1}-k_{r}-1} \dotsb 
(1+\gamma_{r}+\dotsb +\gamma_{2})^{k_{1}-k_{2}-1}
(1+\gamma_{r}+\dotsb +\gamma_{1})^{N-k_{1}-1}\nonumber\\
& \qquad \times (1+\beta_{r})^{k_{r-1}-k_{r}-1}\dotsb 
(1+\beta_{r}+\dotsb +\beta_{1})^{-k_{1}-1}\nonumber\\
& \qquad \times  (1+\alpha_{r})^{k_{r}-k_{r-1}} \dotsb 
(1+\alpha_{r}+\dotsb +\alpha_{1})^{k_{1}}\Big|_{\substack{
\gamma_{1}^{s_{1}}\dotsb \gamma_{r}^{s_{r}}\\
\beta_{1}^{s_{1}}\dotsb \beta_{r}^{s_{r}}}}\ .
\end{align}
When we introduce the notation
\begin{equation}
A_{j} = \alpha_{j}+\dotsb +\alpha_{r} \quad ,\quad
B_{j} = \beta_{j}+ \dotsb + \beta_{r} \quad ,\quad 
C_{j} = \gamma_{j}+\dotsb +\gamma_{r} \ ,
\end{equation}
the expression simplifies to
\begin{align}
\tilde{C} (\{s_{j} \},{\{ \alpha_{j}\}},N) = & \left(\frac{2\pi}{k}\right)^{r} 
\prod_{j\,=\,1}^{r} (s_{j}!)^{2}  (-1)^{s_{j}}
\frac{(1+C_{1})^{N}}{\prod_{j=1}^{r} (1+C_{j}) (1+B_{j})}
\sum_{k_{r}=\,0}^{N-r-S} \dotsb
\sum_{k_{1}=\,0}^{k_{2}}\nonumber\\
& \left(\frac{(1+A_{1})}{(1+B_{1}) (1+C_{1})} \right)^{k_{1}}
\dotsb \left(\frac{(1+A_{r})}{(1+B_{r}) (1+C_{r})} \right)^{k_{r}-k_{r-1}}
\Big|_{\substack{
\gamma_{1}^{s_{1}}\dotsb \gamma_{r}^{s_{r}}\\
\beta_{1}^{s_{1}}\dotsb \beta_{r}^{s_{r}}}}\ .
\end{align}
We can now successively evaluate the geometric sums, starting with the
sum over $k_{1}$ giving
\begin{equation}\label{geometricsum}
\sum_{k_{1}=\,0}^{k_{2}} \left( \frac{(1+A_{1})(1+B_{2})
(1+C_{2})}{(1+B_{1})(1+C_{1})(1+A_{2})}
\right)^{k_{1}} 
= \frac{ \left( \frac{(1+A_{1})(1+B_{2})(1+C_{2})}{(1+B_{1})(1+C_{1})(1+A_{2})}
\right)^{k_{2}+1}-1}{\left(\frac{(1+A_{1}) (1+B_{2}) (1+C_{2})}{(1+B_{1}) (1+C_{1}) (1+A_{2})}\right)-1}\ .
\end{equation}
We expand the denominator in the auxiliary variables except for
$\alpha_{1}$ (the denominator vanishes if all auxiliary variables are
set to zero; since $\alpha_{1}$ appears in all denominators that arise
from the geometric sums, it is enough to keep this parameter finite
while expanding in all others). Thus we have
\begin{multline}
\left(\frac{(1+A_{1}) (1+B_{2}) (1+C_{2})}{(1+B_{1}) (1+C_{1})
(1+A_{2})}-1 \right)^{-1} = \\
\frac{(1+B_{1}) (1+C_{1})
(1+A_{2})}{\alpha_{1} (1+B_{2}) (1+C_{2})} \left\{
1+\frac{1}{\alpha_{1}}\left((1+A_{2})
-\frac{(1+B_{1}) (1+C_{1}) (1+A_{2})}{(1+B_{2}) (1+C_{2})} \right)
\right\}^{-1} \ .
\end{multline} 
From the expansion of the expression in the curly brackets we only
get negative powers of $\alpha_{1}$. Therefore, the term coming from
the $-1$ in the numerator of~\eqref{geometricsum} does not contribute
any non-negative power of $\alpha_{1}$, and we can neglect it if we
later project on non-negative powers of $\alpha_{1}$. Performing all
geometric sums in that way we arrive at
\begin{multline}
\tilde{C} (\{s_{j} \},{\{ \alpha_{j}\}},N) =  \left(\frac{2\pi}{k}\right)^{r} 
\prod_{j\,=\,1}^{r} (s_{j}!)^{2}  (-1)^{s_{j}}
\frac{(1+A_{1})^{N-S} 
(1+B_{1})^{-N+r+S}(1+C_{1})^{r+S}}{\prod_{j=1}^{r} (1+C_{j})
(1+B_{j})} \alpha_{1}^{-r} \\
\times  \prod_{j\,=\,1}^{r}\left\{1+ \frac{1}{\alpha_{1}}\left((1+A_{2})
-\frac{(1+B_{1}) (1+C_{1}) (1+A_{j+1})}{(1+B_{j+1}) (1+C_{j+1})}
\right)\right\}^{-1} \Bigg|_{\substack{
\gamma_{1}^{s_{1}}\dotsb \gamma_{r}^{s_{r}}\\
\beta_{1}^{s_{1}}\dotsb \beta_{r}^{s_{r}}\\
\text{non-neg.\ powers of}\ \alpha_{1}}}\ .
\label{resultbasistransform}
\end{multline}
Here, we have set $S=s_{1}+\dotsb +s_{r}$, and
$A_{r+1}=B_{r+1}=C_{r+1}=0$.  When we expand in the auxiliary
variables (with the expansion in $\alpha_{1}$ done only after all
other expansions have been performed), the coefficients are
polynomials in $N$. Therefore we can replace $N$ by $\lambda$ in the
expression~\eqref{resultbasistransform} and obtain a generating
function for the coefficients of the basis transformation for an
arbitrary value of the parameter $\lambda$. When we do the explicit
expansion (for details see Appendix~\ref{app:proofs}), we obtain our final result,
\begin{align}
& C (\{s_{j} \},\{p_{j} \},\lambda)\, =\,  \left(\frac{2\pi}{k}\right)^{r} 
\prod_{j\,=\,1}^{r} (s_{j}!)^{2}  (-1)^{s_{j}} 
\sum_{\substack{r_{j}^{(1)},\dotsc ,r_{j}^{(6)}\geq 0 \\ 
j=1,\dotsc ,r-1}} \, \sum_{a,b,c\geq 0} \,
\sum_{\substack{b_{1},\dotsc ,b_{r-1}\geq 0\\ c_{1},\dotsc ,c_{r-1}\geq 0}}
\, (-1)^{p_{r}-p_{1}-a}\nonumber\\
&\ \times
\bin{2S+p_{r}-p_{1}-a-b-c-\sum_{j=1}^{r-1}\big(
b_{j}+c_{j}+r_{j}^{(1)}+\dotsb
+r_{j}^{(6)}\big)}{S+p_{r}-p_{1}-a-b-\sum_{j=1}^{r-1}\big(b_{j}+r_{j}^{(2)}+r_{j}^{(3)}+r_{j}^{(5)}+r_{j}^{(6)}
\big)}\nonumber\\
&\ \times \bin{S+p_{r}-p_{1}-a-b-\sum_{j=1}^{r-1}\big(b_{j}+r_{j}^{(2)}+r_{j}^{(3)}+r_{j}^{(5)}+r_{j}^{(6)}
\big)}{S-b-\sum_{j=1}^{r-1}\big(b_{j}+r_{j}^{(2)}+r_{j}^{(5)} \big)}
\nonumber\\
&\ \times
\bin{-\lambda +r+S-1+\sum_{j=1}^{r-1} r_{j}^{(3)}}{b}
\bin{r+2S-1-b-\sum_{j=1}^{r-1}\big(b_{j}-r_{j}^{(3)}+r_{j}^{(5)} \big)}{c}
\nonumber\\
&\ \times
\bin{\lambda -S}{r+2S+p_{r}-b-c-\sum_{j=1}^{r-1} \big( b_{j}+c_{j}\big)}
\bin{r+2S+p_{r}-b-c-\sum_{j=1}^{r-1} \big( b_{j}+c_{j}\big)}{a}
\nonumber\\
& \ \times \prod_{j\,=\,1}^{r-1}
\Bigg[(-1)^{r_{j}^{(3)}+r_{j}^{(4)}+r_{j}^{(5)}}
\bin{r_{j}^{(1)}+\dotsb +r_{j}^{(6)}}{r_{j}^{(1)}} 
\bin{r_{j}^{(2)}+\dotsb +r_{j}^{(6)}}{r_{j}^{(2)}} \dotsb 
\bin{r_{j}^{(5)}+r_{j}^{(6)}}{r_{j}^{(5)}}
 \nonumber\\
&\qquad \qquad  \times 
\bin{-1-r_{j}^{(1)}-\dotsb -r_{j}^{(5)}}{b_{j}} 
\bin{-1-r_{j}^{(1)}-\dotsb -r_{j}^{(4)}}{c_{j}}
\bin{p_{r}-p_{j}
-\sum_{i=j+1}^{r-1}r_{i}^{(3)}}{p_{j+1}-p_{j}}
 \nonumber\\
&\qquad\qquad  \times 
 \bin{s_{r}+\sum_{i=j}^{r-1} \big(
s_{i}-b_{i}-r_{i}^{(5)}\big)}{s_{j}}
\bin{s_{r}+\sum_{i=j}^{r-1} \big(
s_{i}-c_{i}-r_{i}^{(4)}\big)}{s_{j}} 
\Bigg]\ .
\label{finalresultbasistransform}
\end{align}
The sum is finite, because the product of the first two binomial coefficients vanishes
unless
\begin{align}
a+\sum_{j\,=\,1}^{r-1}\left(r_{j}^{(3)}+r_{j}^{(6)}\right) & \leq S+p_{r}-p_{1}\\
b+\sum_{j\,=\,1}^{r-1}\left(b_{j}+r_{j}^{(2)}+r_{j}^{(5)} \right)& \leq S\\
c+\sum_{j\,=\,1}^{r-1}\left(c_{j}+r_{j}^{(1)}+r_{j}^{(4)} \right)& \leq S
\ .
\end{align}
The coefficients~\eqref{finalresultbasistransform} determine the basis
transformation via~\eqref{ubasistransform}, the first terms read
\begin{subequations}
\begin{align}
u_{1} =\  & \frac{2\pi}{k}\bin{\lambda +1}{3}\cW_{1} \\
u_{2} =\  & \frac{8\pi}{k}\bin{\lambda +2}{5}\cW_{2}
+\frac{4\pi}{k}\bin{\lambda +1}{4}\cW_{1}^{(1)} \\
u_{3} =\  & \frac{72\pi}{k}\bin{\lambda
+3}{7}\cW_{3}+\frac{24\pi}{k}\bin{\lambda +2}{6}\cW_{2}^{(1)}
+\frac{6\pi}{k}\bin{\lambda +1}{5}\cW_{1}^{(2)} \nonumber\\
& +\frac{4\pi^{2}}{3k^{2}}\bin{\lambda +1}{5} (5\lambda
+7)\cW_{1}\cW_{1} \\
u_{4} =\  & \frac{1152\pi}{k}\bin{\lambda +4}{9}\cW_{4}+ 
\frac{288\pi}{k}\bin{\lambda +3}{8}\cW_{3}^{(1)} +
\frac{48\pi}{k}\bin{\lambda +2}{7}\cW_{2}^{(2)} \nonumber\\
& + \frac{8\pi}{k}\bin{\lambda +1}{6}\cW_{1}^{(3)} 
+ \frac{16\pi^{2}}{k^{2}}\bin{\lambda +2}{7} (7\lambda +13)\cW_{1}\cW_{2}
\nonumber\\
& + \frac{8\pi^{2}}{k^{2}}\bin{\lambda +1}{6} (5\lambda
+7)\cW_{1}\cW_{1}^{(1)} \\
u_{5} =\  & \frac{28800\pi}{k}\bin{\lambda +5}{11}\cW_{5} 
+ \frac{5760\pi}{k}\bin{\lambda +4}{10}\cW_{4}^{(1)} 
+ \frac{720\pi}{k}\bin{\lambda +3}{9}\cW_{3}^{(2)} \nonumber\\
& + \frac{80\pi}{k}\bin{\lambda +2}{8}\cW_{2}^{(3)} 
+ \frac{10\pi}{k}\bin{\lambda +1}{7}\cW_{1}^{(4)}\nonumber\\
& + \frac{64\pi^{2}}{5k^{2}}\bin{\lambda +2}{8} (44+\lambda
(34+7\lambda))\cW_{2}\cW_{2} \nonumber\\
& + \frac{576\pi^{2}}{k^{2}}\bin{\lambda +3}{9} 
(3\lambda +7) \cW_{1}\cW_{3} \nonumber\\
&  + \frac{64\pi^{2}}{k^{2}}\bin{\lambda +2}{8} (7\lambda +13)
\left( \cW_{1}^{(1)}\cW_{2}+\cW_{1}\cW_{2}^{(1)}\right)\nonumber\\
&  + \frac{10\pi^{2}}{k^{2}}\bin{\lambda +1}{7} (7\lambda +10)
\cW_{1}^{(1)}\cW_{1}^{(1)} + \frac{4\pi^{2}}{k^{2}}\bin{\lambda +1}{7} (21\lambda +29) 
\cW_{1}^{(2)}\cW_{1}\nonumber\\
& +\frac{8\pi^{3}}{63k^{3}}\bin{\lambda -1}{6} (3843+\lambda
(1717+7\lambda (51+5\lambda)))\cW_{1}\cW_{1}\cW_{1} \ .
\end{align}
\end{subequations}
This result reproduces\footnote{Up to the $\cW$-algebra automorphism
in~\eqref{automorphism}.} the basis transformation that was determined
in~\cite{GH} for the first few spins, if one uses the identification
of fields given in~\eqref{identificationGH} and divides our $u$'s by a
factor $-\frac{2\pi}{k}\bin{\lambda +1}{3}$.

\section{Conclusions}\label{sec:conclusions}

In the absence of matter couplings, the interactions of higher-spin
gauge fields in $D=2+1$ can be described by Chern-Simons (CS)
actions. The asymptotic symmetries of asymptotically-AdS solutions of
the field equations are given by the $\cW$-algebras that result from
the Drinfeld-Sokolov (DS) reduction of the gauge algebras. In this
paper we presented a procedure to compute the structure constants of
all classical $\cW$-algebras that can be obtained from the DS
reduction of a (possibly infinite-dimensional) Lie algebra with a
non-degenerate Killing form. We used it to discuss some general
properties of the resulting $\cW$-algebras, and we applied it to a
class of infinite-dimensional Lie algebras, denoted by $hs[\l]$, that
play an important role as higher-spin gauge algebras. A CS action with
gauge algebra $hs[\l] \oplus hs[\l]$ describes indeed the coupling to
gravity of a set of symmetric fields $\vf_{\m_1 \ldots\, \m_s}$ with
ranks $s=3,4,\ldots,\infty$. The field content is thus the same as in
the gauge sector of Vasiliev's models \cite{vas-prok}. The algebra of
asymptotic symmetries is a classical centrally extended
infinite-dimensional $\cW$-algebra, that we denoted by
$\cW_\infty[\l]$. We determined its structure constants in
\eqref{closedFormula1} in a basis where all its generators $\cW_i$ are
Virasoro primaries, and where $\{ \cW_i , \cW_j \}$ is a polynomial in
the generators of the same order as the minimum of the labels $i$ and $j$.

For integer $\l = N$ the Killing form degenerates and the CS action
becomes that of a $sl(N,\mathbb{R}) \oplus sl(N,\mathbb{R})$
theory. The results presented here thus complete the analysis of the
asymptotic symmetries of this class of higher-spin theories that we
initiated in \cite{spin3}, and provide a closed formula for the
structure constants of all classical $\cW_N$ algebras in a Virasoro
primary basis. For $\l = 1/2$ the gauge algebra coincides with the
three-dimensional Fradkin-Vasiliev algebra \cite{Blencowe,frad-vas},
and our formula provides the structure constants of the
infinite-dimensional asymptotic $\cW$-algebra of \cite{HR}. For
generic $\l$ it also reproduces the first few structure constants of
$\cW_\infty[\l]$ that were computed in \cite{GH}. We eventually
presented a way to systematically relate our basis to the non-primary
quadratic basis of \cite{W_infinity}, where $\cW_\infty[\l]$ first
appeared in the context of KP hierarchies.

To stress the relation between $hs[\l] \oplus hs[\l]$ CS theories and
HS gauge theories, in Section \ref{sec:W_metric} we also discussed how
one could express the metric-like fields $\vf_{\m_1 \ldots\, \m_s}$ in
terms of the vielbeine and spin connections entering the CS action. We
were able to establish this relation up to $s=5$, and it would be
interesting to complete the identification following the lines we
proposed. In fact, the relative simplicity of the models we considered
could well shed light on the interplay between the disparate
approaches that were proposed over the years to tackle the difficult
analysis of higher-spin interactions in $D \geq 3+1$. Recovering a
non-linear completion of the metric-like formulation of Fronsdal or
its generalisations (see e.g. \cite{rev-symm}) out of the frame-like
formulation of Vasiliev (see e.g. \cite{vas-rev,exact_review}) could be a
first important step in this direction. Moreover, this would also
allow a reconsideration of our findings along the lines of the
original Brown-Henneaux analysis of the asymptotic symmetries of
Einstein gravity \cite{BH}.

Another context where our results could stimulate further developments
is the study of higher-spin realisations of the AdS/CFT
correspondence. The asymptotic symmetries of the bulk theory should
indeed correspond to global symmetries of the boundary CFT. It is
still not clear whether the pure higher-spin gauge theories that we
discussed admit a CFT dual, but the three-dimensional Vasiliev's
models of \cite{vas-prok} -- describing the coupling to scalar matter
of the same gauge fields we considered -- are also built upon $hs[\l]
\oplus hs[\l]$ gauge algebras. The suggestive asymptotic
$\cW$-symmetries of pure three-dimensional higher-spin gauge theories
already led Gaberdiel and Gopakumar to conjecture a holographic
duality between the large $N$ limit of minimal models with $\cW_N
\times \cW_N$ symmetry and three-dimensional Vasiliev's models
\cite{minimal}. Since $\cW$-symmetries are the cornerstone of this
conjecture, it would be important to reconsider our analysis -- rather
closely related to the CS formulation of the dynamics -- in order to
extend it to Vasiliev's models. In the meantime, we hope that our
detailed description of $\cW_\infty[\l]$ already help further
quantitative checks of this proposal beyond those recently presented
in \cite{GH,minimal2,Yin}.

\section*{Acknowledgements}

We are grateful to X.~Bekaert, D.~Francia, M.~R.~Gaberdiel, J.~Hoppe, E.~Joung, H.~Nicolai, A.~Sagnotti, M.~Taronna, M.~A.~Vasiliev and especially to S.~Theisen for useful discussions. A.C. would like to thank the <<Ettore Majorana>> Foundation and Centre for Scientific Culture of Erice for hospitality while this work was in its final stage of preparation.


\begin{appendix}

\section{Structure constants of $hs[\l]$} \label{app:pope}

The structure constants that appear in the definition of the $\star$-product \eqref{WstarW} can be expressed in terms of those defined in \cite{Pope} (see also \cite{GH}) as
\be \label{flambda}
f_\l \! \left(
\begin{array}{cc|c}
k & \ell & i \\
m & n & m+n
\end{array}
\right) =\, 
q^{k+\ell-i-1} \, g^{\,k-1 \,,\, \ell-1}_{k+\ell-i-1} (m,n;\l) \, ,
\ee
where $q$ is a normalisation factor that must be equal to $1/4$ for
any finite $\l$, but that is useful to discuss the $\l \to \infty$
limit of $hs[\l]$ (see e.g. \cite{GH}). Moreover, in \cite{Pope} it
was proposed that the functions $g^{i\,,\,j}_k(m,n;\l)$ are given by
\begin{subequations} \label{pope_coeff}
\begin{align}
& g^{i\,,\,j}_k(m,n;\l) =\, \frac{1}{2(k+1)!}\, \phi^{i\,,\,j}_k(\l)\, N^{\,i\,,\,j}_k(m,n) \, ,\\[5pt]
& N^{\,i\,,\,j}_k(m,n) =\, \sum_{p\,=\,1}^{k+1}\, (-1)^p \binom{k+1}{p} (2i+2-k)_p[2j+2-p]_{k-p+1}[i+1+m]_{k-p+1}[j+1+n]_p\, , \\
& \phi^{i\,,\,j}_k(\l) = \sum_{p\,=\,0}^{\lfloor k \rfloor} \prod_{q\,=\,1}^p \frac{[(2q-3)(2q+1)-4(\l^2-1)](k-2q+3)(k/2-q+1)}{q(2i-2q+3)(2j-2q+3)(2i+2j-2k+2q+3)}\, .
\end{align}
\end{subequations}
Here $\lfloor k \rfloor$ denotes the integer part of $k$, while $(a)_n$ and $[a]_n$ denote respectively the ascending and descending Pochhammer symbols,
\begin{subequations}
\begin{align}
(a)_n & :=\, a(a+1) \ldots (a+n-1) \, , \label{poch+} \\[2pt]
[a]_n & :=\, a(a-1) \ldots (a-n+1) \, . \label{poch-}
\end{align}
\end{subequations}
See also \cite{Pope} for some alternative rewritings of eqs.~\eqref{pope_coeff}. An expression for the $hs[\l]$ structure constants was also provided in \cite{FL} in terms of Clebsch-Gordan and generalised Wigner - $6j$ - symbols. See \cite{BBS} and \cite{hoppe_proof} for the proof that different values of $\l$ lead to algebras which are not isomorphic. 

\section{Structure constants of $\cW_\infty[\l]$} \label{app:proofs}

In this appendix we prove the expression for the structure constants of $\cW_\infty[\l]$ that we displayed in eq.~\eqref{closedFormula1}. The involved steps closely follow those needed to compute the polynomials in the Virasoro generators that appear in the Drinfeld-Sokolov reduction of a generic algebra. They were presented in eq.~\eqref{var_vir} and we shall begin by proving it, before moving to eq.~\eqref{closedFormula1}. We close this appendix with a proof of eq.~\eqref{finalresultbasistransform} that, through eq.~\eqref{ubasistransform}, determines the basis transformation needed to connect our result with the quadratic basis of \cite{W_infinity}.

\subsection*{Proof of eq.~\eqref{var_vir}}

In order to prove eqs.~\eqref{var_vir} and \eqref{const_vir} it is convenient to omit possible colour indices and to consider the gauge parameter $\l_+ = \e(\theta) W^\ell_\ell$. Using the truncated covariant derivative of \eqref{der_vir}, each summand in \eqref{globalsymmetry} then takes the form
\begin{align}
& (-DRL_-)^n D\l_+ \,=\, (-1)^n\, \bigg\{\, \pr^{\,n+1} \e\, (RL_-)^n \nn \\
& +\, \sum_{r\,=\,1}^{n+1} \left(\frac{2\pi}{k}\right)^{\!\!r}\ \sum_{i_1\,=\,0}^{n-r+1} \, \sum_{i_2\,=\,0}^{(n-r+1)-i_1} \!\ldots\!\! \sum_{i_r\,=\,0}^{(n-r+1)-\sum_1^{r-1} i_t} \!\! \pr^{\,i_1} \big( \cL\, \pr^{\,i_2} \big( \ldots \big( \cL\, \pr^{\,i_r} \big( \cL\, \pr^{(n-r+1)-\sum_1^r i_t}\e \,\big)\big)\big)\big) \nn \\
& \times\, (RL_-)^{i_1}L_-(RL_-)^{i_2+1} \ldots\, L_-(RL_-)^{i_r+1}L_- (RL_-)^{(n-r+1)-\sum_1^r i_t} \bigg\}\, W^\ell_\ell \, , \label{h:op}
\end{align}
where we also omitted the $\theta$ dependence in both ordinary and covariant derivatives. One can check this expression by recursion. To this end, let us compute separately the two summands in
\be \label{p:opstart}
(-DRL_-)^{n+1} D\l_+ \,=\, - \left\{ \pr \,+\, \frac{2\pi}{k}\,\cL\, L_- \right\} (RL_-) (-DRL_-)^n D\l_+ \, .
\ee
The first one reads
\begin{align}
& -\,\pr (RL_-)(-DRL_-)^n D\l_+ \,=\, (-1)^{n+1}\, \bigg\{\, \pr^{\,n+2} \e\, (RL_-)^{n+1} \nn \\
& +\, \sum_{r\,=\,1}^{n+1} \left(\frac{2\pi}{k}\right)^{\!\!r}\ \sum_{i_1\,=\,1}^{n-r+2} \, \sum_{i_2\,=\,0}^{(n-r+2)-i_1} \!\ldots\!\! \sum_{i_r\,=\,0}^{(n-r+2)-\sum_1^{r-1} i_t} \!\! \pr^{\,i_1} \big( \cL\, \pr^{\,i_2} \big( \ldots \big( \cL\, \pr^{\,i_r} \big( \cL\, \pr^{(n-r+2)-\sum_1^r i_t}\e \,\big)\big)\big)\big) \nn \\
& \times\, (RL_-)^{i_1}L_-(RL_-)^{i_2+1} \ldots\, L_-(RL_-)^{i_r+1}L_- (RL_-)^{(n-r+2)-\sum_1^r i_t} \bigg\}\, W^\ell_\ell \, . \label{p:op1}
\end{align}
In order to rebuild \eqref{h:op} with $n \to n+1$ two contributions are missing: the terms with $i_1=0$ in the first sum over derivatives and the term with $r=n+2$ in the sum over the order of non-linearity. They come from the second summand in \eqref{p:opstart} that can be cast in the form
\begin{align}
& -\, \frac{2\pi}{k}\,\cL\, L_-(RL_-) (-DRL_-)^n D\l_+ \nn \\
& =\, (-1)^{n+1} \sum_{r\,=\,1}^{n+2} \left(\frac{2\pi}{k}\right)^{\!\!r}\, \sum_{i_2\,=\,0}^{n-r+2} \ldots\!\! \sum_{i_r\,=\,0}^{(n-r+2)-\sum_2^{r-1} i_t}\!\! \cL\,\pr^{\,i_2} \big( \ldots \big( \cL\, \pr^{\,i_r} \big( \cL\, \pr^{(n-r+2)-\sum_2^r i_t}\e \,\big)\big)\big)\big) \nn \\
& \times\, L_-(RL_-)^{i_2+1} \ldots\, L_-(RL_-)^{i_r+1}L_- (RL_-)^{(n-r+2)-\sum_2^r i_t} \bigg\}\, W^\ell_\ell \label{p:op2}
\end{align}
and thus gives all terms with $i_1=0$. This suffices to conclude because the term with $r=n+2$ does not contain derivatives.

The next step is the elimination of the operators in \eqref{h:op} using
\begin{subequations} \label{tricks}
\begin{align}
& L_-\, W^\ell_{\ell-p} \,=\, -\,(2\ell-p)\, W^\ell_{\ell-(p+1)} \, , \label{LW} \\
& (RL_-)^i\, W^\ell_{\ell-p} \,=\, \frac{p!}{(p+i)!}\, W^\ell_{\ell-(p+i)} \, . \label{RLW}
\end{align}
\end{subequations}
Both identities follow from the commutators \eqref{primary}. In particular, one can get \eqref{RLW} combining $W^\ell_{\ell-p} = (n+1)^{-1}L_+ W^\ell_{\ell-(p+1)}$ with \eqref{comb1}. Taking advantage of \eqref{tricks} we get
\begin{align}
& (-1)^r(n+r)!\,(RL_-)^{i_1}L_-(RL_-)^{i_2+1} \ldots\, L_-(RL_-)^{i_r+1}L_- (RL_-)^{(n-r+1)-\sum_1^r i_t} \, W^\ell_\ell \\
& =\, \prod_{s\,=\,0}^{r-1} \left( (n-r+1) - \sum_{t\,=\,1}^{r-s}i_t + 2s + 1 \right) \left( 2\ell - (n-r+1) + \sum_{t\,=\,1}^{r-s}i_t - 2s \right) W^\ell_{\ell-(n+r)} \, . \nn
\end{align}
In order to make contact with eq.~\eqref{var_vir} we can now evaluate the derivatives in \eqref{h:op},
\be \label{p:deriv}
\begin{split}
& \pr^{\,i_1} \big( \cL\, \pr^{\,i_2} \big( \ldots \big( \cL\, \pr^{\,i_r} \big( \cL\, \pr^{\,(n-r+1)-\sum_1^r i_t}\e \,\big)\big)\big)\big) \\
& =\, \sum_{p_1\,=\,0}^{i_1} \sum_{p_2\,=\,0}^{i_1+i_2-p_1} \ldots\!\! \sum_{p_r\,=\,0}^{\sum_1^r i_k - \sum_1^{r-1} p_t} \prod_{s\,=\,1}^r \binom{\sum_1^s i_t - \sum_1^{s-1} p_t}{p_s}\, \cL^{(p_1)} \ldots\, \cL^{(p_r)} \e^{(n-r+1-\sum_1^r p_t)} \, , 
\end{split}
\ee
and eventually exchange the sums over $i$'s and $p$'s. Here, as in the main body of the text, an exponent between parentheses denotes the action of the corresponding number of derivatives on the field. The rewriting \eqref{p:deriv} leads to
\be \label{p:int_vir}
\begin{split}
& (-DRL_-)^n D\l_+ = \, \frac{(-1)^n}{n!}\, \e^{(n+1)}\, W^\ell_{\ell-n} \ + \ \sum_{r\,=\,1}^{n+1}\, \frac{(-1)^{n+r}}{(n+r)!} \left(\frac{2\pi}{k}\right)^{\!\!r} \\ 
& \times \sum_{p_1\,=\,0}^{n-r+1} \, \ldots\!\!\!\! \sum_{p_r\,=\,0}^{(n-r+1)-\sum_1^{r-1} p_t} \widetilde{C}[n,r]_{p_1\ldots\, p_r}\, \cL^{(p_1)} \ldots\, \cL^{(p_r)} \e^{(n-r+1-\sum_1^r p_t)}\ W^\ell_{\ell-(n+r)} \, . 
\end{split}
\ee
The coefficients $\widetilde{C}[n,r]_{p_1\ldots\, p_r}$ are defined by
\be
\begin{split}
& \widetilde{C}[n,r]_{p_1\ldots\, p_r} = \sum_{i_1\,=\,p_1}^{n-r+1}\, \sum_{i_2\,=\,\langle(p_1+p_2)-i_1\rangle_+}^{(n-r+1)-i_1} \ldots \sum_{i_r\,=\,\langle \sum_1^r p_t - \sum_1^{r-1} i_t \rangle_+}^{(n+r-1)-\sum_1^{r-1} i_t} \prod_{s\,=\,1}^r \binom{\sum_1^s i_t - \sum_1^{s-1} p_t}{p_s} \\
& \times \left( (n-r+1) - \sum_{t\,=\,1}^{r-s+1}i_t + 2s - 1 \right) \left( 2\ell - (n-r+1) + \sum_{t\,=\,1}^{r-s+1}i_t - 2(s-1) \right) ,
\end{split}
\ee
where, as in \eqref{const_vir}, we introduced $\langle a \rangle_+ = \max \left(0,a\right)$. 

To conclude we have to apply the projector $P_-$ to \eqref{p:int_vir} and to sum over $n$ as in \eqref{globalsymmetry}. The projection selects the terms in $W^\ell_{-\ell}$ out of \eqref{p:int_vir}, thus forcing $(n+r)=2\ell$. We recovered in this way the upper bound $n \leq 2\ell$ already discussed after \eqref{globalsymmetry}. The condition $r\leq n+1$ also induces a lower bound on the number of terms that contribute to \eqref{globalsymmetry}. For $\l_+ = \e(\theta) W^\ell_\ell$ one actually has
\be \label{p:sum}
\delta_{\lambda} a (\theta) \,= \sum_{n\,=\,\lfloor\ell\rfloor}^{2\ell} P_- \left(- DRL_{-}\right)^{n} D \e\, W^\ell_{\ell} \, ,
\ee
where $\lfloor \ell \rfloor$ is the integer part of $\ell$.
Reorganising \eqref{p:sum} as a sum over the order of non-linearity (e.g. over $r = 2\ell - n$) eventually leads to \eqref{var_vir}.

\subsection*{Proof of eq.~\eqref{closedFormula1}}

The first step of the proof is the natural extension of \eqref{h:op}. Also in this case we consider the gauge variation induced by a gauge parameter with a definite $L_0$ eigenvalue, say $\l_+ = \e(\theta) W^i_i$. In order to proceed it is convenient to introduce the operators
\be
w_a\, x := \, [\, W^a_{-a} \,,\, x \,] \quad \text{for}\ x\in hs[\l] \, ,
\ee
that enable one to cast the $hs[\l]$-covariant derivative in the form
\be
D \,=\, \pr \,+\, \frac{2\pi}{k}\, \sum_{a\,=\,1}^\infty \cW_a(\th) \, w_a \, .
\ee
As usual we identified $\cL$ with $\cW_1$ and $L_-$ with $w_{1}$. Following \eqref{p:op1} and \eqref{p:op2} one can then prove by recursion that
\begin{align}
& \d_\l a \,=\, P_- \sum_{n\,=\,0}^{2i} \, (-DRL_-)^n D\l_+ \,=\, P_- \sum_{n\,=\,0}^{2i} \, (-1)^n\, \bigg\{\, \pr^{\,n+1} \e\, (RL_-)^n +\, \sum_{r\,=\,1}^{n+1} \ \sum_{a_1\,=\,1}^\infty \ldots\, \sum_{a_r\,=\,1}^\infty \nn \\
& \times \left(\frac{2\pi}{k}\right)^{\!\!r}\ \sum_{q_1\,=\,0}^{n-r+1} \ \ldots\!\! \sum_{q_r\,=\,0}^{(n-r+1)-\sum_1^{r-1} q_t} \pr^{\,q_1} \big( \cW_{a_1} \pr^{\,q_2} \big( \ldots \big( \cW_{a_{r-1}} \pr^{\,q_r} \big( \cW_{a_r} \pr^{(n-r+1)-\sum_1^r q_t}\e \,\big)\big)\big)\big) \nn \\
& \times\, (RL_-)^{q_1}\,w_{a_1}(RL_-)^{q_2+1} \ldots\, w_{a_{r-1}}(RL_-)^{q_r+1}\,w_{a_r} (RL_-)^{(n-r+1)-\sum_1^r q_t} \bigg\}\, W^i_i \, . \label{p:rec}
\end{align}
The first contribution in \eqref{p:rec} gives the central terms that we already discussed in detail in Section \ref{subsec:central}. We shall thus often omit it in the following. 

The $L_0$ eigenvalue of the $hs[\l]$ generator resulting from the application of the chain of operators on $W^i_i$ can be read off quite easily from \eqref{p:rec}. It is
\be
m \,=\, -\, \bigg(\, i \,-\, \sum_{t\,=\,1}^r a_t \,-\, n \,\bigg)
\ee  
since each $w_{a_t}$ insertion lowers it by $-a_t$ and for each $n$ there are $n$ operators $(RL_-)$. On the other hand, as already discussed at the beginning of Section \ref{sec:W_gauge}, the maximum value of the final spin is
\be
\ell \,=\, \sum_{t\,=\,1}^r a_t \,+\, i \,-\, r \, .
\ee
Since $m$ must satisfy $|m| \leq \ell$, the variable $r$ is bound to obey $r \leq \min(n+1,2i-n)$. Reversing the order of the sums over $r$ and $n$ this leads to
\be \label{p:varint}
\begin{split}
\d_\l a \,& =\, \frac{\e^{(2i+1)}}{(2i)!}\, W^i_{-i} \,+\, P_- \sum_{r\,=\,1}^i \left(\frac{2\pi}{k}\right)^{\!\!r}\, \sum_{a_1\,=\,1}^\infty \ldots\, \sum_{a_r\,=\,1}^\infty\ \sum_{n\,=\,r-1}^{2i-r} (-1)^n\, \bigg\{ \ldots \bigg\} \,W^i_i \, ,
\end{split}
\ee
where the terms between braces have the same structure as those appearing in the second and in the third line of \eqref{p:rec}. Note that in getting \eqref{p:varint} we already took into account that $hs[\l]$ only contains generators with integer spin. 

We can now evaluate the action of the operators on $W^i_i$ using \eqref{[W,W]} and \eqref{RLW}: with the identification $b_0 \,=\, i$ we obtain 
\be \label{p:eval}
\begin{split}
& (RL_-)^{q_1}\,w_{a_1}(RL_-)^{q_2+1} \ldots\, w_{a_{r-1}}(RL_-)^{q_r+1}\,w_{a_r} (RL_-)^{(n-r+1)-\sum_1^r q_t} \, W^i_i \\[2pt]
& =\, \frac{1}{(b_r + \sum_1^r a_t - i +n)!} \, \sum_{\substack{b_1\,=\,|a_r-b_0|+1 \\[2pt] a_r+b_0+b_1 \ \textrm{odd}}}^{a_r+b_0-1} \ \, \sum_{\substack{b_2\,=\,|a_{r-1}-b_1|+1 \\[2pt] a_{r-1}+b_1+b_2 \ \textrm{odd}}}^{a_{r-1}+b_1-1} \ \ldots\ \sum_{\substack{b_r\,=\,|a_{1}-b_{r-1}|+1 \\[2pt] a_{1}+b_{r-1}+b_r \ \textrm{odd}}}^{a_{1}+b_{r-1}-1} \\[3pt]
& \times \prod_{s\,=\,1}^r \, \frac{(n - i + b_s + \sum_{r-s+1}^r a_t - \sum_1^{r-s+1} q_t - r + s)!}{(n - i + b_{s-1} + \sum_{r-s+2}^r a_t - \sum_1^{r-s+1} q_t - r + s)!} \\[3pt]
& \times\, f_\l \! \left(
\begin{array}{cc|c}
a_{r-s+1} & b_{s-1} & b_s \\[2pt]
- a_{r-s+1} & i - n - \sum_{r-s+2}^r a_t + \sum_1^{r-s+1} q_t + r - s & \ldots
\end{array}
\right) W^{b_r}_{i-n-\sum_1^r a_t} \, .
\end{split}
\ee
The final spin is $b_r$. As a result, we should bring the sum over $b_r$ in the first position among all other summations. In this fashion we can eventually select a particular $b_r$ to read the gauge variation of $\cW_{b_r}$. Due to its selected role, in the following we shall define $j = b_r$. Bringing the sum over $j$ in the first position casts the summations over $b$'s in \eqref{p:eval} in the form
\be \label{p:bigsum}
\sum_{\substack{j\,=\,\max\left(1\,,\,M(r,i)\right) \\[3pt] \sum_1^r a_t + i + r \ \textrm{even}}}^{\sum_1^r a_t + i - r} \ 
\sum_{\substack{b_1\,=\,\max\left(|a_{r}-b_0|+1\,,\,M(r-1,\,j)\right) \\[3pt] a_{r}+b_0+b_1 \ \textrm{even}}}^{\min\left(a_{r}+b_0-1 \,,\, \sum_1^{r-1} a_t + j - r + 1\right)} \ \ldots \
\sum_{\substack{b_{r-1}\,=\,\max\left(|a_{2}-b_{r-2}|+1\,,\,M(1,\,j)\right) \\[3pt] a_{2}+b_{r-1}+b_r \ \textrm{even}}}^{\min\left(a_{2}+b_{r-2}-1 \,,\, a_1 + j - 1\right)}
\ee
with
\be
M(s,\ell) :=\, 2\max\left(\{a_t\}_{t=1}^s\,,\,\ell\,\right) - \sum_{t\,=\,1}^s a_t - \ell + s \, .
\ee

The sum over $j$ commutes with all other summations with the exception of those on $a_k$. Before performing this last exchange let us evaluate the projector $P_-$. It forces the $L_0$ eigenvalue of the generator appearing in \eqref{p:eval} to coincide with $j$. Therefore it imposes
\be \label{p:ndef}
n \,=\, (i+j) \,-\, \sum_{t\,=\,1}^r a_t \, .
\ee
Both \eqref{p:ndef} and the extrema of the sum over $j$ only depend on the sum of all $a_t$. Let us thus introduce $L = \sum_{1}^r a_k$. For $L \leq i$ the lower bound of the sum over $j$ is given by \eqref{p:bigsum} as $j \geq i - L + r$. On the other hand, for $L > i$ the lower bound comes from \eqref{p:ndef} since $L - i + r - 1 > M(r,i)$. In conclusion we are led to consider
\be
\sum_{L\,=\,1}^\infty\ \sum_{\{a_t\}} \sum_{\substack{j\,=\,|L-i|+r \\[2pt] i+j+L+r\ \textrm{even}}}^{L+i-r} =\ \, \sum_{j\,=\,1}^\infty\, \sum_{\substack{L\,=\,|i-j|+r \\[2pt] i+j+L+r\ \textrm{even}}}^{i+j-r} \sum_{\{a_t\}}
\ee
where the multiple sum over the $a$'s must be such that $\sum_1^r a_t = L$. Substituting everywhere \eqref{p:ndef} and expanding the derivatives as in \eqref{p:deriv} eventually lead to \eqref{closedFormula1}.

\subsection*{Proof of eq.\ \eqref{finalresultbasistransform}}

We start from~\eqref{resultbasistransform}, and first expand the
factors of the last product,
\begin{align}
& \left\{1+ \frac{1}{\alpha_{1}}\left((1+A_{2})
-\frac{(1+B_{1}) (1+C_{1}) (1+A_{j+1})}{(1+B_{j+1}) (1+C_{j+1})}
\right)\right\}^{-1} \nonumber \\
&\qquad = \bigg\{1- \frac{1}{\alpha_{1}(1+B_{j+1})
(1+C_{j+1})}\big[ C_{1}+ B_{1} (1+C_{1})+A_{j+1} (1+B_{1}) (1+C_{1})\nonumber\\
& \qquad \qquad \qquad \qquad  - C_{j+1}-B_{j+1}
(1+C_{j+1})-A_{2} (1+B_{j+1}) (1+C_{j+1}) \big]\bigg\}^{-1} \\
&\qquad = \sum_{r_{j}^{(1)},\dotsc ,r_{j}^{(6)}\geq 0}
\left(\alpha_{1} (1+B_{j+1}) (1+C_{j+1}) \right)^{-r_{j}^{(1)}-\dotsb
-r_{j}^{(6)}}  \nonumber\\
& \qquad \qquad \times \bin{r_{j}^{(1)}+\dotsb +r_{j}^{(6)}}{r_{j}^{(1)}} 
\bin{r_{j}^{(2)}+\dotsb +r_{j}^{(6)}}{r_{j}^{(2)}} \dotsb 
\bin{r_{j}^{(5)}+r_{j}^{(6)}}{r_{j}^{(5)}} \nonumber\\
& \qquad \qquad \times [C_{1}]^{r_{j}^{(1)}} [B_{1} (1+C_{1})]^{r_{j}^{(2)}}
[A_{j+1} (1+B_{1}) (1+C_{1})]^{r_{j}^{(3)}}\nonumber\\
&\qquad \qquad \times  
[-C_{j+1}]^{r_{j}^{(4)}} [-B_{j+1} (1+C_{j+1})]^{r_{j}^{(5)}}
[-A_{2} (1+B_{j+1}) (1+C_{j+1})]^{r_{j}^{(6)}}\ .
\end{align}
For $j=r$ this expression simplifies to
\begin{align}
& \left\{1+ \frac{1}{\alpha_{1}}\left((1+A_{2}) -(1+B_{1}) (1+C_{1})
\right)\right\}^{-1} \nonumber \\
&\qquad = \sum_{r_{r}^{(1)},r_{r}^{(2)},r_{r}^{(6)}\geq 0}
\alpha_{1}^{-r_{r}^{(1)}-r_{r}^{(2)}-r_{r}^{(6)}}  \nonumber\\ 
&\qquad \qquad \times \bin{r_{r}^{(1)}+r_{r}^{(2)} +r_{r}^{(6)}}{r_{r}^{(1)}} 
\bin{r_{r}^{(2)}+r_{r}^{(6)}}{r_{r}^{(2)}} 
 [C_{1}]^{r_{r}^{(1)}} [B_{1} (1+C_{1})]^{r_{r}^{(2)}}
[-A_{2}]^{r_{r}^{(6)}}\ .
\end{align}
In the next step we expand the powers of $(1+A_{1})$, $(1+B_{1})$ and
$(1+C_{1})$ with summation variables $a',b,c$, respectively, and the
powers of $(1+B_{j+1})$ and $(1+C_{j+1})$ with summation variables
$b_{j},c_{j}$ with $j=1,\dotsc ,r-1$. We obtain
\begin{align}
& \tilde{C} (\{s_{j} \},{\{ \alpha_{j}\}},N) \, =\, 
\left(\frac{2\pi}{k}\right)^{r} 
\prod_{j=1}^{r} (s_{j}!)^{2}  (-1)^{s_{j}}
\sum_{\substack{r_{j}^{(1)},\dotsc ,r_{j}^{(6)}\geq 0 \\ 
j=1,\dotsc ,r-1}} \, \sum_{r_{r}^{(1)},r_{r}^{(2)},r_{r}^{(6)}}\, 
\sum_{a',b,c\geq 0} \, 
\sum_{\substack{b_{1},\dotsc ,b_{r-1}\geq 0\\ c_{1},\dotsc
,c_{r-1}\geq 0}} \nonumber \\
& \ \times \prod_{j=1}^{r-1}
\Bigg[(-1)^{r_{j}^{(4)}+r_{j}^{(5)}+r_{j}^{(6)}}
\bin{r_{j}^{(1)}+\dotsb +r_{j}^{(6)}}{r_{j}^{(1)}} 
\bin{r_{j}^{(2)}+\dotsb +r_{j}^{(6)}}{r_{j}^{(2)}} \dotsb 
\bin{r_{j}^{(5)}+r_{j}^{(6)}}{r_{j}^{(5)}}
 \nonumber\\
&\qquad \qquad  \times 
\bin{-1-r_{j}^{(1)}-\dotsb -r_{j}^{(5)}}{b_{j}} 
\bin{-1-r_{j}^{(1)}-\dotsb -r_{j}^{(4)}}{c_{j}}\Bigg]\nonumber\\
&\ \times (-1)^{r_{r}^{(6)}}\bin{r_{r}^{(1)}+r_{r}^{(2)} +r_{r}^{(6)}}{r_{r}^{(1)}} 
\bin{r_{r}^{(2)}+r_{r}^{(6)}}{r_{r}^{(2)}}
\bin{\lambda -S}{a'}\bin{-\lambda +r+S-1+\sum_{j=1}^{r-1}r_{j}^{(3)}}{b}\nonumber\\
& \ \times \bin{r+S-1+\sum_{j=1}^{r-1}\left( r_{j}^{(2)}+r_{j}^{(3)}\right)+r_{r}^{(2)}}{c}
\alpha_{1}^{-r-r_{r}^{(1)}-r_{r}^{(2)}-r_{r}^{(6)}-\sum_{j=1}^{r-1} 
\left( r_{j}^{(1)}+\dots +r_{j}^{(6)}\right)}\nonumber\\
&\ \times
A_{1}^{a'}B_{1}^{b+r_{r}^{(2)}+\sum_{j=1}^{r-1}r_{j}^{(2)}}C_{1}^{c+r_{r}^{(1)}+\sum_{j=1}^{r-1}r_{j}^{(1)}} 
A_{2}^{r_{r}^{(6)}+\sum_{j=1}^{r-1}r_{j}^{(6)}}\nonumber\\
&\ \times 
\prod_{j=1}^{r-1} \left( A_{j+1}^{r_{j}^{(3)}} B_{j+1}^{b_{j}+r_{j}^{(5)}} 
C_{j+1}^{c_{j}+r_{j}^{(4)}}\right)
\Bigg|_{\substack{
\gamma_{1}^{s_{1}}\dotsb \gamma_{r}^{s_{r}}\\
\beta_{1}^{s_{1}}\dotsb \beta_{r}^{s_{r}}\\
\text{non-neg.\ powers of}\ \alpha_{1}}}\ .
\end{align}
We want to extract the coefficients with powers $\beta_{j}^{s_{j}}$ and
$\gamma_{j}^{s_{j}}$. In particular the sum of the exponents
of the $B_{j}$ and of the $C_{j}$ have to match $S=s_{1}+\dotsb
+s_{r}$. This fixes the summation variables $r_{r}^{(1)}$ and $r_{r}^{(2)}$ to
\begin{align}
r_{r}^{(1)} &= S-c-\sum_{j=1}^{r-1}\left(c_{j}+r_{j}^{(1)}+r_{j}^{(4)}
\right) \\
r_{r}^{(2)} &= S-b-\sum_{j=1}^{r-1}\left(b_{j}+r_{j}^{(2)}+r_{j}^{(5)}
\right) \ .
\end{align}
In the next step we expand the power of
$B_{1}=\beta_{1}+B_{2}$,
\begin{equation}
B_{1}^{S-\sum_{j=1}^{r-1}\left(b_{j}+ r_{j}^{(5)}\right)}\Big|_{\beta_{1}^{s_{1}}} = 
\bin{S-\sum_{j=1}^{r-1}\left(b_{j}+ r_{j}^{(5)}\right)}{s_{1}}
B_{2}^{S-s_{1}-\sum_{j=1}^{r-1}\left(b_{j}+ r_{j}^{(5)}\right)}\ ,
\end{equation} 
then the power of $B_{2}=\beta_{2}+B_{3}$, and so on, similarly for
the $C_{j}$. 

To extract the coefficients $C (\{s \},\{p \},\lambda)$ we also have
to project to powers $\alpha_{j}^{p_{j}-p_{j-1}}$. The sum of the
exponents of the $A_{j}$ plus the exponent of $\alpha_{1}$
has to match $p_{r}$, this can be used to fix $a'$ to
\begin{equation}
a'=r+2S+p_{r}-b-c-\sum_{j=1}^{r-1}\left(b_{j}+c_{j} \right) \ .
\end{equation}
Now we expand the power of $A_{1}=\alpha_{1}+A_{2}$,
\begin{multline}
A_{1}^{r+2S+p_{r}-b-c-\sum_{j=1}^{r-1}\left(b_{j}+c_{j} \right)} = \sum_{a\geq 0} \bin{r+2S+p_{r}-b-c-\sum_{j=1}^{r-1}\left(b_{j}+c_{j} \right)}{a} \\
\times \alpha_{1}^{r+2S+p_{r}-a-b-c-\sum_{j=1}^{r-1}\left(b_{j}+c_{j} \right)}A_{2}^{a} \ .
\end{multline}
The sum of exponents of $A_{2},\dotsc ,A_{r}$ has to match
$p_{r}-p_{1}$ leading to the condition
\begin{equation}
r_{r}^{(6)}=p_{r}-p_{1}-a-\sum_{j=1}^{r-1}\left(r_{j}^{(3)}+r_{j}^{(6)}
\right) \ ,
\end{equation}
which fixes $r_{r}^{(6)}$. In the next step the power of $A_{2}$ is
expanded as
\begin{equation}
A_{2}^{p_{r}-p_{1}-\sum_{i=2}^{r-1}r_{i}^{(3)}}
\Big|_{\alpha_{2}^{p_{2}-p_{1}}}
\, = \, \bin{p_{r}-p_{1}-\sum_{i=2}^{r-1}r_{i}^{(3)}}{p_{2}-p_{1}}
A_{3}^{p_{r}-p_{2}-\sum_{i=2}^{r-1}r_{i}^{(3)}} \ ,
\end{equation}
then the power of $A_{3}$, and so on. This leads to the final
result~\eqref{finalresultbasistransform}.

\section{Poisson brackets of $\cW^{(2)}_3$, $\cW^{(2)}_4$ and $\cW_\infty[\l]$}
\label{app:nonPrincipalExp}

In this appendix we present the Poisson brackets of the two
examples of $\cW$-algebras that we discussed in Section \ref{sec:red_examples}. We also present the Poisson brackets of $\cW_\infty[\l]$ for fields of weight $\ell \leq 3$. Imposing $\l = 3$ and rescaling them by $N_2(3)$ they give the $\cW_3$ algebra, while imposing $\l = 4$ and rescaling them by $N_3(4)$ they give the $\cW_4$ algebra. In this fashion they allow to compare the $\cW^{(2)}_3$ and $\cW^{(2)}_4$ algebras with their counterparts associated to a principal $sl(2)$ embedding.
 
\subsection*{Poisson structure of $\cW_3^{(2)}$}

Here we present the full $\cW_3^{(2)}$ algebra of \cite{Polyakov} before implementing the shift \eqref{W4SugawaraShift}, with the convention that all fields on the right-hand side depend on $\th'$ and \mbox{$\d'(\th-\th') \equiv \pr_\th \d(\th-\th')$}. Exponents between square brackets denote colour indices, while exponents between parentheses specify the number of derivatives acting on the corresponding object.
\begin{subequations}\label{poissonBracketExpW3}
\begin{align}
  &\big\{ \cW_0(\th)\, , \, \cW_0(\th')\big\}\, 
  =\, \frac{3k}{4\p} \, \d'(\th-\th') \, , \\[5pt]
  &\big\{ \cW_0(\th)\, , \, \cW_{\frac{1}{2}}^{[a]}(\th') \big\} \, 
  =\, a \, \frac{3}{2} \, \d(\th-\th') \,  \cW_{\frac{1}{2}}^{[a]} \, ,
  \\[5pt]
  &\big\{ \cW_0(\th)\, , \, \cL(\th') \big\} \, 
  =\, 0 \, , \label{nonPrimaryPB1_W3} \\[15pt]
  &\big\{ \cL(\th) \,,\, \cL(\th') \big\}
  \,=\, \d(\th-\th') \, \cL' 
  - 2 \, \d'(\th-\th') \, \cL \,-\, \frac{k}{4\pi}\,\d^{(3)}(\th-\th') \, , \\[5pt]
  &\big\{ \cL(\th)\, , \, \cW_{\frac{1}{2}}^{[a]}(\th') \big\} 
  \,=\, \d(\th-\th')\, \cW_{\frac{1}{2}}^{[a]\,\pe} -\, \frac{3}{2} \, \d'(\th-\th') \, \cW_{\frac{1}{2}}^{[a]}
  \, 
  +\, a\, \frac{2\pi}{k}\, \d(\th-\th') \, \cW_0 \cW_{\frac{1}{2}}^{[a]} \, ,
  \label{nonPrimaryPB_W3} \displaybreak[0]
\\[15pt]
  &\big\{ \cW_{\frac{1}{2}}^{[a]}(\th)\, , \, \cW_{\frac{1}{2}}^{[b]}(\th')\big\}
  \, = \,
  \d_{a+b\,,\,0} \, \bigg(
   -\, \frac{k}{2\p} \, a \, \d''(\th-\th') 
  \, - \, a \, \d(\th-\th')\, \cL \, \label{coloredPB_W3} \\
  &
  \phantom{\big\{ \cW_{\frac{1}{2}}^{[a]}(\th)\, , \, \cW_{\frac{1}{2}}^{[b]}(\th')\big\} \, =\,  }
  + \,  \d(\th-\th')\, \cW_0^{\,\pe} \,- \, 2 \, \d'(\th-\th')\, \cW_0 
   \,- \, a\, \frac{2\pi}{k} \, \d(\th-\th') \, \cW_0 \cW_0 \bigg) \, ,\nn
\end{align}
\end{subequations}
This algebra is written in a non-primary basis, as can be seen explicitly in \eqref{nonPrimaryPB1_W3} and \eqref{nonPrimaryPB_W3}. Redefining $\cL$ as in \eqref{W3SugawaraShift}, the correct transformation in \eqref{nonPrimaryPB1_W3} is recovered and the distracting term in \eqref{nonPrimaryPB_W3} is removed as discussed in Section \ref{subsec:non-principalW3}.

\subsection*{Poisson structure of $\cW_4^{(2)}$}

Here we present the full $\cW_4^{(2)}$ algebra after the shift
\eqref{W4SugawaraShift}.  As before, all fields appearing on the
right-hand side are functions of $\th'$ and $\d'(\th-\th') \equiv \pr_\th \d(\th-\th')$.
\begin{subequations}\label{poissonBracketExpW4}
\begin{align}
  &\big\{ \widehat{\cL}(\th)\, , \, \widehat{\cL}(\th') \big\}\, 
  =\
  \d(\th-\th')\, \widehat{\cL}^{\,\pe} 
  \,  - \
  2 \, \d'(\th-\th') \, \widehat{\cL} 
  \, - \, 
  \frac{k}{4\pi}\, \d^{(3)}(\th-\th') \, ,
  \\
  &
  \big\{ \widehat{\cL}(\th)\, , \, \cW_{\ell}^{[a]}(\th') \big\}\, 
  =\, 
  \d(\th-\th')\, \cW_{\ell}^{[a]\,\pe}
   - \, 
  (\ell+1) \, \d'(\th-\th') \, \cW_{\ell}^{[a]} \, , \quad \ell=0,1,2 \, ,
  \\[10pt]
  &\big\{ \cW_0(\th)\, , \, \cW_0(\th')\big\}\, 
  =\, \frac{8k}{3\p} \, \d'(\th-\th') \ , \\
  &
  \big\{ \cW_0(\th)\, , \, \cW_{1}^{[a]}(\th') \big\} \, 
  = \, a\, \frac{16}{3} \, \d(\th-\th') \, \cW_{1}^{[a]} \, ,
  \\[10pt]
  &
  \big\{ \cW_1^{[a]} (\th) \, , \,  \cW_1^{[b]} (\th')\big\}   \, =\, 
  \d_{a+b\, ,\, 0} \,  
  \bigg(\, \frac{k}{4\p}\, \d^{(3)}(\th - \th') \nn
  \\
  &
  \phantom{\cW_1^{(a)} (\th'),}
  - \, \frac{3a}{2}  \, \d''(\th - \th') \, \cW_0
  \, + \, 
  \frac{3a}{2}  \, \d'(\th - \th') \, \cW_0^{\,\pe}
  \, - \, 
  \frac{a}{2}  \, \d(\th - \th') \, \cW_0^{\,\pe\pe} \nn
  \\
  &
  \phantom{\cW_1^{(a)} (\th'),}
  + \,  2 \, \d'(\th - \th') \, \widehat{\cL}
  \, - \, \d(\th - \th') \, \widehat{\cL}^{\,\pe}
  \, - \, 2\, a \,  \d(\th - \th') \, \cW_2 
  \nn\\
  &
  \phantom{\cW_1^{(a)} (\th'),}
  - \, \frac{2\p}{k} \, 2 \, a \, \d(\th - \th') 
  \, \cW_0 \widehat{\cL}
  \, + \
  \frac{2\p}{k} \, \frac{21}{16} 
  \, \d'(\th - \th')  \, \cW_0 \cW_0
  \nn 
  \\
  &
  \phantom{\cW_1^{(a)} (\th'),}
  - \
  \frac{2\p}{k} \, \frac{27}{16} 
  \, \d(\th - \th') \,  \cW_0 \cW_0^{\,\pe}
  \,  - \
  \left(\frac{2\p}{k}\right)^{\!2}  \frac{11a}{16} \, \d(\th - \th') \,  \cW_0 \cW_0 \cW_0
  \,\bigg) \,, \\[10pt]
  &\big\{ \cW_1^{[a]} (\th) \, , \,  \cW_2 (\th') \big\}\, =\, \nn 
  \\
  &
  \phantom{\cW_1^{[a]}}
  - \frac{5a}{3} \, \d''(\th - \th') \, \cW_1^{[-a]}
  \, +  
   \frac{5a}{6} \, \d'(\th - \th') \, \cW_1^{[-a]\,\pe}
  \, + \
  \frac{a}{6} \, \d(\th - \th') \, \cW_1^{[-a]\,\pe\pe}
  \nn\\
  &
  \phantom{\cW_1^{[a]}}
  + \,  
  \frac{2\p}{k} \, a \, \frac{8}{3} \, \d(\th - \th') \, \widehat{\cL}\, \cW_1^{[-a]} 
  \, - \, 
  \frac{2\p}{k} \, \frac{5}{2} \, \d'(\th - \th') \, \cW_0 \cW_1^{[-a]} 
  \nn\\
  &
  \phantom{\cW_1^{[a]}}
  + \, 
  \frac{2\p}{k} \, \frac{3}{2} \, \d(\th - \th') \, \cW_0^{\,\pe} \cW_1^{[-a]} 
  \, + \, 
  \frac{2\p}{k} \, \frac{a}{2} \, \d(\th - \th') \, \cW_0 \cW_1^{[-a]\,\pe}
  \nn\\
  &
  \phantom{\cW_1^{[a]}}
  + \, \left(\frac{2\p}{k}\right)^{\!2}  
  \frac{5a}{4} \, \d(\th - \th') \, \cW_0 \cW_0 \cW_1^{[-a]} \, ,
  \displaybreak[0] \\[10pt]
  &\big\{ \cW_2 (\th) \, , \,  \cW_2 (\th')\big\} \, =\, 
  \frac{k}{48\pi}  \, \d^{(5)}(\th - \th') \nn
  \\
  &
  \phantom{\cW_2 \, =\, }
  + \, \frac{5}{6} \, \d^{(3)}(\th - \th') \, \widehat{\cL} 
  \, - \, 
  \frac{5}{4} \,  \d''(\th - \th') \, \widehat{\cL}^{\,\pe} 
  \, + \, 
  \frac{3}{4} \, \d'(\th - \th') \, \widehat{\cL}^{\,\pe\pe} \, - \,  
  \frac{1}{6} \,  \d(\th - \th') \, \widehat{\cL}^{(3)}   
  \nn
  \\
  &
  \phantom{ \cW_2 \, =\, }
  + \, \frac{2\pi}{k}\, \, \frac{8}{3} 
  \, \d'(\th - \th') \, \widehat{\cL} \widehat{\cL} 
  \, - \, 
  \frac{2\pi}{k} \, \frac{8}{3} \, \d(\th - \th') \, \widehat{\cL} \widehat{\cL}^{\,\pe}
  \nn
  \\
  &
  \phantom{\cW_2 \, =\, }
  + \, \frac{2\p}{k}\, \frac{5}{64} \, \d^{(3)}(\th - \th') \, \cW_0 \cW_0
  \, - \, 
  \frac{2\p}{k}\, \frac{15}{64} \, \d''(\th - \th') \, \cW_0 \cW_0'
  \nn \\
  &
  \phantom{\cW_2 \, =\, }
  + \, 
  \frac{2\p}{k}\,  \frac{9}{64} \, \d'(\th - \th') 
  \left(\, \cW_0' \cW_0' \, + \, \cW_0 \cW_0'' \,\right)
  \, - \,  
  \frac{2\p}{k}\,   \frac{1}{32} \,  \d(\th - \th') 
  \left(\,3 \, \cW_0' \cW_0'' \, + \, \cW_0 \cW_0''' \,\right)
  \nn
  \\
  &
  \phantom{\cW_2 \, =\, }
  + \, \frac{2\pi}{k}\, \, \frac{1}{2} 
  \, \d'(\th - \th') \, \widehat{\cL} \, \cW_0 \cW_0 
  \, + \, \left(\frac{2\pi}{k}\right)^3  \frac{3}{128} 
  \, \d'(\th - \th') \, \cW_0 \cW_0 \cW_0 \cW_0 
  \nn
  \\
  &
  \phantom{\cW_2 \, =\, }
  - \, 
  \left( \frac{2\pi}{k} \right)^2 \frac{1}{4} \,   \d(\th - \th')  \,
  \left( 2 \, \widehat{\cL} \, \cW_0 \cW_0' \, + \,  \widehat{\cL}^{\,\pe} \,\cW_0 \cW_0    \right)
  \nn \\
  &
  \phantom{\cW_2 \, =\, }
  - \, \left(\frac{2\pi}{k}\right)^3  \frac{3}{64} 
  \, \d(\th - \th') \, \cW_0 \cW_0 \cW_0 \cW_0'
  \nn
  \\
  &
  \phantom{\cW_2 \, =\, }
  + \, 
  \frac{2\pi}{k}\, 8 \, \d'(\th - \th')
  \, \cW_1^{[-1]} \cW_1^{[1]} \, 
  \, - \
  \frac{2\pi}{k}\, 4 \,  \d(\th - \th') \, \left(\cW_1^{[-1]} \cW_1^{[1]} \right)' 
  \, . 
\end{align}
\end{subequations}

\subsection*{Poisson structure of $\cW_\infty[\l]$: the first brackets}

Here we present the first Poisson brackets of $\cW_\infty[\l]$,
i.e. the $\{\cW_i , \cW_j\}$ with $i,j < 4$. As in the previous
examples all fields appearing on the right-hand side are functions of
$\th'$ and $\d'(\th-\th') \equiv \pr_\th \d(\th-\th')$. Moreover,
$N_\ell$ denotes the normalisation factor \eqref{N_l}.
\begin{subequations}\label{poissonBracketExp}
\begin{align}
  &\big\{ \cL(\th) \,,\, \cL(\th') \big\}\, 
  =\, \d(\th-\th') \, \cL'
  - 2 \, \d'(\th-\th') \, \cL \,-\, \frac{k}{4\pi}\, \d^{(3)}(\th-\th') \, , \\[15pt]
  &\big\{ \cL(\th)\, , \, \cW_\ell(\th')\big\}\, 
  =\, \d(\th-\th') \, \cW_\ell^{\,\pe} 
  \, -\,  (\ell+1) \, \d'(\th-\th') \, \cW_\ell\ ,  \quad \ell>1 \, , \\[15pt]
  &\big\{ \cW_2(\th)\, , \, \cW_2(\th')\big\} \,
    =\, -\frac{2N_3}{\left(N_2\right)^2}\, \big[\, \d(\th-\th') \, \cW_3^{\,\pe} \, 
    -\, 2 \, \d'(\th-\th') \, \cW_3 \,\big] \nn \\
  & -\, \frac{1}{12 N_2}\,
    \big[\, 2\, \d(\th-\th')\, \cL^{(3)}  
    -\, 9\, \d'(\th-\th')\, \cL'' 
    \, +\,15\, \d''(\th-\th')\, \cL' 
    -\, 10\, \d^{(3)}(\th-\th')\, \cL
      \,\big]\nn\\
  &  -\, \frac{16\pi}{3kN_2}\, \big[\, \d(\th-\th')\, \cL \cL'
    -\, \d'(\th-\th')\, \cL^2
      \,\big] \,-\, \frac{k}{48\pi N_2}\, \d^{(5)}(\th-\th') \, ,
    \\[15pt]
  &\big\{ \cW_2(\th)\, , \, \cW_3(\th') \big\} \,=\, -\, \frac{N_4}{N_2 N_3}\, 
    \big[\,2\, \d(\th-\th') \, \cW_4^{\,\pe} 
    \,-\, 5 \, \d'(\th-\th') \, \cW_4\,\big]  \nn\\
  & -\,\frac{1}{15 N_2}\, \big[\, \d(\th-\th')\, \cW_2^{(3)}
    \,-\,6\, \d'(\th-\th')\, \cW_2^{\,\pe\pe} \,
    +\,14\, \d''(\th-\th')\, \cW_2^{\,\pe} 
    \,-\, 14\,  \d^{(3)}(\th-\th')\, \cW_2
    \,\big]\nn\\
  & -\, \frac{4\pi}{15k N_2}\, 
    \big[\, 25\,  \d(\th-\th')\, \cW_2\,  \cL' \,+\,18\,  \d(\th-\th')\, \cL\,  \cW_2^{\,\pe}
    \,-\,52\, \d'(\th-\th')\, \cL\, \cW_2
    \,\big] \, , \displaybreak[0] \\[15pt]
  &\big\{ \cW_3(\th)\, , \, \cW_3(\th')\big\} \, =\, -\, \frac{3N_5}{\left(N_3\right)^2}\,
    \big[\, \d(\th-\th') \, \cW_5^{\,\pe} \,-\, 2 \, 
      \d'(\th-\th') \, \cW_5 \,\big]  \nn\\
  & +\, \frac{(\l^2-19)}{30 N_3}\,
    \big[\, \d(\th-\th')\, \cW_3^{(3)}
    -\,5\, \d'(\th-\th')\, \cW_3^{(2)} +\,9\,  \d''(\th-\th')\, \cW_3^{\,\pe} 
    -\, 6\, \d^{(3)}(\th-\th')\, \cW_3 \,\big]\nn\\[5pt]
  & -\,\frac{1}{360 N_3}\,
    \big[\,3\,  \d(\th-\th')\, \cL^{(5)}
      -\, 20\, \d'(\th-\th')\, \cL^{(4)} +\, 56\, \d''(\th-\th')\,  \cL^{(3)} -\,84\, 
    \d^{(3)}(\th-\th')\,  \cL'' \nn\\
  & \phantom{-\,\frac{1}{360 N_3}\,
    \big[} +\, 70\, \d^{(4)}(\th-\th')\, \cL'
    -\, 28\, \d^{(5)}(\th-\th')\, \cL \,\big]\nn\\[5pt]
  & +\,\frac{2\pi(29\l^2-284)}{15k N_3}\,
    \big[\, \d(\th-\th')\, \cW_2\cW_2^{\,\pe}
    \,-\, \d'(\th-\th') \left( \cW_2\right)^2
    \,\big]\nn\\[5pt]
  & -\,\frac{\pi}{90k N_3}\, 
    \big[\,177\,  \d(\th-\th')\, \cL'\cL^{(2)} \,+\,78\,  \d(\th-\th')\, \cL\cL^{(3)} 
     -\, 295\, \d'(\th-\th')  \left(\cL'\right)^2 \nn \\
  &\phantom{-\,\frac{\pi}{90k N_3}\, 
    \big[}
   -\, 352\, \d'(\th-\th')\,  \cL\cL''  +\,588\, \d''(\th-\th')\, \cL \cL'
    \,-\,196\, \d^{(3)}(\th-\th')\, \cL^2 \,\big]\nn\\[5pt]
  & +\, \frac{28\pi(\l^2-19)}{15k N_3}\,
    \big[\,\d(\th-\th')\, \cW_3\cL' \,+\, \d(\th-\th')\, \cL\cW_3^{\,\pe}
    \,-\,2\, \d'(\th-\th')\, \cL\cW_3
    \big]\nn\\[5pt]
   & -\, \frac{32\pi^2}{5k^2 N_3}\,
    \big[\,3\, \d(\th-\th')\, \cL^2 \cL' 
    \,-\, 2\, \d'(\th-\th')\, \cL^3
    \,\big]
   \,-\,  \frac{k}{1440\pi N_3}\, \d^{(7)}(\th-\th')  \, . 
\end{align}
\end{subequations}

\end{appendix}

\newpage


\end{document}